\documentclass[%
 reprint,
nofootinbib,
 amsmath,amssymb,
 aps,
floatfix,
]{revtex4-2}

\usepackage{amsmath,amssymb}
\usepackage[scr=boondox]{mathalpha}
\usepackage{amsfonts}
\usepackage{physics}
\usepackage{indentfirst}
\usepackage{graphicx}
\usepackage[caption=false]{subfig}
\usepackage[colorlinks=true, allcolors=blue]{hyperref}
\usepackage[normalem]{ulem}
\usepackage{multirow}
\usepackage{diagbox}
\usepackage{makecell}
\usepackage{adjustbox}
\usepackage{booktabs}
\usepackage{tikz}

\usepackage{dcolumn}% Align table columns on decimal point
\usepackage{bm}% bold math
\begin{document}

\preprint{APS/123-QED}

\title{Precise Determination of the Long-Time Asymptotics of the Diffusion Spreadability of Two-Phase Media}% Force line breaks with \\

\author{Shaobing Yuan}

\affiliation{%
Department of Chemistry and Princeton Institute for the Science and Technology of Materials, Princeton University, Princeton, New Jersey, 08544, USA
}%

\author{Salvatore Torquato}%
 \email{torquato@electron.princeton.edu }
\affiliation{
Department of Chemistry, Department of Physics, Princeton Center for Theoretical Science, Princeton Institute for the Science and Technology of Materials, and Program in Applied and Computational Mathematics, Princeton University, Princeton, New Jersey, 08544 USA% with \\
}%
% \affiliation{
%  Third institution, the second for Charlie Author
% \affiliation{%
%  Authors' institution and/or address\\
%  This line break forced with \textbackslash\textbackslash
% }%

% \collaboration{CLEO Collaboration}%\noaffiliation

\date{\today}% It is always \today, today,

\begin{abstract}
The time-dependent diffusion spreadability $\mathcal{S}(t)$ is a powerful dynamical probe of the microstructure of two-phase heterogeneous media across length scales [Torquato, S., \emph{Phys. Rev. E.}, 104 054102 (2021)]. The spreadability can be exactly represented as a
certain functional of the spectral density  $\tilde{\chi}_{_V}(\mathbf{k})$, where $\bf k$ is the wave vector. 
Experimentally, it is intimately related to nuclear magnetic resonance (NMR) measurements in fluid-saturated media.
The short-, intermediate-, and long-time behavior of the spreadability reflects structural features at small, intermediate, and large length scales, respectively. It has been shown that when the spectral density takes the power-law form $\tilde{\chi}_{_V}(\mathbf{k})\sim |\mathbf{k}|^\alpha$ as the wavenumber $|\mathbf{k}|$ tends to zero, the normalized excess spreadability $\mathscr{s}^{ex}(t)$ [proportional to $\mathcal{S}(\infty)-\mathcal{S}(t)$] scales as $\mathscr{s}^{ex}(t)\sim t^{-\frac{d+\alpha}{2}}$ in the long-time limit $t\to\infty$, enabling one to determine the infinite-wavelength scaling exponent $\alpha$.
An algorithm that allows one to reliably extract the exponent $\alpha$ from long-time spreadability data was previously devised [Wang, H., Torquato, S., \emph{Phys. Rev. Appl.}, 17 034022 (2022)].
In this paper, we further improve this procedure to obtain $\alpha$ even more accurately by incorporating higher-order correction terms to the long-time asymptotics and by utilizing analyticity properties of $\tilde{\chi}_{_V}(k)$ at the origin. We illustrate our procedure by analyzing hyperuniform ($\alpha> 0$), typical nonhyperuniform ($\alpha=0$), and antihyperuniform ($-d < \alpha <0$) models of two-phase media.
In addition, by combining the large-$t$ asymptotic expansion of $\mathscr{s}^{ex}(t)$ with the small-$t$ expansion, we have devised a two-point Pad\'e approximant to accurately approximate $\mathscr{s}^{ex}(t)$ for all $t$ with just a few parameters.
Our findings facilitate the characterization of the microstructure of two-phase media across length scales as obtained from numerical spreadability data or experimental data obtained from NMR relaxation measurements.
Our work can also be applied in the inverse design of two-phase microstructures with targeted spreadability behaviors.
\end{abstract}

\maketitle

\section{Introduction}\label{sec:intro}

Time-dependent interphase diffusion processes in heterogeneous media play an important role in a myriad of contexts, including MRI \cite{wedeenMappingComplexTissue2005a,novikovRevealingMesoscopicStructural2014a,leeVivoObservationBiophysical2020,novikovQuantifyingBrainMicrostructure2019,wuTimeDependentDiffusionMRI2022}, geological media \cite{torquatoRandomHeterogeneousMaterials2002}, biological cells \cite{brownsteinImportanceClassicalDiffusion1979}, chemical catalysis \cite{sosnaProbingLocalDiffusion2020}, material design \cite{tahmasebiAccurateModelingEvaluation2018}, and drug delivery \cite{langerPresentFutureApplications1981}. Multiphase heterogeneous media are ubiquitous in nature and synthetic materials; examples include, among others, composites, foams, gels, porous media, and biological tissues \cite{torquatoRandomHeterogeneousMaterials2002,torquatoMorphologyEffectiveProperties1998,klattPhoamtonicDesignsYield2019,huangCircularSwimmingMotility2021,Sa03}. The effective properties of heterogeneous media as a whole, including transport properties like the effective conductivity, depend on and reflect the underlying microstructure, which is fully statistically characterized by an infinite set of $n$-point correlation functions \cite{torquatoRandomHeterogeneousMaterials2002, senEffectiveConductivityAnisotropic1989,torquatoNonlocalEffectiveElectromagnetic2021} defined in Sec. \ref{sec:background}.

Torquato developed the concept of the time-dependent \emph{diffusion spreadability} $\mathcal{S}(t)$ \cite{torquatoDiffusionSpreadabilityProbe2021}, making a direct link between the interphase diffusion problem and the microstructure of the system across length scales. Consider the time-dependent mass transfer of a solute in a two-phase medium where phases 1 and 2 occupy volume fractions $\phi_1$ and $\phi_2$, respectively, under a diffusion coefficient $D$ that is constant at any time $t$ in either phase, such that all of the solute is uniformly distributed in phase 2 at an initial time $t=0$. The mass fraction of the total solute that has diffused from phase 2 to phase 1 at time $t$ is defined as the \emph{diffusion spreadability} $\mathcal{S}(t)$. Torquato provided the exact formula for $\mathcal{S}(t)$ in any dimension $d$ in terms of the autocovariance function $\chi_{_V}(\mathbf{r})$ in direct space (Generalizing Prager's 3D solution \cite{pragerInterphaseTransferStationary1963}), or equivalently, the spectral density $\tilde{\chi}_{_V}(\mathbf{k})$ in Fourier space, as follows:
\begin{equation}\label{eq:excess-spreadability}
        \begin{split}
            \mathscr{s}^{ex}(t)&\equiv1-\frac{\mathcal{S}(t)}{\mathcal{S}(\infty)}=\phi_1^{-1}[\mathcal{S}(\infty)-\mathcal{S}(t)]\\
            &=\frac{1}{(4\pi D t)^{d/2}}\int_{\mathbb{R}^d}\frac{\chi_{_V}(\mathbf{r})}{\phi_1\phi_2}\mathrm{e}^{-\frac{r^2}{4Dt}}\mathrm{d}\mathbf{r}\\
            &=\frac{1}{(2\pi)^d}\int_{\mathbb{R}^d}\frac{\tilde{\chi}_{_V}(\mathbf{k})}{\phi_1\phi_2}\mathrm{e}^{-k^2Dt}\mathrm{d}\mathbf{k},
        \end{split}
\end{equation}
where $\mathcal{S}(\infty)-\mathcal{S}(t)$ is called the \emph{excess spreadability}, i.e., the spreadability in excess of its infinite-time limit $\mathcal{S}(\infty)=\phi_1$, and $\mathscr{s}^{ex}(t)$ is the \emph{normalized excess spreadability}, which monotonically decreases from $1$ to $0$ as time $t$ goes from $0$ to $\infty$.
This is one of the rare cases where the effective transport property is exactly determined by up to two-point correlations rather than the whole infinite set of correlation functions. Equation \eqref{eq:excess-spreadability} implies that small-, intermediate-, and long-time behaviors of $\mathcal{S}(t)$ are equivalent to small-, intermediate-, and large-scale features of the microstructure \cite{torquatoDiffusionSpreadabilityProbe2021,wangDynamicMeasureHyperuniformity2022}. 

Torquato \cite{torquatoDiffusionSpreadabilityProbe2021} showed that, assuming a power-law scaling of the spectral density $\tilde{\chi}_{_V}(\mathbf{k})\sim B|\mathbf{k}a|^\alpha$ at the infinite-wavelength limit $|\mathbf{k}|\to0$, the long-time excess spreadability scales as
\begin{equation}\label{eq:asym-pw}
        \begin{split}
            \mathcal{S}(\infty)-\mathcal{S}(t)=\frac{B\Gamma(\frac{d+\alpha}{2})}{(4\pi)^{d/2}\Gamma(d/2)\phi_2}(Dt/a^2)^{-\frac{d+\alpha}{2}}\\+o[(Dt/a^2)^{-\frac{d+\alpha}{2}}],\hspace{2em}t\to\infty,
        \end{split}
\end{equation}
where $a$ is some characteristic heterogeneity length scale of the two-phase random media.
The value of the exponent $\alpha$ distinguishes nonhyperuniform systems ($\alpha\le0$) from hyperuniform ones ($\alpha>0$).
Disordered hyperuniform systems are of particular importance because of their anomalous suppression of volume fraction fluctuations relative to ordinary disordered media \cite{torquatoLocalDensityFluctuations2003a,zacharyHyperuniformityPointPatterns2009,torquatoHyperuniformStatesMatter2018a}, which endows them with novel properties of fundamental interest and practical use, relevant contexts including glass formation \cite{marcotteNonequilibriumStaticGrowing2013,coniglioUniversalBehaviourGlass2017,zhangApproachHyperuniformityMetallic2023,wangHyperuniformDisorderedSolids2025,mitraHyperuniformityCyclicallyDriven2021}, maximally random jammed sphere packings \cite{atkinsonCriticalSlowingHyperuniformity2016,rissoneLongRangeAnomalousDecay2021,torquatoStructuralCharacterizationManyparticle2021,torquatoJammingThresholdSphere2007,donevUnexpectedDensityFluctuations2005,dreyfusDiagnosingHyperuniformityTwodimensional2015,daleHyperuniformJammedSphere2022}, photonic and phononic materials \cite{florescuDesignerDisorderedMaterials2009a,froufe-perezBandGapFormation2017,manPhotonicBandGap2013,klattWavePropagationBand2022,christogeorgosExtraordinaryDirectiveEmission2021,aubryExperimentalTuningTransport2020,gkantzounisHyperuniformDisorderedPhononic2017}, mathematics \cite{torquatoHiddenMultiscaleOrder2019,brauchartHyperuniformPointSets2019,brauchartHyperuniformPointSets2020,torquatoUncoveringMultiscaleOrder2018,bjorklundHyperuniformityRandomMeasures2026,bylehnHyperuniformityRegularTrees2025,ghoshGeneralizedStealthyHyperuniform2018}, biology and ecology \cite{jiaoAvianPhotoreceptorPatterns2014,mayerHowWelladaptedImmune2015,zhengHyperuniformityDensityFluctuations2020,huCausesConsequencesDisordered2025,houVegetationPatternsStructures2026}, self-organization \cite{hexnerEnhancedHyperuniformityRandom2017,hexnerHyperuniformityCriticalAbsorbing2015a,maHyperuniformityGeneralizedRandom2019}, disordered classical ground states \cite{torquatoEnsembleTheoryStealthy2015,zacharyAnomalousLocalCoordination2011,gangulyGroundStatesHyperuniformity2020,torquatoExistenceNonequilibriumGlasses2025,dawleyEvolutionVariousInitial2025,morseGeneratingLargeDisordered2023,chenDisorderedHyperuniformSolid2023}, and quantum systems \cite{torquatoPointProcessesArbitrary2008a,crowleyQuantumCriticalityIsing2019,abreuWeylHeisenbergEnsemble2017,wangCorrelationsInteractingElectron2024,chenAnomalousSuppressionLargescale2025,deglinnocentiHyperuniformDisorderedTerahertz2016,sakaiQuantumPhaseTransition2022,vanoniQuantifyingWhenHyperuniformity2025}. 

\begin{figure}
    \centering
    \includegraphics[width=\linewidth]{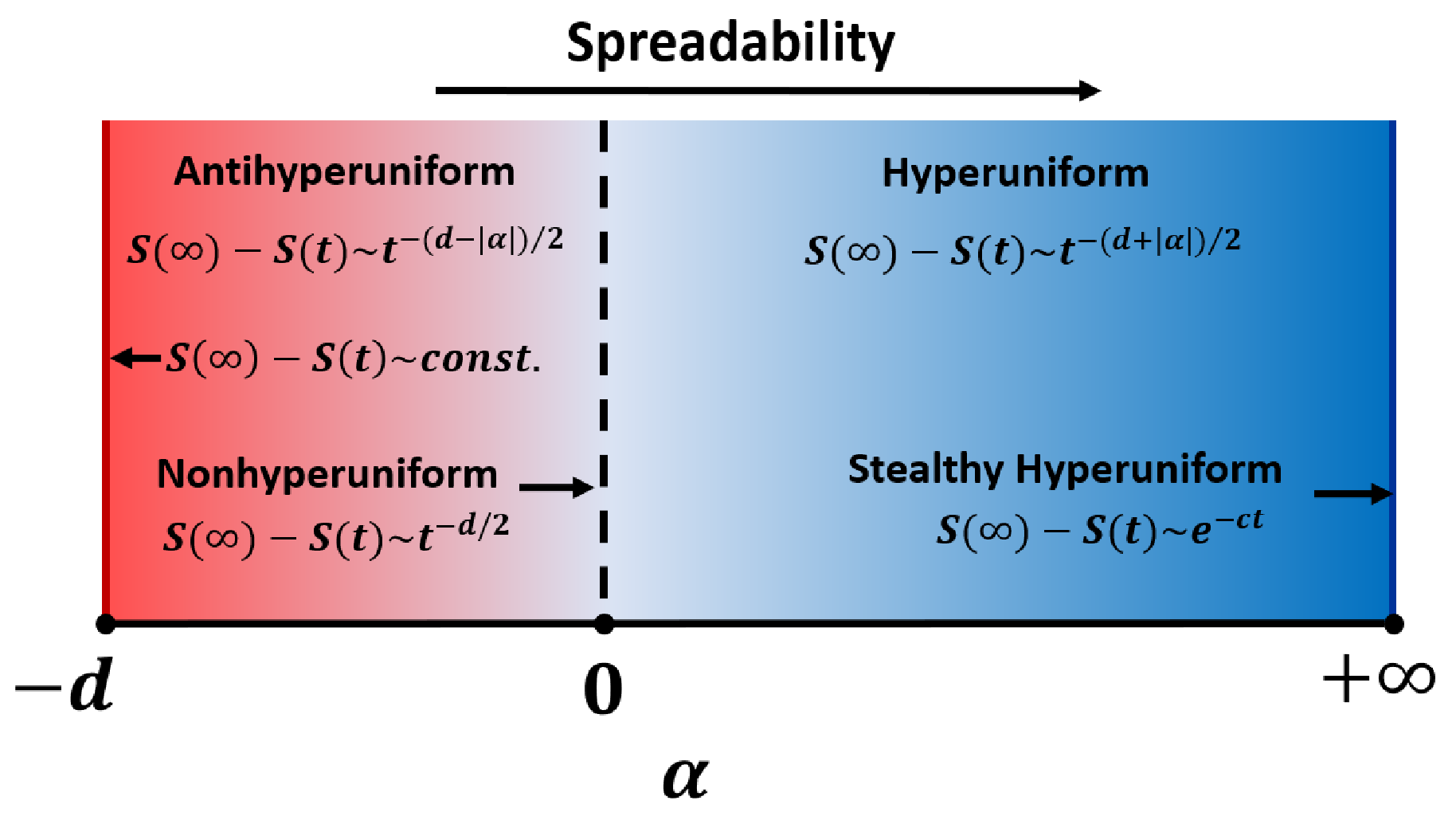}
    \caption{“Phase diagram” that schematically shows the spectrum of spreadability regimes in terms of the long-distance scaling exponent $\alpha$. As $\alpha$ increases from the extreme antihyperuniform limit of $\alpha\to-d$, the spreadability decay rate at long times decays faster, that is, the excess spreadability follows the inverse power law $1/t^{(d+\alpha)/2}$, except when $\alpha\to\infty$, which corresponds to stealthy hyperuniform media with a decay rate that is exponentially fast. This figure is reproduced from the one presented in Ref. \cite{torquatoDiffusionSpreadabilityProbe2021}}
    \label{fig1}
\end{figure}

Figure \ref{fig1} schematically shows the ``phase diagram'' of spreadability regimes in terms of the exponent $\alpha$ \cite{torquatoDiffusionSpreadabilityProbe2021}. The long-time excess spreadability for antihyperuniform media ($-d < \alpha < 0$) has the slowest decay among all translationally invariant media. A long-time decay rate of $\mathscr{s}^{ex}(t) \sim t^{-d/2}$ corresponds to a ``typical" nonhyperuniform medium in which the spectral density is a bounded positive number at the origin, i.e., $\alpha=0$. The excess spreadability for nonstealthy hyperuniform media ($0<\alpha<\infty$) has the aforementioned power-law decay $\mathscr{s}^{ex}(t)\sim t^{-(d+\alpha)/2}$. The limit $\alpha\rightarrow +\infty$ corresponds to media in which the decay rate of $\mathscr{s}^{ex}(t)$ is faster than any inverse power law, which is the case for stealthy hyperuniform media, disordered or ordered \cite{torquatoDiffusionSpreadabilityProbe2021}. The reader is referred to Sec. \ref{sec:background} for terminology.

The concept of diffusion spreadability is intimately
related to magnetic relaxation measurements, as demonstrated by Torquato \cite{torquatoDiffusionSpreadabilityProbe2021}. He identified precise mappings between the long-time asymptotic formulas for the spreadability and the nuclear magnetic resonance (NMR) pulsed field gradient spin-echo (PFGSE) amplitude via a diffusion propagator that also only involves the two-point correlation function \cite{mitraDiffusionPropagatorProbe1992a} as well as diffusion-weighted magnetic resonance imaging (dMRI) measurements in biological tissue  \cite{novikovRevealingMesoscopicStructural2014a,leeVivoObservationBiophysical2020}.
Furthermore,  Wilken \textit{et al.} \cite{wilkenSpatialOrganizationPhaseSeparated2023} showed that phase-separated DNA droplets spontaneously form hyperuniform structures and suggested that the long-time diffusion spreadability behaviors can be exploited in chemical reaction schemes analogous to those present in biomolecular condensates.
To better model real experimental systems, Skolnick and Torquato \cite{skolnickSimulatedDiffusionSpreadability2023} generalized the spreadability problem by considering different phase diffusion coefficients and showed that when time is rescaled by the effective diffusion coefficient $D_e$, the simulated spreadability for all times closely follows the spreadability curve for equal diffusion coefficients and hence is effectively controlled by the two-point correlation function only.

Apart from its connection to the NMR and other experiments, the spreadability has been shown to be a powerful diagnostic to characterize the large-scale structure of both hyperuniform and nonhyperuniform media by determining the scaling exponent $\alpha$. Such examples include the analysis of various hard particle packing problems \cite{maherCharacterizationVoidSpace2022,maherHyperuniformityMaximallyRandom2023,maherHyperuniformityScalingMaximally2024a}, the conserved directed percolation problem \cite{wieseHyperuniformityMannaModel2024}, and certain quasicrystalline systems \cite{hitin-bialusHyperuniformityClassesQuasiperiodic2024}.
The spreadability offers a superior means
to extract  $\alpha$ accurately and robustly from fitting its long-time asymptotics compared to fitting noisy scattering data, since the spreadability can be regarded as a Gaussian smoothing of the spectral density, as indicated by Eq. \eqref{eq:excess-spreadability}, and is thus less susceptible to pointwise variations in spectral density $\tilde{\chi}_{_V}(k)$ \cite{maherCharacterizationVoidSpace2022}.

Wang and Torquato \cite{wangDynamicMeasureHyperuniformity2022} developed a systematic spreadability fitting algorithm that
enables one to reliably extract the scaling exponent $\alpha$ in the asymptotic power law \eqref{eq:asym-pw} from the long-time data of $\mathscr{s}^{ex}(t)$, which can be further improved to estimate $\alpha$ even more accurately.
The precision of the Wang-Torquato algorithm depends on whether the true long-time behavior is achieved; the larger the time $t$, the closer one is to the asymptotic power law \eqref{eq:asym-pw}, and hence the more accurate is the estimate of the exponent $\alpha$.
Furthermore, since the residual $o[(Dt/a^2)^{-\frac{d+\alpha}{2}}]$ does not change the sign at large $t$, the fitting procedure inevitably leads to a systematic error in the value of the exponent $\alpha$. 
In this paper, we demonstrate how we can enhance the fitting precision by incorporating higher-order terms in the large-$t$ asymptotic expansion of $\mathscr{s}^{ex}(t)$ in addition to exploiting the mathematical properties of $\tilde{\chi}_{_V}(\mathbf{k})$ and $\chi_{_V}(\mathbf{r})$.
Additionally, we obtain two-point Pad\'e approximants by combining the large-$t$ fits of the asymptotic expansion of $\mathscr{s}^{ex}(t)$ with the small-$t$ expansion, and show that they provide a good approximation of $\mathscr{s}^{ex}(t)$ for all times $t$.

The paper is organized as follows. In Sec. \ref{sec:background}, we provide basic definitions and preliminaries. We then demonstrate the mathematical properties and the physical implications of the large-$t$ asymptotic expansion of the spreadability $\mathscr{s}^{ex}(t)$ (Sec. \ref{sec:theory}). Section \ref{sec:fitting} presents how the fitting is improved based on the theory. Its applications to a few representative models are shown in Sec. \ref{sec:results}. Finally, we discuss the implications of the results and conclude in Sec. \ref{sec:discussion-and-conclusion}.

\section{Background}\label{sec:background}

\subsection{Correlation Functions}

A two-phase  random medium is a domain of space $\mathcal{V} \subseteq \mathbb{R}^d$ of volume $V$ that is partitioned into two disjoint regions that make up  $\mathcal{V}$:
a phase 1 region $\mathcal{V}_1$ of volume fraction $\phi_1$ and a phase 2 region $\mathcal{V}_2$ of volume fraction $\phi_2$ \cite{torquatoRandomHeterogeneousMaterials2002}. 
The microstructure of a two-phase medium is fully statistically characterized by an infinite set of $n$-point correlation functions \cite{torquatoRandomHeterogeneousMaterials2002, senEffectiveConductivityAnisotropic1989,torquatoNonlocalEffectiveElectromagnetic2021} defined by
\begin{equation}
    \begin{split}
        S_n^{(i)}(\mathbf{x}_1,...,\mathbf{x}_n)\equiv\expval{\mathcal{I}^{(i)}(\mathbf{x}_1)...\mathcal{I}^{(i)}(\mathbf{x}_n)},\\i=1,2,\hspace{1em}n=1,2,3,...,
    \end{split}
\end{equation}
where $\mathcal{I}^{(i)}(\mathbf{x})$ is the \emph{indicator function} for phase $i=1,2$ defined as
\begin{equation}
{\cal I}^{(i)}({\bf x}) = \left\{
{\begin{array}{*{20}c}
{1, \quad\quad {\bf x} \in {\cal V}_i,}\\
{0, \quad\quad {\bf x} \notin {\cal V}_i},
\end{array} }\right.
\label{phase-char}
\end{equation}
for a given realization, and angular brackets denote an ensemble average.
Physically, the value of $S_n^{(i)}(\mathbf{x}_1,...,\mathbf{x}_n)$ corresponds to the probability of finding the vertices of a polyhedron located at $\mathbf{x}_1,...,\mathbf{x}_n$ all in phase $i$.
Assuming statistical translational symmetry, $S_n^{(i)}(\mathbf{x}_1,...,\mathbf{x}_n)$ depends only on the relative positions, i.e., $S_n^{(i)}(\mathbf{x}_1,\mathbf{x}_2,...,\mathbf{x}_n)=S_n^{(i)}(\mathbf{x}_{12},...,\mathbf{x}_{1n})$, where $\mathbf{x}_{ij}=\mathbf{x}_j-\mathbf{x}_i$.
Particularly, we have $S_1^{(i)}(\mathbf{x}_1)=\expval{\mathcal{I}^{(i)}(\mathbf{x}_1)}=\phi_i$.

In this paper, our central focus is on the two-point correlation function $S_2^{(i)}(\mathbf{r})$, which exactly determines $\mathscr{s}^{ex}(t)$ via formula \eqref{eq:excess-spreadability}.
The \emph{autocovariance} function $\chi_{_V}(\mathbf{r})$ is an equivalent expression of the two-point correlation function $S_2^{(i)}(\mathbf{r})$ defined by $\chi_{_V}(\mathbf{r})\equiv S_2^{(1)}(\mathbf{r})-\phi_1^2=S_2^{(2)}(\mathbf{r})-\phi_2^2$.
The nonnegative \emph{spectral density} $\tilde{\chi}_{_V}(\mathbf{k})$, which can be obtained from scattering experiments \cite{debyeScatteringInhomogeneousSolid1949,debyeScatteringInhomogeneousSolid1957}, is the Fourier transform of $\chi_{_V}(\mathbf{r})$ at wave vector $\mathbf{k}$, i.e., $\tilde{\chi}_{_V}(\mathbf{k})=\int_{\mathbb{R}^d}\chi_{_V}(\mathbf{r})\mathrm{e}^{-\mathrm{i\mathbf{k}\cdot\mathbf{r}}}\mathrm{d}\mathbf{r}\ge0$.
Without loss of generality, we can always define the isotropic radial functions $\chi_{_V}(r)$ and $\tilde{\chi}_{_V}(k)$ by averaging the vector-dependent $\chi_{_V}(\mathbf{r})$ and $\tilde{\chi}_{_V}(\mathbf{k})$ over all angles, i.e.,  $\chi_{_V}(r)=\frac{1}{\Omega}\int_{\Omega}\chi_{_V}(\mathbf{r})\mathrm{d}\Omega$, $\tilde{\chi}_{_V}(k)=\frac{1}{\Omega}\int_{\Omega}\tilde{\chi}_{_V}(\mathbf{k})\mathrm{d}\Omega$ \cite{torquatoDiffusionSpreadabilityProbe2021}.

\subsection{Classification of Hyperuniform and Nonhyperuniform Media}

For two-phase heterogeneous media in $d$-dimensional Euclidean space $\mathbb{R}^d$, \emph{hyperuniformity} is defined by the following infinite-wavelength condition on the spectral density ${\tilde \chi}_{_V}({\bf k})$ \cite{zacharyHyperuniformityPointPatterns2009, torquatoHyperuniformStatesMatter2018a}, i.e., $\lim\limits_{|{\bf k}|\to 0 }{\tilde \chi}_{_V}({\bf k}) = 0$, which corresponds to $\alpha>0$. Hyperuniformity generalizes the traditional notion of long-range order in many-particle systems to not only include all perfect crystals and perfect quasicrystals, but also exotic amorphous states of matter \cite{torquatoLocalDensityFluctuations2003a,torquatoHyperuniformStatesMatter2018a}.

The hyperuniformity concept has led to a unified means of classifying equilibrium and nonequilibrium states of matter, whether hyperuniform or not, according to their large-scale fluctuation characteristics \cite{torquatoDiffusionSpreadabilityProbe2021,wangDynamicMeasureHyperuniformity2022}. The hyperuniform two-phase media  \cite{zacharyHyperuniformityPointPatterns2009,torquatoHyperuniformStatesMatter2018a} are further divided into three different scaling regimes (classes) based on the associated large-$R$ behaviors of the volume-fraction variance determined by the value of the scaling exponent $\alpha$:
\begin{align}  
\sigma^2_{_V}(R) \sim 
\begin{cases}
R^{-(d+1)}, \quad\quad\quad \alpha >1 \qquad &\text{(Class I)}\\
R^{-(d+1)} \ln R, \quad \alpha = 1 \qquad &\text{(Class II)}\\
R^{-(d+\alpha)}, \quad 0 < \alpha < 1\qquad  &\text{(Class III).}
\end{cases}
\label{eq:classes}
\end{align}
Classes I and III are the strongest and weakest forms of hyperuniformity, respectively. Class I media with $\alpha\to\infty$, such as crystals, are known to be stealthy hyperuniform for possessing zero-scattering intensity for a set of wavevectors around the origin \cite{torquatoDisorderedHyperuniformHeterogeneous2016b}, i.e.,
\begin{equation}
{\tilde \chi}_{_V}({\bf k})=0 \qquad \mbox{for}\; 0 \le |{\bf k}| \le K.
\label{stealth}
\end{equation}
By contrast, for any nonhyperuniform two-phase system, the local variance has the following large-$R$ scaling behaviors \cite{torquatoStructuralCharacterizationManyparticle2021}:
\begin{align}  
\begin{split}
\sigma^2_{_V}(R) \sim \begin{cases}
R^{-d}, &\alpha =0 \, \text{(typical nonhyperuniform)}\\
R^{-(d+\alpha)}, &-d <\alpha < 0 \, \text{ (antihyperuniform)}.\\
\end{cases}
\end{split}
\label{sigma-nonhyper}
\end{align}
For a  ``typical" nonhyperuniform system, ${\tilde \chi}_{_V}(0)=B>0$ is positive and bounded \cite{torquatoHyperuniformStatesMatter2018a}. In antihyperuniform systems,
$\lim\limits_{|{\bf k}| \to 0} {\tilde \chi}_{_V}({\bf k})=+\infty$ is unbounded \cite{torquatoStructuralCharacterizationManyparticle2021},
and hence are diametrically opposed to hyperuniform systems. Prototypical examples of ``typical" nonhyperuniform and antihyperuniform systems involve equilibrium hard spheres and systems at thermal critical points (e.g., liquid-vapor and magnetic critical points) \cite{stanleyIntroductionPhaseTransitions1987,binneyTheoryCriticalPhenomena1992}, respectively.
The ``phase diagram" with respect to the exponent $\alpha$ in Fig. \ref{fig1} summarizes this classification.

\subsection{Long-Time Asymptotic Expansion of $\mathscr{s}^{ex}(t)$}

Suppose all of the moments $M_n(\chi_{_V})=\int_0^\infty r^n\chi_{_V}(r)\mathrm{d}r$ exist, Torquato \cite{torquatoDiffusionSpreadabilityProbe2021} showed that $\mathscr{s}^{ex}(t)$ can be expanded as 
\begin{equation}\label{eq:s-moments-expan}
    \begin{split}
        \mathscr{s}^{ex}(t)&=\frac{d\omega_d}{(4\pi Dt)^{d/2}}\sum_{i=0}^\infty\frac{(-1)^i}{i!}\frac{M_{d-1+2i}\left(\chi_{_V}\right)}{\phi_1\phi_2(4Dt)^i}\\
        &=(Dt/a^2)^{-\frac{d+\alpha}{2}}\sum_{i=0}^\infty C_i(Dt/a^2)^{-i},
    \end{split}
\end{equation}
where $\alpha=0,2,4,...$, and $\omega_d=\frac{\pi^{d/2}}{\Gamma(1+d/2)}$ is the volume of a $d$-dimensional sphere of unit radius.
On the other hand, suppose the spectral density can be expanded as $\tilde{\chi}_{_V}(k)=(ka)^{\alpha}[B+B_{\beta_1/2}(ka)^{\beta_1}+B_{\beta_2/2}(ka)^{\beta_2}+...]$, then $\mathscr{s}^{ex}(t)$ should be expanded as
\begin{equation}\label{eq:s-expan}
    \mathscr{s}^{ex}(t)=(Dt/a^2)^{-\frac{d+\alpha}{2}}\sum_{i=0}^\infty C_{\beta_i/2}(Dt/a^2)^{-\beta_i/2},
\end{equation}
with
\begin{equation}
    C_{\beta_i/2}=\frac{d\omega_d}{(2\pi)^d}\frac{B_{\beta_i/2}\Gamma[(d+\alpha+\beta_i)/2]}{2\phi_1\phi_2}.
\end{equation}

Clearly, the asymptotic expansion of $\mathscr{s}^{ex}(t)$ obtained from either the real-space or Fourier-space expansion must agree with each other; however, the Fourier-space expansion \eqref{eq:s-expan} apparently takes a more general form than the real-space expansion \eqref{eq:s-moments-expan}. In other words, if all of the moments $M_l(\chi_{_V})$ exist, then $\tilde{\chi}_{_V}(k)$ is an analytic function at the origin \cite{torquatoDiffusionSpreadabilityProbe2021}, and vice versa. By contrast, if $\tilde{\chi}_{_V}(k)$ is nonanalytic at the origin, some of the moments $M_l(\chi_{_V})$ are bound to diverge, and vice versa. The large-$r$ behavior of $\chi_{_V}(r)$, or equivalently, the analyticity of $\tilde{\chi}_{_V}(k)$ at the origin, is thus encoded in the large-$t$ asymptotic expansion of $\mathscr{s}^{ex}(t)$.

Note that the long-time asymptotic expansion \eqref{eq:s-moments-expan} or \eqref{eq:s-expan} may be convergent or divergent. For any fixed time $t$, the expansion \eqref{eq:s-moments-expan} is convergent in the $n\to\infty$ limit only when $\lim\limits_{n\to\infty}|(C_{n+1}t^{-n-1})/(C_nt^{-n})|<1$, i.e., $t>\lim\limits_{n\to\infty}|C_{n+1}/C_n|$, and divergent if $\lim\limits_{n\to\infty}|C_{n+1}/C_n|=\infty$. However, even if the series is divergent, the $n^{\text{th}}$-order truncated asymptotic expansion $\sum_{j=0}^n C_j(Dt/a^2)^{-\frac{d+\alpha}{2}-j}$ still approximates well the exact $\mathscr{s}^{ex}(t)$ for some sufficiently long time, i.e., $\mathscr{s}^{ex}(t)-\sum_{j=0}^n C_j(Dt/a^2)^{-\frac{d+\alpha}{2}-j}=O(t^{-\frac{d+\alpha}{2}-n-1}), t\to\infty$. The accuracy of the truncated series can be improved by increasing the order $n$ for either the convergent series or the divergent series. For the latter instance, the improvement is optimized at a certain order $n_p(t)$, depending on the time $t$.

In the stealthy hyperuniform case where $\alpha\to\infty$, the predicted infinitely fast inverse-power decay rate implies that the infinite-time asymptote is approached exponentially fast \cite{torquatoDiffusionSpreadabilityProbe2021}. For periodic media, the long-time excess spreadability scales as $\mathscr{s}^{ex}(t)\sim C\mathrm{e}^{-K^2Dt}$; for disordered stealthy hyperuniform media, the long-time excess spreadability scales as $\mathscr{s}^{ex}(t)\sim C\frac{\exp(-K^2Dt)}{K^2Dt}$.

\subsection{Nonanalytic Spectral Density and Irregular Spreadability Expansion}

\begin{table*}[]
\centering
\begin{tabular}{!{\vrule width 1.2pt}ccc!{\vrule width 1.2pt}cc|cc!{\vrule width 1.2pt}}
\Xhline{1.2pt}
\multicolumn{3}{!{\vrule width 1.2pt}c!{\vrule width 1.2pt}}{\multirow{2}{*}{\makecell[c]{\textbf{Leading Term in Long-Time}\\\textbf{Asymptotic Expansion of Spreadability}\\($\gamma\equiv\frac{d+\alpha}{2}$, $D=a=1$, $t\to\infty$)}}}                                                                                               & \multicolumn{2}{c|}{Disordered}         & \multicolumn{2}{c!{\vrule width 1.2pt}}{Ordered}                \\ \cline{4-7} 
\multicolumn{3}{!{\vrule width 1.2pt}c!{\vrule width 1.2pt}}{}                                                                                                                & \multicolumn{1}{c|}{\makecell[c]{Regular\\\footnotesize($\alpha/2\in\mathbb{N}$)}} & \makecell[c]{Singular\\\footnotesize($\alpha=\zeta$)} & \multicolumn{1}{c|}{Crystal} & Quasicrystal \\ \Xhline{1.2pt}
\multicolumn{1}{!{\vrule width 1.2pt}c|}{\multirow{4}{*}{\makecell[c]{\\\\Hyperuniform\\\footnotesize($\alpha>0$)}}}    & \multicolumn{1}{c|}{\multirow{2}{*}{\makecell[c]{\\Class I\\\footnotesize($\alpha>1$)}}} & \makecell[c]{Stealthy\\Hyperuniform\\\footnotesize($\alpha\to\infty$)}                     & \multicolumn{1}{c|}{$C\frac{\mathrm{e}^{-K^2t}}{K^2t}$}        & -        & \multicolumn{1}{c|}{$C\mathrm{e}^{-K^2t}$}        & -            \\ \cline{3-7} 
\multicolumn{1}{!{\vrule width 1.2pt}c|}{}                                 & \multicolumn{1}{c|}{}                         & \multirow{3}{*}{\makecell[c]{\\Nonstealthy\\Hyperuniform\\\footnotesize($0<\alpha<\infty$)}} & \multicolumn{1}{c|}{$Ct^{-\gamma}$}        &     \makecell[c]{$Ct^{-\gamma}$\\\footnotesize (if $\alpha/2\notin\mathbb{N}_+$),\\$Ct^{-\gamma}\ln \frac{t}{t_0}$\\\footnotesize (if $\alpha/2\in\mathbb{N}_+$)}     & \multicolumn{1}{c|}{-}       &       \makecell[c]{oscillate\\between\\$C_\pm t^{-\gamma}$}       \\ \cline{2-2} \cline{4-7} 
\multicolumn{1}{!{\vrule width 1.2pt}c|}{}                                 & \multicolumn{1}{c|}{\makecell[c]{Class II\\\footnotesize($\alpha=1$)}}                 &                              & \multicolumn{1}{c|}{-}       &     $Ct^{-\gamma}$     & \multicolumn{1}{c|}{-}       &      \makecell[c]{oscillate\\between\\$C_\pm t^{-\gamma}$}        \\ \cline{2-2} \cline{4-7} 
\multicolumn{1}{!{\vrule width 1.2pt}c|}{}                                 & \multicolumn{1}{c|}{\makecell[c]{Class III\\\footnotesize($0<\alpha<1$)}}                &                              & \multicolumn{1}{c|}{-}       &     $Ct^{-\gamma}$     & \multicolumn{1}{c|}{-}       & -            \\ \hline
\multicolumn{1}{!{\vrule width 1.2pt}c|}{\multirow{2}{*}{\makecell[c]{Nonhyperuniform\\\footnotesize($-d<\alpha\le0$)}}} & \multicolumn{2}{c!{\vrule width 1.2pt}}{\makecell[c]{Typical Nonhyperuniform\\\footnotesize($\alpha=0$)}}                                                 & \multicolumn{1}{c|}{$Ct^{-\gamma}$}        &    $Ct^{-\gamma}\ln \frac{t}{t_0}$      & \multicolumn{1}{c|}{-}       & -            \\ \cline{2-7} 
\multicolumn{1}{!{\vrule width 1.2pt}c|}{}                                 & \multicolumn{2}{c!{\vrule width 1.2pt}}{\makecell[c]{Antihyperuniform\\\footnotesize($-d<\alpha<0$)}}                                        & \multicolumn{1}{c|}{-}       &     $Ct^{-\gamma}$     & \multicolumn{1}{c|}{-}       & -            \\ \Xhline{1.2pt}
\end{tabular}
\caption{Classification of two-phase media and the corresponding long-time normalized excess spreadability $\mathscr{s}^{ex}(t)$ behaviors. We only list the leading order terms, but the general form of the expansion can be inferred from the linear combination of these terms due to the linearity of Eq. \eqref{eq:excess-spreadability}. Both the diffusion coefficient $D$ and the characteristic heterogeneity length scale $a$ are set to be unity, i.e., $D=a=1$.}
\label{tab1}
\end{table*}

Consider a system of identical impenetrable particles of radius $R$ in a matrix interacting via a pair potential $v(\mathbf{r})$ that scales as $v(\mathbf{r})\sim E/|\mathbf{r}|^{d+\zeta}, |\mathbf{r}|\to\infty$ for a constant $E$ and $\zeta>0$. The effective potential between the centroids is $v_{\text{eff}}(\mathbf{r})=v(\mathbf{r})+u(\mathbf{r})$, where $u(\mathbf{r})=\infty$ for $|\mathbf{r}|<R$ and $u(\mathbf{r})=0$ otherwise is the hard-core potential, which does not affect the large-distance scaling $v_{\text{eff}}(\mathbf{r})\sim v(\mathbf{r})\sim E/|\mathbf{r}|^{d+\zeta}$. Torquato and Wang \cite{torquatoPreciseDeterminationPair2022} show that for odd integers $\zeta$, the \emph{structure factor}\footnote{The structure factor is defined by the Fourier transform of the pair correlation function \cite{torquatoHyperuniformStatesMatter2018a,hansenTheorySimpleLiquids2013}.} $S(\mathbf{k})$ associated with the positions of particle centers of the equilibrium system away from the critical point is given by
\begin{equation}\label{eq:sf-division}
    S(\mathbf{k})\sim-\frac{E}{k_BT} c_1(\zeta,d)\rho S(\mathbf{0})^2|\mathbf{k}|^\zeta+f(\mathbf{k}),\hspace{1em}|\mathbf{k}|\to0,
\end{equation}
where $c_1(\zeta,d)$ is a function of $\zeta$ and dimension $d$, and $f(\mathbf{k})$ is an analytic function of $\mathbf{k}$ at the origin. As for even integers $\zeta$, the expansion contains not just the $|\mathbf{k}|^\zeta$ term but also a $|\mathbf{k}|^\zeta\ln|\mathbf{k}|$ term \cite{torquatoPreciseDeterminationPair2022}. The corresponding spectral density $\tilde{\chi}_{_V}(\mathbf{k})$ when such a packing is mapped to a two-phase medium (identical impenetrable spheres
in a matrix) is directly related to the structure factor $S(\mathbf{k})$ by \cite{torquatoHyperuniformStatesMatter2018a}
\begin{equation}
    \tilde{\chi}_{_V}(\mathbf{k})=\phi_2\tilde{\alpha}_2(k;R)S(\mathbf{k}),
\end{equation}
where $\phi_2$ is the volume fraction of the spheres and $\tilde{\alpha}_2(k;R)=\Gamma(d/2+1)k^{-d}J_{d/2}^2(kR)$ is the Fourier transform of the scaled intersection volume of two spherical windows. $\tilde{\alpha}_2(k;R)$ is analytic at the origin, so the $|\mathbf{k}|^\zeta$ and $|\mathbf{k}|^\zeta\ln|\mathbf{k}|$ singularity in $S(\mathbf{k})$ is preserved in the spectral density $\tilde{\chi}_{_V}(\mathbf{k})$, which then enters the asymptotic expansion of $\mathscr{s}^{ex}(t)$ via Eq. \eqref{eq:s-expan}.

For instance, the Lennard-Jones potential $v_{_\text{LJ}}(\mathbf{r})=4\varepsilon[(\sigma/r)^{12}-(\sigma/r)^6]$ scales as $v_{_\text{LJ}}(\mathbf{r})\sim-4\varepsilon(\sigma/r)^6$, $r\to\infty$, i.e., $E=-4\varepsilon\sigma^6$ and $\zeta=3$ for $d=3$. Its structure factor near the origin can be expanded as
\begin{equation}
    S(\mathbf{k})=s_0+s_2|\mathbf{k}|^2+4\rho\beta\varepsilon\sigma^6c_1(3,3)s_0^2|\mathbf{k}|^3+s_4|\mathbf{k}|^4+...,
\end{equation}
which contains no first-order term but a third-order term, a signature of the $\zeta=3$ interaction and correlation in the system. Correspondingly, the asymptotic expansion of the spectral density also contains a singular $|\mathbf{k}|^3$ term, i.e.,
\begin{equation}
    \tilde{\chi}_{_V}(\mathbf{k})=B_0+B_1|\mathbf{k}\sigma|^2+B_{3/2}|\mathbf{k}\sigma|^3+B_2|\mathbf{k}\sigma|^4+...,
\end{equation}
which results in an irregular spreadability expansion in contrast to the regular expansion \eqref{eq:s-moments-expan}:
\begin{equation}
    \begin{split}
        \mathscr{s}^{ex}(t)&=t^{-3/2}[C_0+C_1(Dt/\sigma^2)^{-1}\\
        &\phantom{=}+C_{3/2}(Dt/\sigma^2)^{-3/2}+C_2(Dt/\sigma^2)^{-2}+...]
    \end{split}
\end{equation}

\section{Theory of Long-Time Spreadability Asymptotic Expansion}\label{sec:theory}
\subsection{Classification of the Long-Time Spreadability Asymptotics}

In this subsection, we provide a unified classification of the basic asymptotic behaviors of the long-time normalized excess spreadability $\mathscr{s}^{ex}(t)$ under certain conditions, e.g., hyperuniform or nonhyperuniform media, media with spectral densities analytic or non-analytic at the origin, and media with or without long-range orders. Some of the results have already been reported in previous works \cite{torquatoDiffusionSpreadabilityProbe2021,torquatoPreciseDeterminationPair2022}. For clarity and simplicity, we henceforth set both the diffusion coefficient $D$ and the characteristic heterogeneity length scale $a$ to be unity, i.e., $D=a=1$, without loss of generality.

For disordered two-phase media without any long-range order, the spectral density $\tilde{\chi}_{_V}(k)$ contains no Bragg peaks. If $\tilde{\chi}_{_V}(k)$ is analytic at the origin so that its expansion contains only regular powers $1,k^2,k^4,...$, the autocovariance function $\chi_{_V}(r)$ decays faster than any power law at large distance $r$ so that all of the moments $M_n(\chi_{_V})$ exist, and the regular expansion \eqref{eq:s-moments-expan} applies. On the other hand, the singular powers $k^\zeta, \zeta\ne0,2,4...$ as well as the generalized powers $k^\zeta\ln k, \zeta=0,2,4...$ are proportional to the Fourier transform of $r^{-d-\zeta}$ \cite{hormanderAnalysisLinearPartial2003,kanwalGeneralizedFunctions2004,torquatoPreciseDeterminationPair2022}. More detailed information about the Fourier transform of the power law is presented in Appendix \ref{app:fourier-pw}.

There are some exceptions when the spectral density $\tilde{\chi}_{_V}(k)$ cannot be expanded in terms of powers at the origin: stealthy hyperuniform media with $\tilde{\chi}_{_V}(k)\equiv0$ near the origin, periodic media with discontinuous $\tilde{\chi}_{_V}(k)$ due to the Bragg peaks \cite{wangDynamicMeasureHyperuniformity2022}, as well as quasicrystals with dense Bragg peaks \cite{hitin-bialusHyperuniformityClassesQuasiperiodic2024}. For quasicrystals, the long-time spreadability $\mathscr{s}^{ex}(t)$ oscillates between two asymptotic power laws of the same exponent $\alpha$ but different coefficients $C_\pm$. This completes our classification of the long-time asymptotic behaviors of $\mathscr{s}^{ex}(t)$ of all known kinds of two-phase media as summarized in Table \ref{tab1}. The general form of the asymptotic expansion series can be inferred from Table \ref{tab1} based on the linearity of Eq. \eqref{eq:excess-spreadability}.

\begin{figure*}
    \centering
    \subfloat[]{%
      \includegraphics[width=0.33\linewidth]{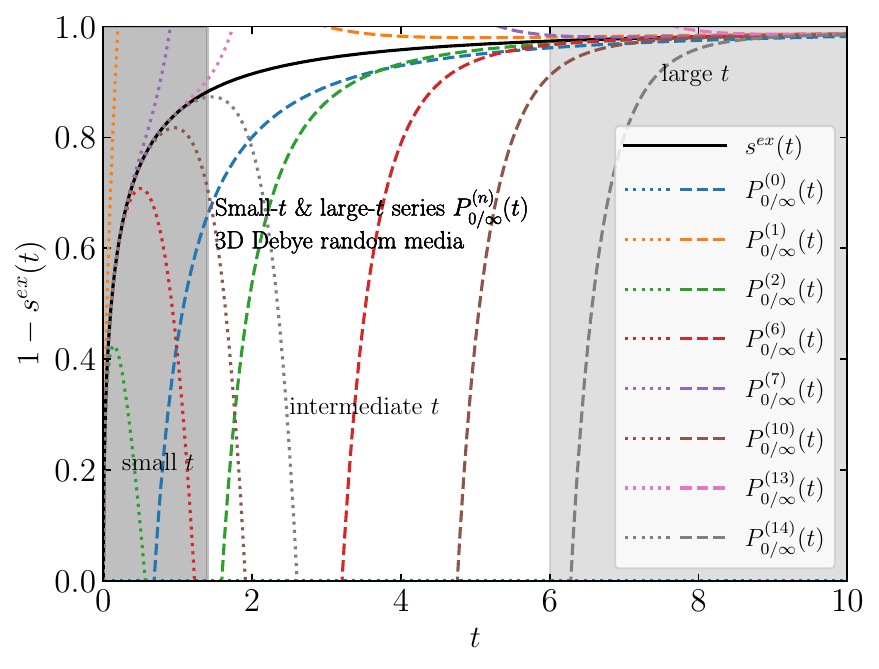}%
      \label{fig2a}%
    }
    \subfloat[]{%
      \includegraphics[width=0.33\linewidth]{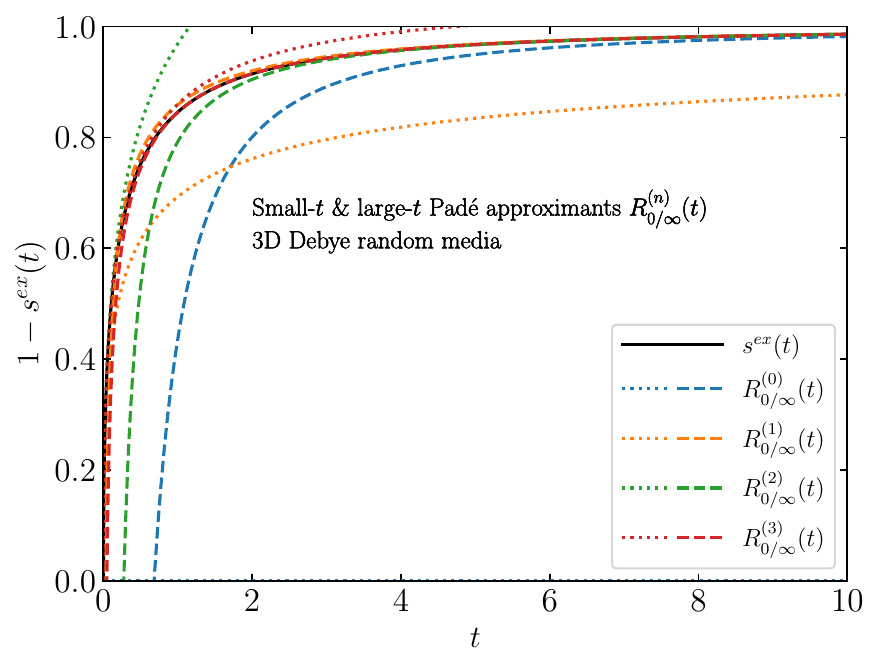}%
      \label{fig2b}%
    }
    \subfloat[]{%
      \includegraphics[width=0.33\linewidth]{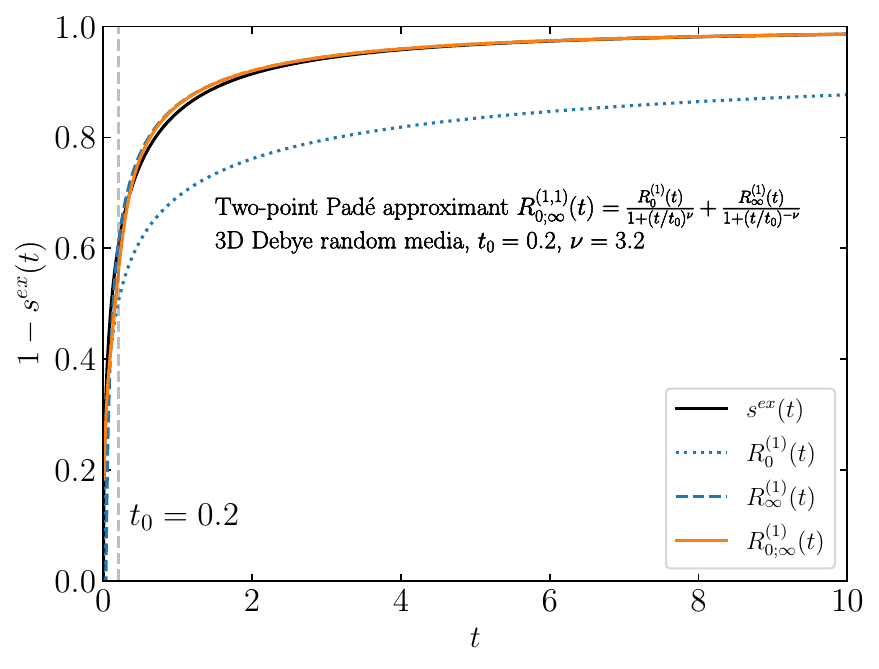}%
      \label{fig2c}%
    }
    \caption{\textbf{(a)} The numerical results of the truncated series $P_0^{(n)}(t)$ and $P_\infty^{(n)}(t)$ [Eq. \eqref{eq:poly0} and \eqref{eq:polyinf}] against the exact $\mathscr{s}^{ex}(t)$. The small-, intermediate-, and large-$t$ regions distinguished by different shadows are settled semi-quantitatively based on the agreement between $P_0^{(n)}(t)$, $P_\infty^{(n)}(t)$, and $\mathscr{s}^{ex}(t)$. \textbf{(b)} Pad\'e approximants $R_0^{(n)}(t)$ and $R_\infty^{(n)}(t)$ [Eq. \eqref{eq:pade0} and \eqref{eq:padeinf}] in contrast to the exact $\mathscr{s}^{ex}(t)$. \textbf{(c)} The two-point Pad\'e approximant $R^{(1,1)}_{0;\infty}(t)$ [Eq. \eqref{eq:2pt-pade}] that resumes the correct short- and large-$t$ behaviors up to the same order as $R_0^{(1)}(t)$ and $R_\infty^{(1)}(t)$. In all three plots \textbf{(a)}, \textbf{(b)}, and \textbf{(c)}, both the diffusion coefficient $D$ and the characteristic heterogeneity length scale $a$ are set to be unity, i.e., $D=a=1$.}
    \label{fig:approximants}
\end{figure*}

By incorporating correction terms in the asymptotic expansion to the fitting function, we can extract information from the long-time $\mathscr{s}^{ex}(t)$ data much more accurately and efficiently, which is the main focus of Secs. \ref{sec:fitting} and \ref{sec:results}.
In the next subsection, we extend our long-time spreadability analysis to other times to obtain accurate approximants of $\mathscr{s}^{ex}(t)$ for all times.

\subsection{Approximants of the Spreadability for all Times}\label{sec:approximants}

\begin{figure}
    \centering
    \includegraphics[width=0.75\linewidth]{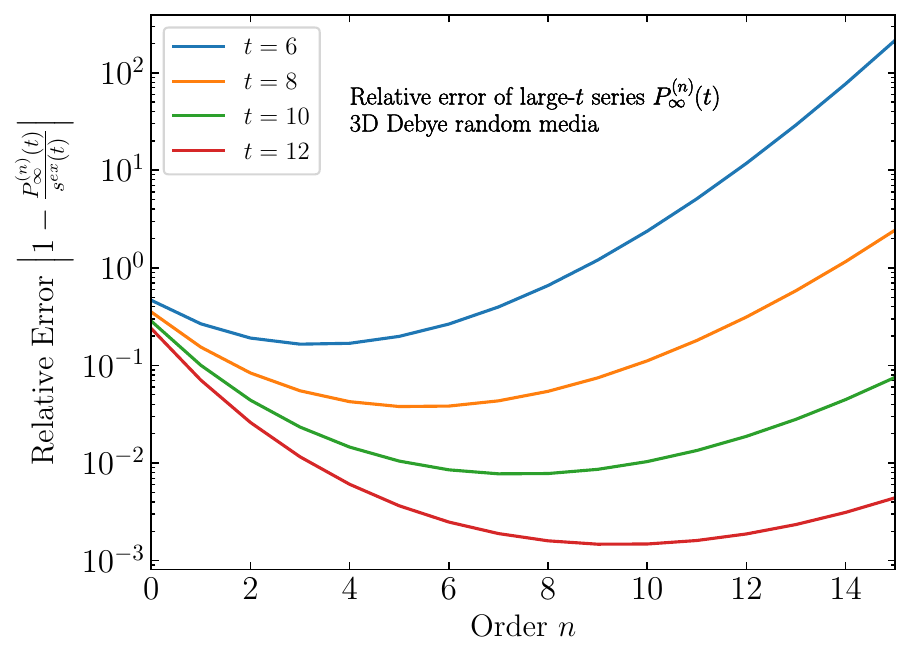}
    \caption{The relative error of the truncated long-time asymptotic expansion $P_\infty^{(n)}(t)$ [Eq. \eqref{eq:polyinf}] against the exact $\mathscr{s}^{ex}(t)$ of different order $n$ at different times $t$. Both the diffusion coefficient $D$ and the characteristic heterogeneity length scale $a$ are set to be unity, i.e., $D=a=1$.}
    \label{fig3}
\end{figure}

The long-time asymptotic expansion of the spreadability $\mathscr{s}^{ex}(t)=t^{-\frac{d+\alpha}{2}}\sum_{i=0}^\infty C_{\beta_i/2}t^{-\beta_i/2}$ together with its short-time counterpart $\mathscr{s}^{ex}(t)=\sum_{j=0}^\infty A_jt^{j/2}$ allow us to construct accurate approximants for all times.
The most straightforward approximants are the one-point polynomials $P_\infty^{(n)}(t)$ and $P_0^{(n)}(t)$, defined as simple truncations of the long- and short-time expansions of $\mathscr{s}^{ex}(t)$ at the order $n$, i.e.,
\begin{align}
    P_\infty^{(n)}(t)&=t^{-\frac{d+\alpha}{2}}\sum_{i=0}^n C_{\beta_i/2}t^{-\beta_i/2}\hspace{1em} \text{(long times)},\label{eq:polyinf}\\
        P_0^{(n)}(t)&=\sum_{j=0}^n A_jt^{j/2}\hspace{5.8em} \text{(short times)},\label{eq:poly0}
\end{align}
where the coefficient $A_0\equiv1$, since $P_0^{(n)}(0)=\mathscr{s}^{ex}(0)\equiv1$.
Clearly, $P_\infty^{(n)}(t)$ approximates $\mathscr{s}^{ex}(t)$ well at long times, while $P_0^{(n)}(t)$ approximates $\mathscr{s}^{ex}(t)$ well at short times.
Note that neither of the polynomials is very accurate at the intermediate times.
Particularly, the asymptotic expansion may sometimes be divergent, causing the deviation of $P_\infty^{(n)}(t)$ or $P_0^{(n)}(t)$ to grow rather than diminish with the order $n$ after the order exceeds an optimal order $n_p(t)$ at time $t$.

One well-known way to improve the accuracy of the polynomial approximants is to use them to form Pad\'e approximants \cite{bakerEssentialsPadeApproximants1975}, i.e., rational functions.
Specifically, the one-point Pad\'e approximants $R_\infty^{(n)}(t)$ and $R_0^{(n)}(t)$ are given as 
\begin{align}
    R_\infty^{(n)}(t)&=t^{-\frac{d+\alpha}{2}}\mathcal{R}_{t\to\infty}\left[C_{\beta_i/2}t^{-\beta_i/2}\right]\hspace{0.5em} \text{(long times)},\label{eq:padeinf}\\
        R_0^{(n)}(t)&=\mathcal{R}_{t=0}\left[\sum_{j=0}^n A_jt^{j/2}\right]\hspace{3.5em} \text{(short times)},\label{eq:pade0}
\end{align}
where  $\mathcal{R}_{x=x_0}[p(x)]$ denotes the operation of converting the polynomial $p(x)$ into a Pad\'e expression around $x=x_0$.
More details about the Pad\'e approximants are presented in Appendix \ref{app:pade}.
The key conclusion is that the one-point Pad\'e approximants $R_\infty^{(n)}(t)$ and $R_0^{(n)}(t)$ can present much better approximation for the intermediate-time behaviors of $\mathscr{s}^{ex}(t)$ while maintaining the correct long-/short-time behaviors, respectively.

Importantly, we can approximate $\mathscr{s}^{ex}(t)$ for all times by combining the long-time Pad\'e approximant $R_\infty^{(n)}(t)$ and the short-time Pad\'e approximant $R_0^{(n)}(t)$ into the following two-point Pad\'e approximant:
\begin{equation}\label{eq:2pt-pade}
    \begin{split}
        R_{0;\infty}^{(m,n)}(t)=\frac{R_0^{(m)}(t)}{1+(t/t_0)^\nu}+\frac{R_\infty^{(n)}(t)}{1+(t/t_0)^{-\nu}},\\
        \nu>\frac{d+\alpha+m+\beta_n}{2}.
    \end{split}
\end{equation}
Similarly, one can further improve these estimates in the intermediate-time region by incorporating intermediate-time terms to get a three-point Pad\'e approximant. In this way, we can approximate $\mathscr{s}^{ex}(t)$ with just a few parameters while constraining it to match the most important short-, intermediate-, and long-time features.
Our parametrization scheme of the spreadability can improve the efficiency of time-dependent diffusive transport property engineering \cite{shiThreedimensionalConstructionHyperuniform2025}.
Specifically, it can be useful in the inverse design of self-assembly patterns that manifest desired spreadability properties.

Figure \ref{fig:approximants} contrasts the one-point polynomials $P_{\infty}^{(n)}(t)$ and $P_{0}^{(n)}(t)$, the corresponding one-point Pad\'e approximants $R_{\infty}^{(n)}(t)$ and $R_{0}^{(n)}(t)$, as well as the two-point Pad\'e approximant $R_{0;\infty}^{(1,1)}(t)$ for Debye random media (DRM). (Additional information about this specific example model can be found in Appendix \ref{app:drm}.)
Not surprisingly, due to the divergent nature of $P_{\infty}^{(n)}(t)$ for DRM, its deviation from the exact $\mathscr{s}^{ex}(t)$ decreases at first but eventually grows with the order $n$ at any fixed $t$ in Fig. \ref{fig2a}, limiting its accuracy as an approximant, especially at the short and intermediate time range.
This divergence effect of $P_{\infty}^{(n)}(t)$ is demonstrated more explicitly in Fig. \ref{fig3}, where we show the decrease and increase of its deviation from the exact $\mathscr{s}^{ex}(t)$ explicitly.
Despite being a convergent series, $P_{0}^{(n)}(t)$ converges rather slowly and also deviates drastically from $\mathscr{s}^{ex}(t)$ at the intermediate times, not to mention the long-time region.
By contrast, the corresponding Pad\'e expression $R_{\infty}^{(n)}(t)$ exhibits superior convergence and accuracy in Fig. \ref{fig2b}, especially at the intermediate-time region.
Finally and importantly, the two-point Pad\'e approximant $R^{(1,1)}(t)$ in Fig. \ref{fig2c} provides an excellent approximation of the exact $\mathscr{s}^{ex}(t)$ at any time $t$, despite involving only three independent coefficients $A_1$, $C_0$, and $C_1$ ($A_0\equiv1$).

\section{Long-Time Asymptotic Fitting Scheme}\label{sec:fitting}

Given the lessons learned about getting accurate approximations of the asymptotic behavior of $\mathscr{s}^{ex}(t)$ (Sec. \ref{sec:theory}), we now proceed to estimating such long-time behaviors when $\mathscr{s}^{ex}(t)$ is extracted from numerical or experimental data, which often occurs in practice.
The ensuing fitting procedure applies to the restricted class of microstructure described by long-time spread that involves the power law, i.e, nonhyperuniform as well as nonstealthy hyperuniform disordered microstructure described in Table \ref{tab1}.
Clearly, such estimates require fitting the spreadability data at sufficiently long times.
All such previous works \cite{wangDynamicMeasureHyperuniformity2022,hitin-bialusHyperuniformityClassesQuasiperiodic2024} estimate the exponent $\alpha$ in the asymptotic power law $\mathscr{s}^{ex}(t)\sim Ct^{-\frac{d+\alpha}{2}}$, $t\to\infty$ by taking the logarithm of both sides to get
\begin{equation}\label{eq:fit0}
    \ln[\hat{\mathscr{s}}^{ex}_0(t)]=\ln\hat{C}-\frac{d+\hat{\alpha}}{2}\ln t
\end{equation}
and then fitting this form at sufficiently large times $t$ using a standard least-squares regression, i.e., minimizing the sum of $\Big(\ln[\mathscr{s}^{ex}(t)]-\ln[\hat{\mathscr{s}}^{ex}_0(t)]\Big)^2$ for all data points. We use the hat symbol above quantities to distinguish the fitting function and fitting parameters from the raw data and fixed parameters. In what follows, we propose three types of fitting functions by incorporating correction terms to the original fitting function $\ln[\hat{\mathscr{s}}^{ex}_0(t)]$.

\subsection{Three Types of Improved Fitting Functions}

We intend to improve the fitting function $\ln[\hat{\mathscr{s}}^{ex}_0(t)]$ by incorporating correction terms in the long-time asymptotic expansion of $\mathscr{s}^{ex}(t)$. However, the most general form \eqref{eq:s-expan} of the asymptotic expansion is unstable as a fitting function if all the exponents $\alpha$ and $\beta_i$ are set to be free fitting parameters. In order to strike a balance between the generality of the fitting procedure and numerical stability, we propose three types of fitting functions by adding three different types of corrections to the fitting function $\ln[\hat{\mathscr{s}}^{ex}_0(t)]$ based on different assumptions about the exponents $\alpha$ and $\beta_i$.
\begin{itemize}
    \item \textbf{Type I}

Assume that $\beta_i=i=1,2,3,...$ and thus the spectral density expansion takes the form 
\begin{equation}
    \tilde{\chi}_{_V}(k)=k^\alpha(B+B_{1/2}k+B_1k^2+...),
\end{equation}
which, if $\alpha$ is not an integer, corresponds to an autocovariance function of the form
\begin{equation}
    \chi_{_V}(r)=r^{-d-\alpha}(K+K_{1/2}r^{-1}+K_1r^{-2}+...).
\end{equation}
The corresponding spreadability asymptotic expansion is given by $\mathscr{s}^{ex}(t)=t^{-\frac{d+\alpha}{2}}(C+C_{1/2}t^{-1/2}+C_1t^{-1}+...)$, and so we have the first type of fitting function
\begin{equation}\label{eq:fit1}
    \ln[\hat{\mathscr{s}}^{ex}_1(t)]=-\frac{d+\hat{\alpha}}{2}\ln t+\ln\left(\sum_{i=0}^n\hat{C}_{i/2}t^{-i/2}\right).
\end{equation}
When $n=0$, $\ln[\hat{\mathscr{s}}^{ex}_1(t)]$ reduces to the original fitting function $\ln[\hat{\mathscr{s}}^{ex}_0(t)]$ [Eq. \eqref{eq:fit0}]. We can improve the fitting by considering $n\ge1$.
\item \textbf{Type II}

Still assuming that $\beta_i=i=1,2,3,...$, if $\alpha$ takes an integer value, the Fourier transform of the same autocovariance function form
\begin{equation}
    \chi_{_V}(r)=r^{-d-\alpha}(K+K_{1/2}r^{-1}+K_1r^{-2}+...)
\end{equation}
involves generalized power-law terms like $\ln k$, $k^2\ln k$, $k^4\ln k,...$, which shall thus be taken into account in the spectral density expansion, i.e., 
\begin{equation}
    \begin{split}
        \tilde{\chi}_{_V}(k)&=k^\alpha\sum_{i=0}^\infty B_{i/2}k^i+k^{2\lceil\alpha/2\rceil_+}\ln k\sum_{j=0}^\infty F_jk^{2j},
    \end{split}
\end{equation}
where $\alpha\in\mathbb{Z}$. Correspondingly, the spreadability asymptotic expansion is given by 
\begin{equation}
    \begin{split}
        \mathscr{s}^{ex}(t)&=t^{-\frac{d+\alpha}{2}}\sum_{i=0}^\infty C_{i/2}t^{-i/2}\\&\phantom{=}+t^{-d/2-\lceil\alpha/2\rceil_+}\ln t\sum_{j=0}^\infty E_{j}t^{-j}
    \end{split}
\end{equation}
The second type of fitting function is then given by
\begin{equation}\label{eq:fit2}
    \begin{split}
        \ln[\hat{\mathscr{s}}^{ex}_2(t)]=-\frac{d+\alpha}{2}\ln t+\ln\Big(\sum_{i=0}^n \hat{C}_{i/2}t^{-i/2}\\+t^{-\Delta}\ln t\sum_{j=0}^m \hat{E}_{j}t^{-j}\Big),
    \end{split}
\end{equation}
where $\Delta=\lceil\alpha/2\rceil_+-\alpha/2$ and $m=\lfloor(\alpha+n)/2\rfloor-\lceil\alpha/2\rceil_+$. The integer $\alpha$ assumption eliminates the possibility of class III hyperuniform media.
\item \textbf{Type III}

If we assume $\beta_i=2i=2,4,6,...$, meaning that the spectral density is analytic at the origin, i.e.
\begin{equation}
    \tilde{\chi}_{_V}(k)=k^\alpha(B+B_1k^2+B_2k^4+...),\hspace{1em}\alpha/2\in\mathbb{N},
\end{equation}
then the autocovariance function $\chi_{_V}(r)$ decays faster than any power law at large distance $r$ so that all of the moments $M_n(\chi_{_V})$ exist, and the regular expansion \eqref{eq:s-moments-expan} applies, i.e. $\mathscr{s}^{ex}(t)=t^{-\frac{d+\alpha}{2}}(C+C_1t^{-1}+C_2t^{-2}+...)$. Thus, the third type of fitting function takes the form
\begin{equation}\label{eq:fit3}
    \ln[\hat{\mathscr{s}}^{ex}_3(t)]=-\frac{d+\alpha}{2}\ln t+\ln\Big(\sum_{i=0}^n \hat{C}_{i}t^{-i}\Big),
\end{equation}
where $\alpha/2\in\mathbb{N}$. The nonnegative even integer $\alpha$ assumption narrows the possibilities down to typical nonhyperuniform and class I hyperuniform media.
\end{itemize}
For convenience, we summarize the three types of fitting functions in Table \ref{tab2} and the classes of microstructures to which they apply in Table \ref{tab3}.

\begin{table}[]
\adjustbox{width=\linewidth}{
\begin{tabular}{!{\vrule width 1.2pt}cl!{\vrule width 1.2pt}}
\Xhline{1.2pt}
\multicolumn{2}{!{\vrule width 1.2pt}c!{\vrule width 1.2pt}}{\textbf{The Three Types of Fitting Functions}}    \\ \Xhline{1.2pt}
\multicolumn{1}{!{\vrule width 1.2pt}c!{\vrule width 1.2pt}}{Type I} &  $\ln[\hat{\mathscr{s}}^{ex}_1(t)]=-\frac{d+\hat{\alpha}}{2}\ln t+\ln\left(\sum_{i=0}^n\hat{C}_{i/2}t^{-i/2}\right)$\\ \hline
\multicolumn{1}{!{\vrule width 1.2pt}c!{\vrule width 1.2pt}}{Type II} &  \makecell[l]{$\ln[\hat{\mathscr{s}}^{ex}_2(t)]=-\frac{d+\alpha}{2}\ln t+\ln\Big(\sum_{i=0}^n \hat{C}_{i/2}t^{-i/2}$\\$\hspace{9em}+t^{-\Delta}\ln t\sum_{j=0}^m \hat{E}_{j}t^{-j}\Big)$}\\ \hline
\multicolumn{1}{!{\vrule width 1.2pt}c!{\vrule width 1.2pt}}{Type III} &  $\ln[\hat{\mathscr{s}}^{ex}_3(t)]=-\frac{d+\alpha}{2}\ln t+\ln\Big(\sum_{i=0}^n \hat{C}_{i}t^{-i}\Big)$\\ \Xhline{1.2pt}
\end{tabular}}
\caption{The expressions of the three types of fitting functions, where $\hat{\alpha}$ and $\hat{C}_{\beta_i/2}$ denote the fitting parameters to be determined from the fitting, $n$ is the fitting order, while $t$, $d$, $\alpha$, $\Delta=\lceil\alpha/2\rceil_+-\alpha/2$, and $m=\lfloor(\alpha+n)/2\rfloor-\lceil\alpha/2\rceil_+$ are fixed parameters set to be the input of the fitting.}
\label{tab2}
\end{table}

\begin{table}[]
\adjustbox{width=\linewidth}{
\begin{tabular}{!{\vrule width 1.2pt}cc!{\vrule width 1.2pt}cc!{\vrule width 1.2pt}}
\Xhline{1.2pt}
\multicolumn{2}{!{\vrule width 1.2pt}c!{\vrule width 1.2pt}}{\multirow{2}{*}{\makecell[c]{\vspace{-0.5em}\\\textbf{Classes of Microstructures to which}\\ \textbf{the Fitting Functions Apply}}}}                            & \multicolumn{2}{c!{\vrule width 1.2pt}}{Disordered}         \\ \cline{3-4} 
\multicolumn{2}{!{\vrule width 1.2pt}c!{\vrule width 1.2pt}}{}                                             & \multicolumn{1}{c|}{\makecell[c]{Regular\\\footnotesize($\alpha/2\in\mathbb{N}$)}} & \makecell[c]{Singular\\\footnotesize($\alpha=\zeta$)} \\ \Xhline{1.2pt}
\multicolumn{1}{!{\vrule width 1.2pt}c|}{\multirow{3}{*}{\makecell[c]{\\Nonstealthy\\Hyperuniform\\\footnotesize($0<\alpha<\infty$)}}}     & \makecell[c]{Class I\\\footnotesize($\alpha>1$)}   & \multicolumn{1}{c|}{Type III}        &     Type I\&II     \\ \cline{2-4} 
\multicolumn{1}{!{\vrule width 1.2pt}c|}{}                                 & \makecell[c]{Class II\\\footnotesize($\alpha=1$)}  & \multicolumn{1}{c|}{-}       &    Type II      \\ \cline{2-4} 
\multicolumn{1}{!{\vrule width 1.2pt}c|}{}                                 & \makecell[c]{Class III\\\footnotesize($0<\alpha<1$)} & \multicolumn{1}{c|}{-}       &    Type I      \\ \hline
\multicolumn{1}{!{\vrule width 1.2pt}c|}{\multirow{2}{*}{\makecell[c]{Nonhyperuniform\\\footnotesize($-d<\alpha\le0$)}}} & \multicolumn{1}{c!{\vrule width 1.2pt}}{\makecell[c]{Typical\\Nonhyperuniform\\\footnotesize($\alpha=0$)}}   & \multicolumn{1}{c|}{Type III}        &    Type II      \\ \cline{2-4} 
\multicolumn{1}{!{\vrule width 1.2pt}c|}{}                                 & \multicolumn{1}{c!{\vrule width 1.2pt}}{\makecell[c]{Antihyperuniform\\\footnotesize($-d<\alpha<0$)}}      & \multicolumn{1}{c|}{-}       &       Type I\&II   \\ \Xhline{1.2pt}
\end{tabular}}
\caption{Applicability of the three types of fitting functions for different two-phase media. Note that only a subset of the classification given by Table \ref{tab1} is covered, namely, nonhyperuniform disordered media as well as hyperuniform disordered media that are not stealthy. }
\label{tab3}
\end{table}

\subsection{Fitting Procedure}

In practice, we often have no prior knowledge about the functional form of the expansion of the spectral density $\tilde{\chi}_{_V}(k)$ of the heterogeneous medium of interest. In those cases, we propose the following systematic procedure:
\begin{enumerate}
    \item Fit with Type I $\ln[\hat{\mathscr{s}}^{ex}_1(t)]$ \eqref{eq:fit1} first, which makes the weakest assumption and thus applies to the broadest range of media among the three;
    \item According to the fitting results $\hat{\alpha}$ and $\hat{C}_i$ of $\ln[\hat{\mathscr{s}}^{ex}_1(t)]$, attempt to fit with Type II or III if one of the following criteria is met:
    \begin{enumerate}
        \item fit with Type II $\ln[\hat{\mathscr{s}}^{ex}_2(t)]$ \eqref{eq:fit2} if $|\hat{\alpha}-\alpha|\ll1$ for an integer $\alpha$
        \item fit with Type III $\ln[\hat{\mathscr{s}}^{ex}_3(t)]$ \eqref{eq:fit3} if $|\hat{\alpha}-\alpha|\ll1$ for a nonnegative even integer $\alpha$ and $|\hat{C}_ {i/2}|\approx0$ for odd $i$ (Quantitatively, if $|\hat{C}_ {1/2}|\ll\sqrt{|\hat{C}\hat{C}_1|}$, $|\hat{C}_ {3/2}|\ll\sqrt{|\hat{C}_1\hat{C}_2|}$, ...)
    \end{enumerate}
\end{enumerate}
The fitting procedure is summarized in the following flow chart.
\begin{equation*}
    \adjustbox{width=\linewidth}{
    \boxed{
    \begin{aligned}
        &\mathscr{s}^{ex}(t)\text{ data}\implies \text{Type I fitting}\implies\text{fitting results $\hat{\alpha}$\&$\hat{C}_{i/2}$}\\
        &\implies
        \begin{cases}
            \left.\makecell[l]{\text{if $\hat{\alpha}\approx0,2,4,...$}\\\text{\&$\hat{C}_{\frac{1}{2},\frac{3}{2},...}\approx0$}}\right\}&\to\text{switch to Type III fitting}\\
            \text{else if $\hat{\alpha}\approx$ integer}& \to\text{switch to Type II fitting}
            \\
            \text{else}&\to\text{keep the Type I fitting}
        \end{cases}
    \end{aligned}
    }}
\end{equation*}

For each of the three fitting functions we propose, the fitting order $n$ determines the number of fitting parameters. Additional fitting parameters should be able to improve the accuracy of the fits; still, too many tend to lead to overfitting, i.e., the fitting accuracy degrades rather than improves by incorporating additional fitting parameters. Consequently, there should be an optimal fitting order $n_o$ separating the underfitting and overfitting regions, which shall be settled by comparing the fits across different orders. In the underfitting region ($n<n_o$), as the order $n$ increases to approach the optimal fitting order $n_o$, the fits should exhibit a converging trend, which stabilizes in the vicinity of the optimal order $n=n_o$. In the overfitting region ($n>n_o$), as the order $n$ further increases, the fits demonstrate a (usually abruptly) different trend than in the underfitting region, or simply become unstable. 

\section{Applications to Nonhyperuniform, Hyperuniform, and Antihyperuniform Media}\label{sec:results}

Here, we demonstrate the accuracy of our fitting procedure to ascertain the long-time
asymptotic behaviors of the spreadability for three different models: (1) Debye random media, (2) hyperuniform media, and (3) antihyperuniform media. 
These models are purposely chosen because they span the possible range of large-scale 
structural properties of two-phase media. 
Debye random media are an example of typical nonhyperuniform media with $\alpha=0$, the disordered hyperuniform model is class I media with $\alpha=2$ (possessing some vanishing coefficients ($C_{2i+1}=0$ for 2D and $C_{3i+2}=0$ for 3D), and the antihyperuniform model has $\alpha=-1$. 
We begin by generating exact (noise-free) discrete data sets of the spreadabilities for these models
by substituting exact analytical forms for their two-point statistics, given in Appendix \ref{app:models},
into Eq. (\ref{eq:excess-spreadability}). Such fits yield the highest possible accuracy that we can achieve with our fitting scheme. Then, we add Gaussian noise of varying magnitudes to the data to test its robustness to random noise in real data.
  In what follows, $\mathscr{s}^{ex}(t)$ is computed for 1000 data points evenly distributed on the log scale from $t_{\text{min}}=100$ to $t_{\text{max}}=10000$, where, as noted earlier, we set both the diffusion coefficient $D$ and the characteristic heterogeneity length scale $a$ to be unity, i.e., $D=a=1$.

\subsection{Fitting with Exact Representations of $\mathscr{s}^{ex}(t)$}

\subsubsection{Debye random media}

\begin{figure}
    \centering
    \subfloat[]{\includegraphics[width=0.79\linewidth]{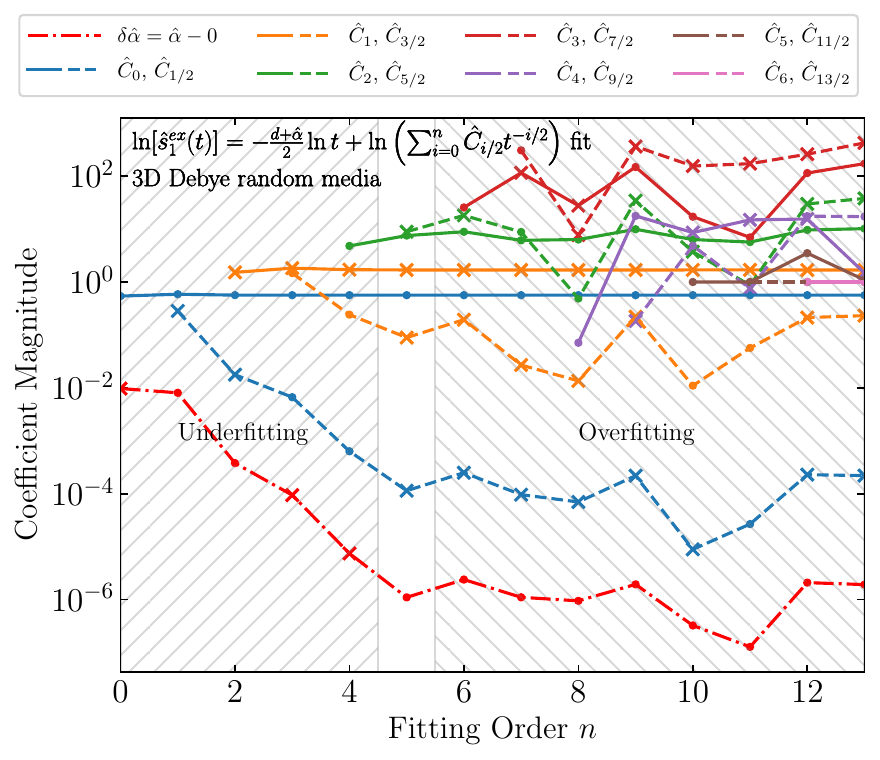}\label{fig4a}}\\
    \subfloat[]{\includegraphics[width=0.79\linewidth]{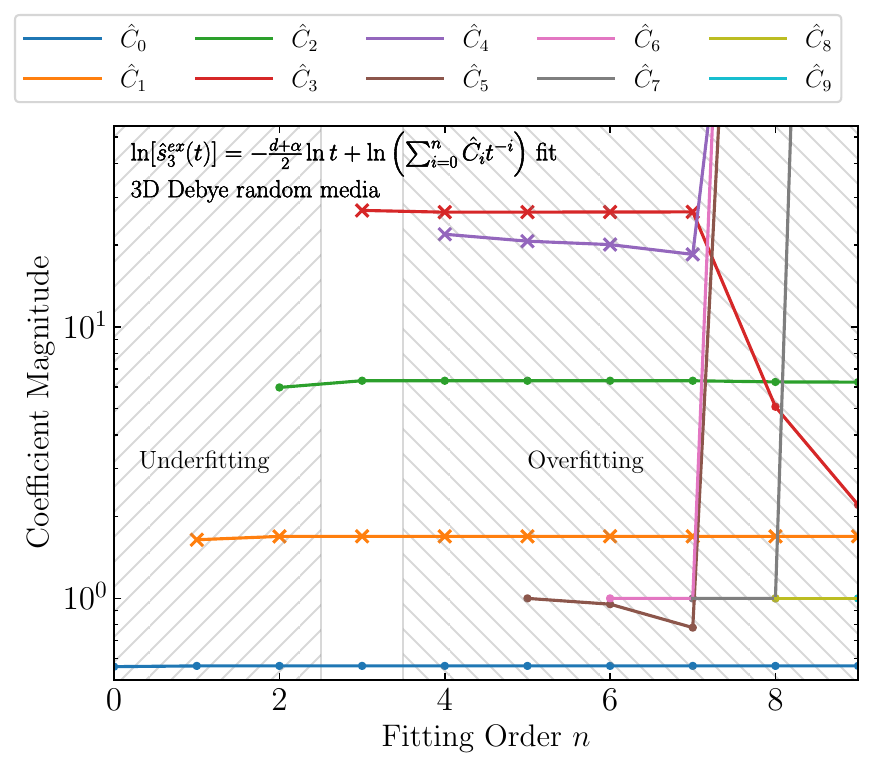}\label{fig4b}}
    \caption{Results of the fitting functions \textbf{(a)} $\ln[\hat{\mathscr{s}}^{ex}_1(t)]$ [Eq. \eqref{eq:fit1}] and \textbf{(b)} $\ln[\hat{\mathscr{s}}^{ex}_3(t)]$ [Eq. \eqref{eq:fit3}] for DRM spreadability [Eq. \eqref{eq:large-t-Debye}]. Positive and negative coefficients are indicated by filled circles and crosses, respectively. The underfitting region ($n<n_o$) and the overfitting region ($n>n_o$) are forward slash and backslash hatched, respectively, while the optimal fitting ($n=n_o$) is left blank.}
\end{figure}

\begin{figure}
    \centering
    \subfloat[]{\includegraphics[width=0.79\linewidth]{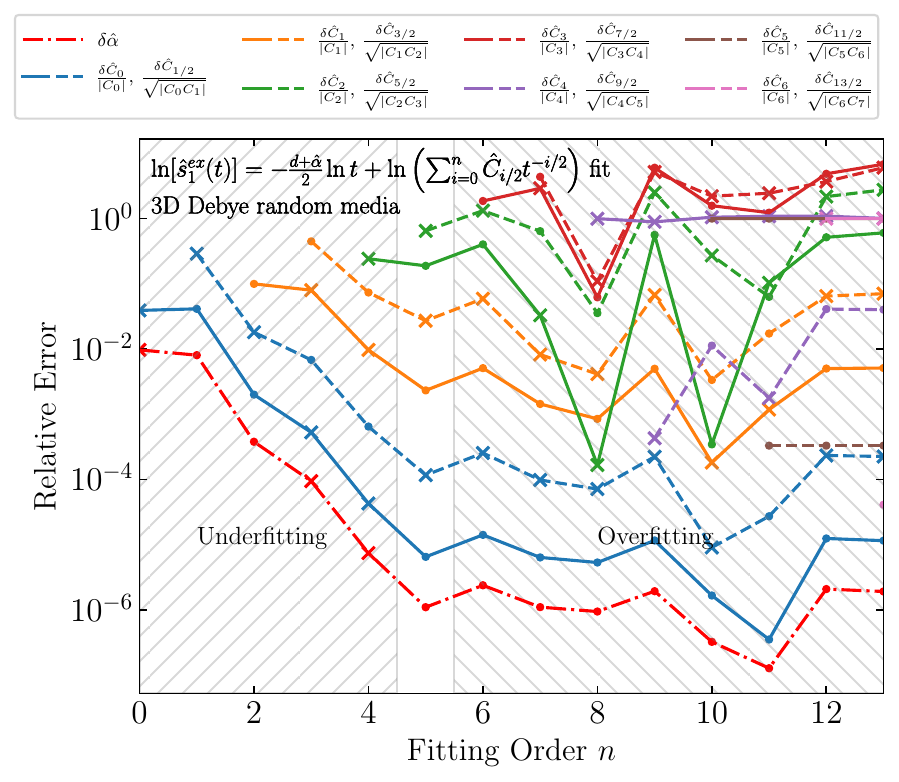}}\\
    \subfloat[]{\includegraphics[width=0.79\linewidth]{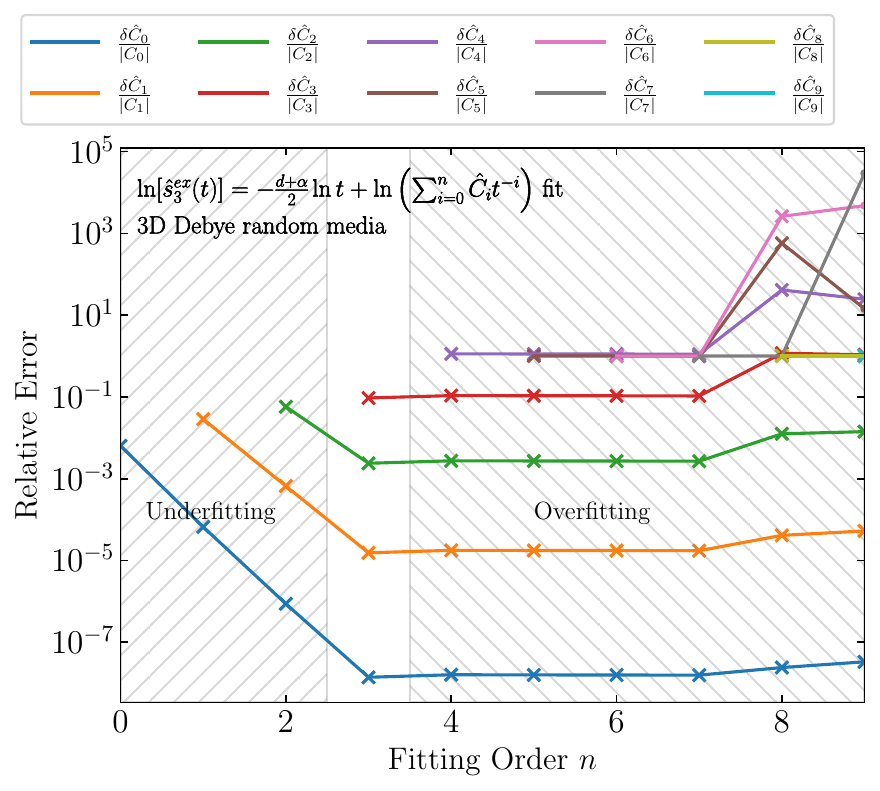}}
    \caption{Relative fitting errors of \textbf{(a)} $\ln[\hat{\mathscr{s}}^{ex}_1(t)]$ [Eq. \eqref{eq:fit1}] and \textbf{(b)} $\ln[\hat{\mathscr{s}}^{ex}_3(t)]$ [Eq. \eqref{eq:fit3}] compared to the exact values for Debye random media [Eq. \eqref{eq:large-t-Debye}]. Positive and negative relative errors are indicated by filled circles and crosses, respectively. The underfitting region ($n<n_o$) and the overfitting region ($n>n_o$) are forward slash and backslash hatched, respectively, while the optimal fitting ($n=n_o$) is left blank. The fit with $n=0$ in \textbf{(a)} is equivalent to Wang and Torquato's fitting method proposed in \cite{wangDynamicMeasureHyperuniformity2022}.}
    \label{fig5}
\end{figure}

We first implement the Type I fitting procedure for Debye random media, and the results are shown in Fig. \ref{fig4a}.
As can be seen from the figure, $|\delta\hat{\alpha}|=|\hat{\alpha}-0|$ continuously decreases down to $\sim10^{-6}$ as the order $n$ increases up to the optimal order $n_o=5$ and keeps decreasing even in the overfitting region $n>n_o$, so one can conclude that $\alpha=0$ within the error range. Also, the absolute values of half-integer-indexed coefficients (namely $\hat{C}_{1/2}$ and $\hat{C}_{3/2}$ plotted in dashed lines) are systematically smaller than those of integer-indexed coefficients ($\hat{C}_{0}$, $\hat{C}_{1}$, and $\hat{C}_2$ plotted in solid lines); quantitatively, we can verify the dimensionless $|\hat{C}_{1/2}|/\sqrt{|\hat{C}_{0}\hat{C}_{1}|}\sim10^{-4}\ll1$ and $|\hat{C}_{3/2}|/\sqrt{|\hat{C}_{1}\hat{C}_{2}|}\sim10^{-1}\ll1$. We can thus conclude that $C_{1/2}=C_{3/2}=0$ within the prescribed margin of error, and then reasonably assume the analyticity of the spectral density at the origin. 

Consequently, we then fit the same data with the Type III fitting function $\ln[\hat{\mathscr{s}}^{ex}_3(t)]$, the results of which are shown in Fig. \ref{fig4b}. As the fitting order $n$ increases from $0$ to $n_o=3$, the fits steadily converge. Then the overfitting effect kicks in, and the fits become increasingly inaccurate. The final optimal fit ($n_o=3$) is given by
\begin{equation}
    \hat{\mathscr{s}}^{ex}(t)=t^{-\frac{d}{2}}(\hat{C}+\hat{C}_1t^{-1}+\hat{C}_2t^{-2}+\hat{C}_3t^{-3}),
\end{equation}
agreeing perfectly with the exact expansion \eqref{eq:large-t-Debye} truncated at the same order. In addition to the exponent $\alpha=0$ and the coefficients $\hat{C}_i$, whose relative errors compared to the exact values are plotted in Fig. \ref{fig5}, this fit indicates the analyticity of the spectral density $\tilde{\chi}_{_V}(k)$ at the origin up through $k^{\alpha+\beta_3}=k^{6}$, or equivalently, the existence of the moment $M_{d-1+6/2}(\chi_{_V})=M_{5}(\chi_{_V})$, i.e., $\chi_{_V}(r)=o(r^{-6})$ as $r\to\infty$.

\subsubsection{Disordered hyperuniform media}

\begin{figure}
    \centering
    \subfloat[]{\includegraphics[width=0.79\linewidth]{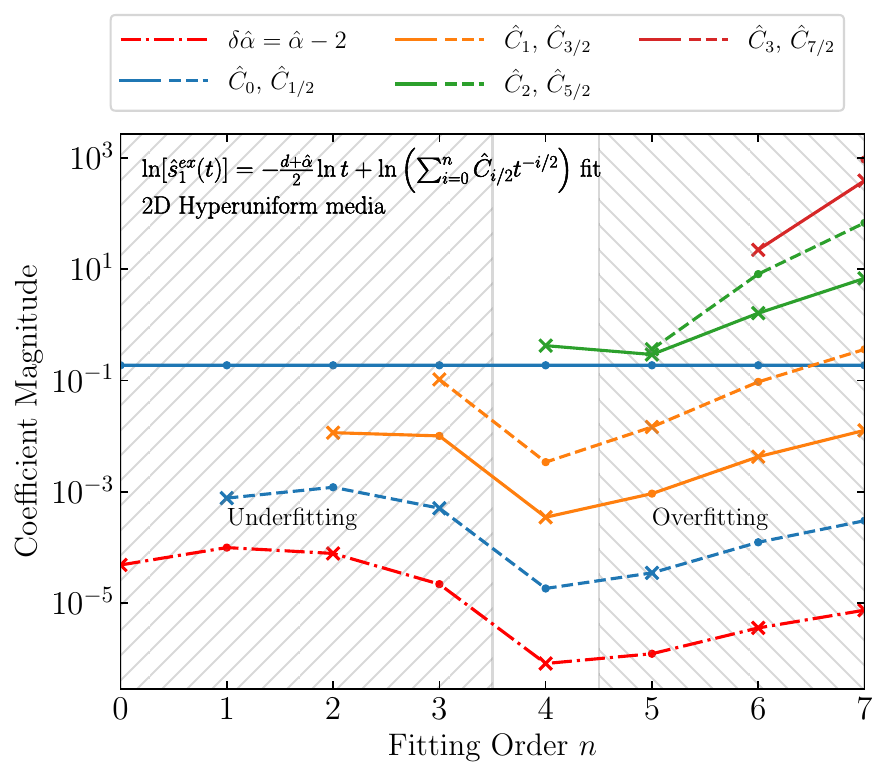}\label{fig6a}}\\
    \subfloat[]{\includegraphics[width=0.79\linewidth]{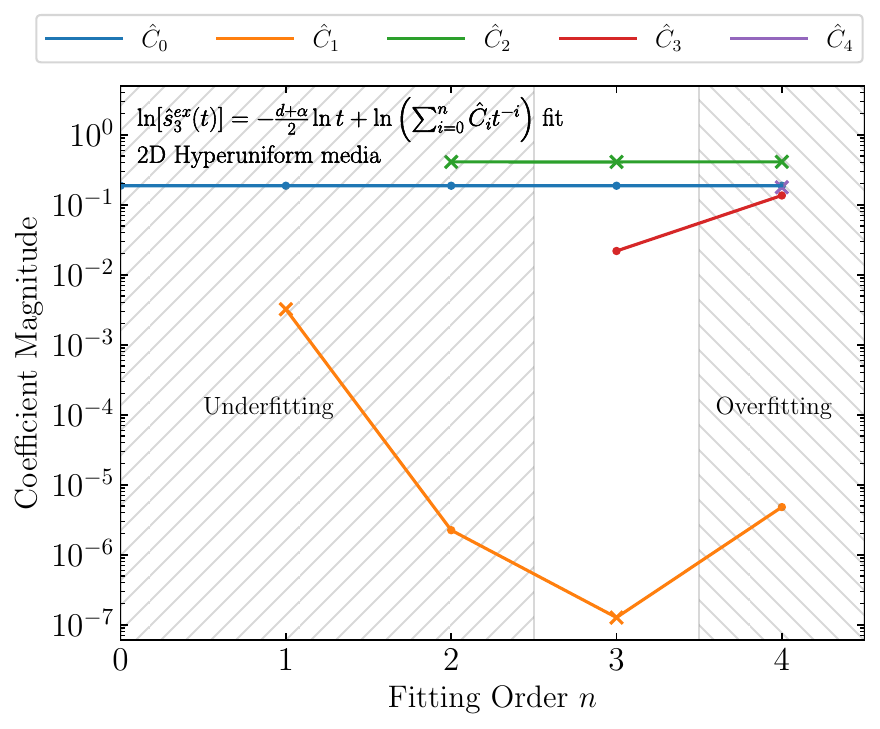}\label{fig6b}}
    \caption{\textbf{(a)} Type I [Eq. \eqref{eq:fit1}] and \textbf{(b)} Type III [Eq. \eqref{eq:fit3}] fits of $\mathscr{s}^{ex}(t)$ for 2D disordered hyperuniform media [Eq. \eqref{eq:large-t-hyper}]. Positive and negative coefficients are indicated by filled circles and crosses, respectively. The underfitting region ($n<n_o$) and the overfitting region ($n>n_o$) are forward slash and backslash hatched, respectively, while the optimal fitting ($n=n_o$) is left blank.}
\end{figure}

\begin{figure}
    \centering
    \subfloat[]{\includegraphics[width=0.79\linewidth]{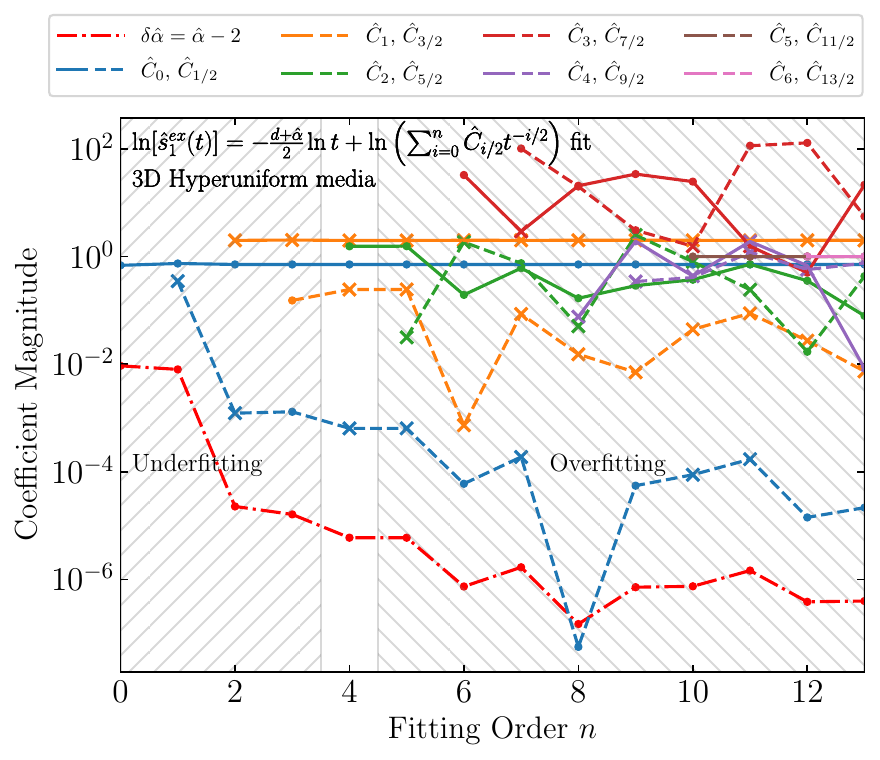}\label{fig7a}}\\
    \subfloat[]{\includegraphics[width=0.79\linewidth]{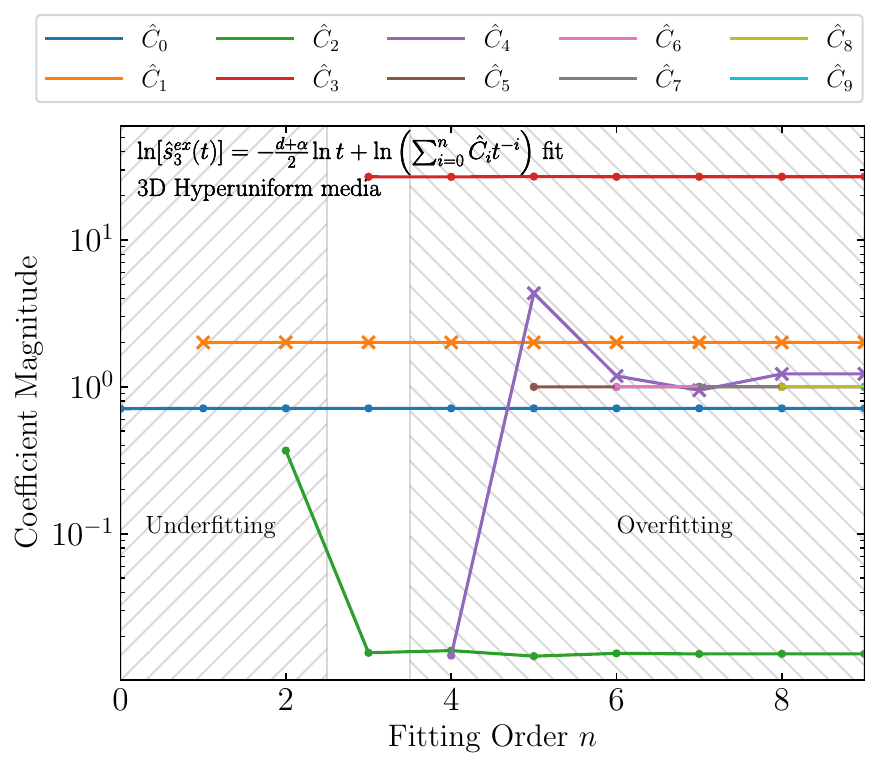}\label{fig7b}}
    \caption{\textbf{(a)} Type I [Eq. \eqref{eq:fit1}] and \textbf{(b)} Type III [Eq. \eqref{eq:fit3}] fits of $\mathscr{s}^{ex}(t)$ for 3D disordered hyperuniform media [Eq. \eqref{eq:large-t-hyper}]. Positive and negative coefficients are indicated by filled circles and crosses, respectively. The underfitting region ($n<n_o$) and the overfitting region ($n>n_o$) are forward slash and backslash hatched, respectively, while the optimal fitting ($n=n_o$) is left blank.}
\end{figure}

Figure \ref{fig6a} plots the results of the Type I fitting function $\ln[\hat{\mathscr{s}}^{ex}_1(t)]$ for $d=2$, where $|\delta\hat{\alpha}|=|\hat{\alpha}-2|$ continuously decreases down to $\lesssim10^{-5}$ as the order $n$ increases up to the optimal order $n_o=4$, implying $\alpha=2$ within the error range.
Notice that the magnitudes of $\hat{C}_{1/2}$, $\hat{C}_{1}$, and $\hat{C}_{3/2}$ are systematically much smaller than those of $\hat{C}_{0}$ and $\hat{C}_{2}$, and their signs constantly flip across different orders $n$ in contrast to order coefficients, implying that $\hat{C}_{1/2}=\hat{C}_{1}=\hat{C}_{3/2}=0$ within the error range.
We thus assume that the spectral density is analytic at the origin with $\alpha=2$ and fit the $\mathscr{s}^{ex}(t)$ data with the Type III fitting function $\ln[\hat{\mathscr{s}}^{ex}_3(t)]$, results of which are plotted in Fig. \ref{fig6b}. 

At the optimal fitting order $n_o=3$, we have $|\hat{C}_{1}|/\sqrt{|\hat{C}_{0}\hat{C}_{2}|}\sim10^{-6}\ll1$, agreeing again with our assumption that $C_{1/2}=C_{1}=C_{3/2}=0$. The final optimal fit ($n_o=3$) is given by
\begin{equation}
    \hat{\mathscr{s}}^{ex}(t)=t^{-\frac{d+2}{2}}(\hat{C}+\hat{C}_1t^{-1}+\hat{C}_2t^{-2}+\hat{C}_3t^{-3}),
\end{equation}
where $|\hat{C}_{1}|/\sqrt{|\hat{C}_{0}\hat{C}_{2}|}\sim10^{-6}\ll1$ and $|\hat{C}_{3}|/\sqrt{|\hat{C}_{2}^3/\hat{C}_{0}|}\sim10^{-1}<1$, agreeing
with the exact expansion at the same order
\begin{equation}
    \mathscr{s}^{ex}(t)=t^{-\frac{d+2}{2}}(C+C_2t^{-2}+...).
\end{equation}

In addition to the exponent $\alpha=2$ and the coefficients $\hat{C}_i$ themselves, we can infer that the spectral density $\tilde{\chi}_{_V}(k)$ is analytic at the origin up to the order of $k^{\alpha+\beta_3}=k^{8}$. Equivalently, the moment $M_{d-1+8/2}(\chi_{_V})=M_5(\chi_{_V})$ exists, i.e., $\chi_{_V}(r)=o(r^{-6})$ as $r\to\infty$. The inferences that $\alpha=2$, $|\hat{C}_{1}|=0$, and $|\hat{C}_{3}|\approx0$ further imply that $M_{d-1}(\chi_{_V})=0$, $M_{d+1}(\chi_{_V})=0$, and $M_{d+3}(\chi_{_V})\approx0$.

Similarly, we fit the spreadability data of the 3D hyperuniform medium to the Type I fitting function $\ln[\hat{\mathscr{s}}^{ex}_1(t)]$ and plot the results in Fig. \ref{fig7a}. We find that $|\delta\hat{\alpha}|=|\hat{\alpha}-2|\sim10^{-5}$ and $|\hat{C}_{1/2}|/\sqrt{|\hat{C}_{0}\hat{C}_{1}|}\sim10^{-4}\ll1$ at the optimal order $n_o=4$, and thus infer that the spectral density is analytic at the origin with $\alpha=2$. Figure \ref{fig7b} shows the results of the Type III fitting function $\ln[\hat{\mathscr{s}}^{ex}_3(t)]$. The large variance of $\hat{C}_2$ compared to $\hat{C}_0$, $\hat{C}_1$, and $\hat{C}_3$ around the optimal order $n_o=3$ and the fact that $|\hat{C}_{2}|/\sqrt{|\hat{C}_{1}\hat{C}_{3}|}\sim10^{-3}\ll1$ indicates that $C_2=0$. The final optimal fit ($n_o=3$) is given by
\begin{equation}
    \hat{\mathscr{s}}^{ex}(t)=t^{-\frac{d+2}{2}}(\hat{C}+\hat{C}_1t^{-1}+\hat{C}_2t^{-2}+\hat{C}_3t^{-3}),
\end{equation}
where $|\hat{C}_{2}|/\sqrt{|\hat{C}_{1}\hat{C}_{3}|}\sim10^{-3}\ll1$, agreeing
perfectly with the exact expansion at the same order
\begin{equation}
    \mathscr{s}^{ex}(t)=t^{-\frac{d+2}{2}}(C+C_1t^{-1}+C_3t^{-3}+...).
\end{equation}

In addition to the exponent $\alpha=2$ and the coefficients $\hat{C}_i$ themselves, we can conclude that the spectral density $\tilde{\chi}_{_V}(k)$ is analytic at the origin up to the order of $k^{\alpha+\beta_3}=k^{8}$, or equivalently, the moment $M_{d-1+8/2}(\chi_{_V})=M_6(\chi_{_V})$ exists, i.e., $\chi_{_V}(r)=o(r^{-7})$ as $r\to\infty$. The inferences that $\alpha=2$ and $|\hat{C}_{1}|\approx0$ further imply that $M_{d-1}(\chi_{_V})=0$ and $M_{d+1}(\chi_{_V})=0$.

\subsubsection{Antihyperuniform media}
\begin{figure}
    \centering
    \subfloat[]{\includegraphics[width=0.79\linewidth]{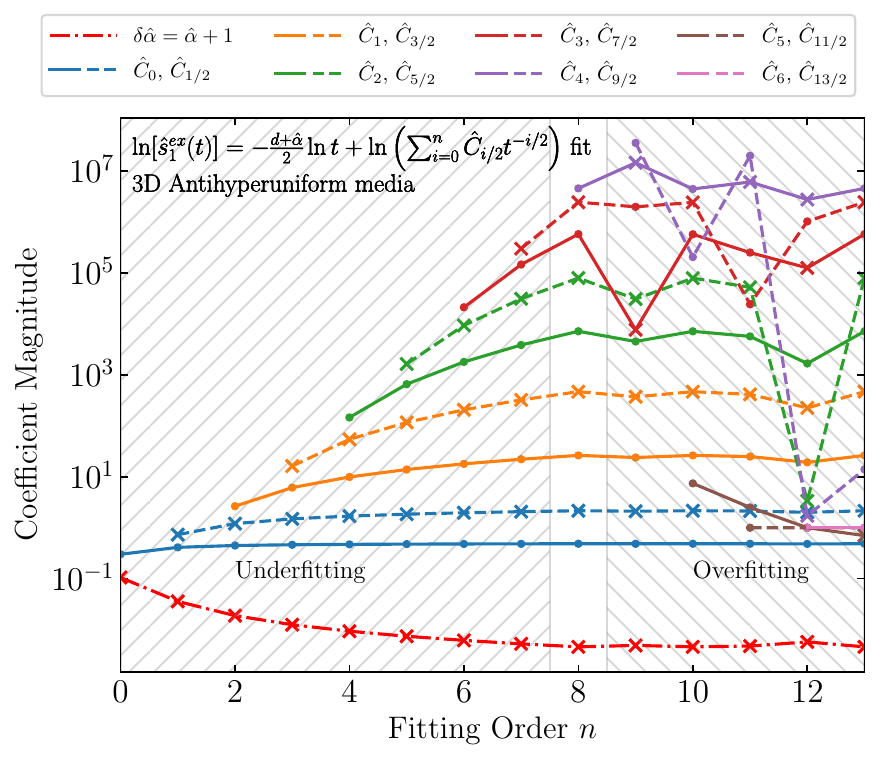}\label{fig8a}}\\
    \subfloat[]{\includegraphics[width=0.79\linewidth]{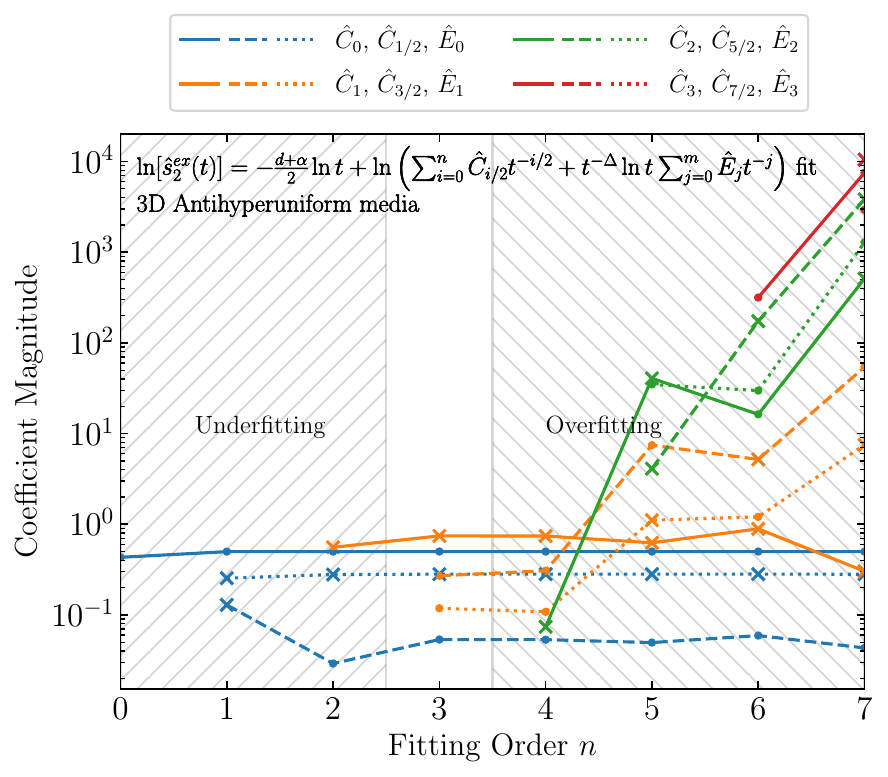}\label{fig8b}}
    \caption{\textbf{(a)} Type I [Eq. \eqref{eq:fit1}] and \textbf{(b)} type II [Eq. \eqref{eq:fit2}] fits of $\mathscr{s}^{ex}(t)$ for 3D antihyperuniform media  [Eq. \eqref{eq:large-t-anti}]. Positive and negative coefficients are indicated by filled circles and crosses, respectively. The underfitting region ($n<n_o$) and the overfitting region ($n>n_o$) are forward slash and backslash hatched, respectively, while the optimal fitting ($n=n_o$) is left blank.}
\end{figure}
 Figure \ref{fig8a} plots the results of the Type I fits of $\ln[\hat{\mathscr{s}}^{ex}_1(t)]$, where $|\delta\hat{\alpha}|=|\hat{\alpha}+1|$ continuously decreases down to $\lesssim10^{-2}$ as the order $n$ increases up to the optimal order $n_o=8$, implying $\alpha=-1$ within the error range. Since $\alpha=-1$ is negative and odd, the spectral density cannot be analytic at the origin. We then fit the same sample data to the Type II fitting function $\ln[\hat{\mathscr{s}}^{ex}_2(t)]$ and plot the results in Fig. \ref{fig8b}. The final optimal fit ($n_o=3$) is given by
\begin{equation}
    \begin{split}
        \hat{\mathscr{s}}^{ex}(t)&=t^{-\frac{d-1}{2}}(\hat{C}+\hat{C}_{1/2}t^{-1/2}+\hat{C}_1t^{-1}\\&\phantom{=}+\hat{C}_{3/2}t^{-3/2})+t^{-\frac{d}{2}}\ln t(\hat{E}+\hat{E}_1t^{-1}),
    \end{split}
\end{equation}
matching the exact expansion truncated at the same order perfectly. In addition to the exponent $\alpha=-1$ and the coefficients $\hat{C}_{i/2}$ and $\hat{E}_i$, we can also infer from the fitting that the autocovariance function behaves like $\chi_{_V}(r)\approx r^{-d-\alpha}(K+K_{1/2}r^{-1}+K_1r^{-2}+K_{3/2}r^{-3})\sim Kr^{-2}$ as $r\to\infty$.

\subsection{Robustness to Random Noise}\label{subsec:noise}

Next, we carry out our analysis by adding random noise to the exact spreadability data to mimic real data.
Specifically, we add independent and identically distributed truncated Gaussian noise $\xi(t)$
for each exact spreadability $\mathscr{s}^{ex}(t)$ datum, i.e., the noise amplitude at each time step is independently sampled from a normal distribution 
$\mathcal{N}(0, \sigma²)$, but truncated to lie in the interval $[-3\sigma, 3\sigma]$. This ensures that the noise at each datum does not exceed three standard deviations from the mean, preventing unrealistic outliers in the simulation.
For each $\sigma$, we run 50 independent realizations and compute the mean and standard deviation of the fits.

\begin{figure}
    \centering
    \includegraphics[width=0.79\linewidth]{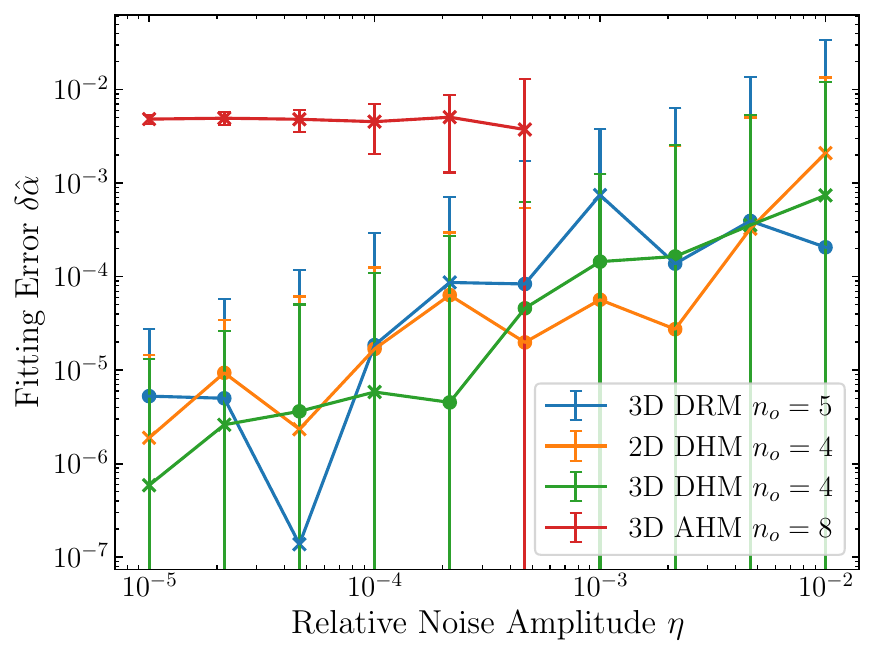}
    \caption{The fitting error $\delta\hat{\alpha}=\hat{\alpha}-\alpha$ with respect to the relative noise amplitude $\eta=\frac{\sigma}{\mathscr{s}^{ex}(t_{\text{max}})}$ for Debye random media (DRM), disordered hyperuniform media (DHM), and antihyperuniform media (AHM), at their respective optimal order $n_o$. Positive and negative errors are indicated by filled circles and crosses, respectively.}
    \label{fig9}
\end{figure}

The error of the exponent $\hat{\alpha}$ computed this way for the Type I fit of the examples in the last subsection is shown in Fig. \ref{fig9}.
Clearly, the fitting error $\delta\hat{\alpha}=\hat{\alpha}-\alpha$ is about the same order of magnitude as the relative noise amplitude $\eta=\frac{\sigma}{\mathscr{s}^{ex}(t_{\text{max}})}$, i.e., $|\delta\hat{\alpha}|\sim\eta$, unless the {\it intrinsic error} (without noise) is greater by itself. As can be seen from Fig. \ref{fig5}, the fitting errors of the coefficients essentially synchronize with $\delta\hat{\alpha}$.
Therefore, as long as the magnitude of random noise is sufficiently smaller than the normalized excess spreadability, i.e., $\eta=\frac{\sigma}{\mathscr{s}^{ex}(t_{\text{max}})}\ll1$, it only adds to the fitting error but, importantly, does not affect the fitting procedure and the qualitative conclusions.

For an ensemble of $\mathscr{s}^{ex}(t)$ data obtained experimentally or numerically, the noise amplitude $\sigma$ can be estimated by the standard deviation of the measured $\mathscr{s}^{ex}(t)$ value at the fixed time $t$. The relative error $\eta=\frac{\sigma}{\mathscr{s}^{ex}(t_{\text{max}})}$ can be calculated accordingly and used for error analysis. If $\eta$ is comparable to the estimated error of $\hat{\alpha}$, the random noise will be the main limitation of the precision. On the other hand, if the estimated $\delta\hat{\alpha}$ is much greater than $\eta$, the precision is likely to be improved simply by acquiring data for longer times.

\section{Discussion and Conclusion}\label{sec:discussion-and-conclusion}

In summary, we have developed an improved asymptotic fitting scheme based on our theory of the long-time asymptotics of the normalized excess spreadability $\mathscr{s}^{ex}(t)$ for disordered two-phase random media that are not stealthy hyperuniform.
We theoretically classified the asymptotics of all known kinds of two-phase media based on the mathematical properties of their correlation functions $\chi_{_V}(r)$ and $\tilde{\chi}_{_V}(k)$, and numerically show the accuracy of the asymptotic expansion truncated at different orders for the DRM test case.
Based on the analyticity properties of $\tilde{\chi}_{_V}(k)$ at the origin [equivalently, the long-distance decay rate of $\chi_{_V}(r)$], we proposed three types of fitting functions for the long-time spreadability fitting.
We then demonstrated their ability to accurately extract large-scale two-point correlation information from spreadability data with typical nonhyperuniform (Debye), hyperuniform, and antihyperuniform example models.
From our asymptotic fitting procedure, we not only obtained accurate values of the exponent $\alpha$ and the coefficients $C_{\beta_i/2}$ with high precision, but also gained insights into certain mathematical constraints of the correlation functions, including analyticity of $\tilde{\chi}_{_V}(k)$ at the origin, existence of the moment $M_l(\chi_{_V})$, and long-distance decay rate of $\chi_{_V}(r)$.

Our asymptotic fitting scheme can be readily implemented to analyze real data of $\mathscr{s}^{ex}(t)$ obtained from NMR experiments \cite{torquatoDiffusionSpreadabilityProbe2021} and numerical simulations \cite{skolnickSimulatedDiffusionSpreadability2023}, and thus helps characterize microstructures in these realistic and practical cases.
There are also microstructures of interest requiring separate considerations that are not fully covered in this work, including those whose spectral density cannot be expanded as powers of $k$ at the origin.
Examples involve quasicrystalline media due to the dense Bragg peaks in their spectral density \cite{hitin-bialusHyperuniformityClassesQuasiperiodic2024}, and stealthy hyperuniform media characterized by $\tilde{\chi}_{_V}(k)\equiv0$ for $k<K$.
Disordered stealthy hyperuniform materials have unusual structural characteristics (hidden order at large length scales) and physical properties, including desirable photonic and transport properties \cite{zhangTransportGeometricalTopological2016a,torquatoDisorderedHyperuniformHeterogeneous2016b}.
Generalizing our asymptotic fitting procedure to these exotic microstructures is an open and intriguing problem. 
% \sy{For example, based on the robustness against the random noise (Sec. \ref{subsec:noise}), we would expect the fitting procedure to apply conditionally to the fluctuating asymptotics of quasicrystalline media.}
Indeed, in future work, we will use our fitting procedure that includes correction terms to estimate the exponent $\alpha$ for quasicrystals, going beyond the use of just the leading term as done in \cite{hitin-bialusHyperuniformityClassesQuasiperiodic2024}.

A useful byproduct of our long-time asymptotic analysis is the two-point Pad\'e approximants of  $\mathscr{s}^{ex}(t)$ given by Eq. \eqref{eq:2pt-pade}, which approximate the spreadability well for all times.  
We note that the accuracy of such expressions for all times can be further improved by forming three-point Pad\'e approximants, as noted in Sec. \ref{sec:approximants}. 
These accurate approximants can then be used to target desired spreadability behaviors at short, intermediate, and long times.
This parametrization scheme for the spreadability for all times could be fruitfully applied in the inverse design of microstructures \cite{torquatoInverseOptimizationTechniques2009} 
with targeted spreadabilities \cite{shiThreedimensionalConstructionHyperuniform2025}, which can then be fabricated 
using modern additive manufacturing techniques \cite{bhushanOverviewAdditiveManufacturing2017,shiraziReviewPowderbasedAdditive2015,tumblestonContinuousLiquidInterface2015,wongReviewAdditiveManufacturing2012} and colloids
with tailored interactions \cite{chenBinaryMixturesCharged2018,maOptimizedLargeHyperuniform2020}.

\section*{Acknowledgements}

The authors were supported by the Army Research Office under Cooperative Agreement No. W911NF-22-2-0103.

\section*{Data availability}

The data that support the findings of this article are not publicly available because they contain commercially sensitive information. The data are available from the authors upon reasonable request.

\appendix

\section{Fourier Transform of Power-Law Terms}\label{app:fourier-pw}

When $-d<\Re[\zeta]<0$, it can be proven that \cite{hormanderAnalysisLinearPartial2003,kanwalGeneralizedFunctions2004}
\begin{equation}
    \begin{split}
        \mathcal{F}[r^{-d-\zeta}](\mathbf{k})\equiv\int_{\mathbb{R}^d}r^{-d-\zeta}\mathrm{e}^{-\mathrm{i}\mathbf{k}\cdot\mathbf{r}}\mathrm{d}\mathbf{r}=G_d(\zeta)k^\zeta,\\G_d(\zeta)=\frac{\pi^{d/2}\Gamma(-\zeta/2)}{2^{\zeta}\Gamma[(d+\zeta)/2]}.
    \end{split}
\end{equation}
$G_d(\zeta)k^\zeta$ is a meromorphic function on the complex plane, so we can easily extend this result to any $\zeta>-d$ and $\zeta\ne0,2,4,...$ via analytic continuation. Around the poles $\zeta_0=2l$, $l=0,1,2,...$, we have the Laurent expansion
\begin{equation}
    \begin{split}
        G_d(\zeta)=\frac{g^{(-1)}_{d,l}}{\zeta-\zeta_0}+g^{(0)}_{d,l}+O[\zeta-\zeta_0],\\k^\zeta=k^{\zeta_0}\Big[1+(\zeta-\zeta_0)\ln k+O[(\zeta-\zeta_0)^2]\Big],
    \end{split}
\end{equation}
so that
\begin{equation}
    \begin{split}
        G_d(\zeta)k^\zeta=\frac{g^{(-1)}_{d,l}k^{\zeta_0}}{\zeta-\zeta_0}+g^{(0)}_{d,l}k^{\zeta_0}+g^{(-1)}_{d,l}k^{\zeta_0}\ln k\\+O[\zeta-\zeta_0].
    \end{split}
\end{equation}
At the poles $\zeta_0=0,2,4,...$, the finite part gives the meaningful result
\begin{equation}
    \begin{split}
        \mathcal{F}[r^{-d-\zeta_0}](\mathbf{k})&=\text{f.p.}\left[G_d(\zeta_0)k^\zeta_0\right]\\
        &=g^{(0)}_{d,l}k^{\zeta_0}+g^{(-1)}_{d,l}k^{\zeta_0}\ln k,
    \end{split}
\end{equation}
and it thus contains a generalized power-law term $k^{\zeta_0}\ln k$. 

\section{Pad\'e Approximant}\label{app:pade}

The Pad\'e approximant technique is a well-known procedure
to approximate a function $f(z)$ as a rational function and often provides
a superior approximation than its corresponding Taylor series
\cite{bakerEssentialsPadeApproximants1975}. Specifically, given the Taylor series $f(z)=\sum_{k=0}^\infty c_kz^k$, one constructs the rational function $R(z)$ that matches the Taylor series
up to that order, i.e.,
\begin{equation}
    R(z)=\frac{\sum_{k=0}^Ma_kz^k}{1+\sum_{k=1}^Nb_kz^k}
\end{equation}
such that it obeys the condition
\begin{equation}
    R^{(k)}(0)=f^{(k)}(0),\hspace{1em}k=0,1,...,M+N.
\end{equation}
We refer to $R(z)$ as the $[M,N]$ Pad\'e approximant of $f(z)$. We can generalize its use by substituting $z\to(x-x_0)^s$.

\section{Exact Two-Phase Medium Models}\label{app:models}

\subsection{Debye Random Media}\label{app:drm}

The Debye random medium (DRM) is a prototypical model of ``typical" nonhyperuniform two-phase media \cite{yeongReconstructingRandomMedia1998a}, for which the autocovariance function
\begin{equation}
    \chi_{_V}(r)=\phi_1\phi_2\exp(-r/a)
\end{equation}
and the spectral density
\begin{equation}
    \tilde{\chi}_{_V}(k) = \frac{\phi_1\phi_22^d\pi^{d-1}a^d}{\omega_{d-1}[1+(k a)^2]^{(d+1)/2}}
\end{equation}
are known exactly \cite{torquatoDiffusionSpreadabilityProbe2021}.
Without loss of generality, we set $D=a=1$.
The exponential decay of the autocovariance function guarantees the existence of all its moments, i.e., $M_n(\chi_{_V})=\phi_1\phi_2\Gamma(n+1)<\infty$, and the spectral density is clearly analytic at the origin, i.e., $\tilde{\chi}_{_V}(k) =\sum_{n=0}^\infty B_nk^{2n}$ containing only even powers of $k$.
Correspondingly, the asymptotic expansion of $\mathscr{s}^{ex}(t)$ takes the form of
\begin{equation}\label{eq:large-t-Debye}
    \begin{split}
        \mathscr{s}^{ex}(t)&=\sum_{n=0}^{\infty}2\frac{(-1)^n}{n!}\frac{\Gamma(d+2n)}{\Gamma(d/2)}(4t)^{-d/2-n}\\
        &=\sum_{n=0}^{\infty}C_nt^{-d/2-n}\hspace{2em}(t\to\infty).
    \end{split}
\end{equation}
Similarly, we can find its small-$t$ expansion
\begin{equation}\label{eq:small-t-Debye}
    \begin{split}
        \mathscr{s}^{ex}(t)&=\sum_{n=0}^{\infty}\frac{(-1)^n}{n!}\frac{\Gamma(\frac{n+d}{2})}{\Gamma(d/2)}(4t)^{n/2}\\
        &=\sum_{n=0}^\infty A_nt^{n/2}\hspace{2em}(t\to0).
    \end{split}
\end{equation}

\subsection{Disordered Hyperuniform Media}\label{app:dhu}

To model hyperuniform two-phase media in $\mathbb{R}^d$, Torquato \cite{torquatoDisorderedHyperuniformHeterogeneous2016b} proposed the following  family of autocovariance functions:
\begin{equation}
 \chi_{_V}(r)=\phi_1\phi_2\exp({-r/a})\frac{\cos(qr +\theta)}{\cos\theta},
\label{eq:Xr_DHU}
\end{equation}
where the parameters $q$ and $\theta$ are the wavenumber and phase associated with the oscillations of $\chi_{_V}(r)$, respectively, and $a$ is a correlation length.  In the special case in which $\theta=0$, Torquato showed that the corresponding autocovariance function satisfies all of the necessary realizability conditions and hyperuniformity constraint for $d=2$ if $(qa)^2=1$ and for $d=3$ if $(qa)^2=1/3$. Thus, the spectral densities for $d=2$ and $d=3$ are respectively given by
\begin{equation}\label{eq:Xk_DHU}
    \frac{{\tilde \chi}_{_V}(k)}{\phi_1\phi_2}=
    \begin{cases}
         \frac{ 2\pi a^2[\sqrt{(ka)^4+4}+4]}{[(ka)^4+4]^{3/2} }[A(k)-A^{-1}(k)]&(d=2),\\
          \frac{216\pi a^3[3 (ka)^2+8](ka)^2}{[9 (ka)^4+ 12 (ka)^2+16]^2}&(d=3),
    \end{cases}
\end{equation}
where $A(k)=\sqrt{\frac{\sqrt{(ka)^4+4}+(ka)^2}{2}}$.

We again set $D=a=1$. As with DRM, the exponentially decaying autocovariance function \eqref{eq:Xr_DHU} guarantees the existence of all its moments and the analyticity of the spectral density \eqref{eq:Xk_DHU} at the origin. Specifically, we have
\begin{equation}
    \frac{M_n(\chi_{_V})}{\phi_1\phi_2}=
    \begin{cases}
        \frac{n!}{2^{(n+1)/2}}\cos[(n+1)\pi/4]&(d=2),\\
        \frac{n!3^{(n+1)/2}}{2^{n+1}}\cos[(n+1)\pi/6]&(d=3),
    \end{cases}
\end{equation}
and
\begin{equation}
    \begin{split}
        \frac{{\tilde \chi}_{_V}(k)}{\phi_1\phi_2}&=
        \begin{cases}
            k^2\left(\frac{3 \pi}{4}-\frac{35 \pi}{128}k^4+\frac{693 \pi}{8192}k^{8}+...\right)&(d=2),\\
            k^2\left(\frac{27 \pi}{4}-\frac{243 \pi}{32}k^2+\frac{3645 \pi}{512}k^6+...\right)&(d=3),
        \end{cases}\\
        &=
        \begin{cases}
            k^2\sum_{\begin{subarray}{l}
                i=0\\i \text{ even}
            \end{subarray}}^\infty B_ik^{2i}&(d=2),\\
            k^2\sum_{\begin{subarray}{l}
                i=0\\i \,\mathrm{mod}\, 3 \ne 2
            \end{subarray}}^\infty B_ik^{2i}&(d=3).
        \end{cases}
    \end{split}
\end{equation}
Correspondingly, the long-time asymptotic expansion of $\mathscr{s}^{ex}(t)$ is given by
\begin{equation}\label{eq:large-t-hyper}
    \begin{split}
        \mathscr{s}^{ex}(t)&=
        \begin{cases}
            t^{-2}\left(\frac{3}{16} - \frac{105}{256}t^{-2} + \frac{10395}{4096}t^{-4}+...\right)&(d=2),\\
            t^{-\frac{5}{2}}\big(\frac{81}{64 \sqrt{\pi }}-\frac{3645}{1024 \sqrt{\pi }}t^{-1}+...\big)&(d=3),
        \end{cases}\\
        &=
        \begin{cases}
            t^{-\frac{d+2}{2}}\sum_{\begin{subarray}{l}
                i=0\\i \text{ even}
            \end{subarray}}^\infty C_{i}t^{-i}&(d=2),\\
            t^{-\frac{d+2}{2}}\sum_{\begin{subarray}{l}
                i=0\\i \,\mathrm{mod}\, 3 \ne 2
            \end{subarray}}^\infty C_{i}t^{-i}&(d=3).
        \end{cases}
    \end{split}
\end{equation}
Note that we have $C_{2i+1}=0$ for $d=2$ and $C_{3i+2}=0$ for $d=3$ because $M_{d+2i+1}(\chi_{_V})=0$ for $d=2$ and $M_{d+3i+2}(\chi_{_V})=0$ for $d=3$, respectively.

\subsection{Antihyperuniform Media}\label{app:ahu}

Here, we examine a 3D antihyperuniform model proposed by Torquato \cite{torquatoDiffusionSpreadabilityProbe2021}, which is defined by the autocovariance function
\begin{equation}
    \frac{\chi_{_V}(r)}{\phi_1\phi_2}=(1+r/a)^{-2},
\end{equation}
which meets all known necessary realizability conditions on a valid autocovariance function \cite{torquatoDisorderedHyperuniformHeterogeneous2016b}. The corresponding spectral density is given by
\begin{equation}
    \begin{split}
        \frac{\tilde{\chi}_{_V}(k)}{\phi_1\phi_2}=\frac{4\pi a^2}{ka}\{\mathrm{Ci}(ka)[ka\cos(ka)+\sin(ka)]\\+\mathrm{Ssi}(ka)[ka\sin(ka)-\cos(ka)]\},
    \end{split}
\end{equation}
where $\mathrm{Ci}(x)\equiv-\int_x^\infty\mathrm{d}t\cos(t)/t$ is the cosine integral, $\mathrm{Ssi}(x) \equiv \mathrm{Si}(x) -\pi/2$ is the shifted sine integral and $\mathrm{Si}(x) \equiv \int_0^x dt \sin(t)/t$ is the sine integral. We see that ${\tilde \chi}_{_V}(k)\sim2\pi^2a/k$ in the limit $k\to 0$, which is consistent with the power-law decay $a^2/r^2$  of the $\chi_{_V}(r)$ in the limit $r \to \infty$.

We again set $D=a=1$. Clearly, any $n$-th order moment $M_n(\chi_{_V})$ for $n \ge 1$ is unbounded. This result implies that the spectral density is not analytic at the origin, but this function can be expanded about $k=0$ as
\begin{equation}
    \begin{split}
        \frac{\tilde{\chi}_{_V}(k)}{\phi_1\phi_2}&=\frac{2 \pi ^2}{k}+4 \pi  (2 \ln k+2 \gamma_e -1)-3 \pi ^2 k\\&\phantom{=}-\frac{2}{9}\pi k^2 (12 \ln k+12 \gamma_e -19)+\frac{5 }{12}\pi ^2 k^3+...\\
        &=k^{-1}(B+B_{1/2}k+B_1k^2+...)\\&\phantom{=}+\ln k(F+F_1k^2+F_2k^4+...).
    \end{split}
\end{equation}
This, in combination with Eq. \eqref{eq:excess-spreadability} means that the long-time $\mathscr{s}^{ex}(t)$ can be expanded as
\begin{equation}\label{eq:large-t-anti}
    \begin{split}
        \mathscr{s}^{ex}(t)&=\frac{1}{2 t}+\frac{-\ln t+\gamma_e +1-2 \ln2}{2 \sqrt{\pi }t^{3/2}}-\frac{3}{4 t^2}\\&\phantom{=}+\frac{2 \ln t-2 \gamma_e +1+4\ln2}{8 \sqrt{\pi }t^{5/2}}+\frac{5}{24t^3}+...\\
        &=t^{-1}(C+C_{1/2}t^{-1/2}+C_1t^{-1}+...)\\&\phantom{=}+t^{-3/2}\ln t(E+E_1t^{-1}+E_2t^{-2}+...),
    \end{split}
\end{equation}
where $\gamma_e$ is the Euler-Mascheroni constant.


\begin{thebibliography}{99}%
\makeatletter
\providecommand \@ifxundefined [1]{%
 \@ifx{#1\undefined}
}%
\providecommand \@ifnum [1]{%
 \ifnum #1\expandafter \@firstoftwo
 \else \expandafter \@secondoftwo
 \fi
}%
\providecommand \@ifx [1]{%
 \ifx #1\expandafter \@firstoftwo
 \else \expandafter \@secondoftwo
 \fi
}%
\providecommand \natexlab [1]{#1}%
\providecommand \enquote  [1]{``#1''}%
\providecommand \bibnamefont  [1]{#1}%
\providecommand \bibfnamefont [1]{#1}%
\providecommand \citenamefont [1]{#1}%
\providecommand \href@noop [0]{\@secondoftwo}%
\providecommand \href [0]{\begingroup \@sanitize@url \@href}%
\providecommand \@href[1]{\@@startlink{#1}\@@href}%
\providecommand \@@href[1]{\endgroup#1\@@endlink}%
\providecommand \@sanitize@url [0]{\catcode `\\12\catcode `\$12\catcode `\&12\catcode `\#12\catcode `\^12\catcode `\_12\catcode `\%12\relax}%
\providecommand \@@startlink[1]{}%
\providecommand \@@endlink[0]{}%
\providecommand \url  [0]{\begingroup\@sanitize@url \@url }%
\providecommand \@url [1]{\endgroup\@href {#1}{\urlprefix }}%
\providecommand \urlprefix  [0]{URL }%
\providecommand \Eprint [0]{\href }%
\providecommand \doibase [0]{https://doi.org/}%
\providecommand \selectlanguage [0]{\@gobble}%
\providecommand \bibinfo  [0]{\@secondoftwo}%
\providecommand \bibfield  [0]{\@secondoftwo}%
\providecommand \translation [1]{[#1]}%
\providecommand \BibitemOpen [0]{}%
\providecommand \bibitemStop [0]{}%
\providecommand \bibitemNoStop [0]{.\EOS\space}%
\providecommand \EOS [0]{\spacefactor3000\relax}%
\providecommand \BibitemShut  [1]{\csname bibitem#1\endcsname}%
\let\auto@bib@innerbib\@empty
%</preamble>
\bibitem [{\citenamefont {Wedeen}\ \emph {et~al.}(2005)\citenamefont {Wedeen}, \citenamefont {Hagmann}, \citenamefont {Tseng}, \citenamefont {Reese},\ and\ \citenamefont {Weisskoff}}]{wedeenMappingComplexTissue2005a}%
  \BibitemOpen
  \bibfield  {author} {\bibinfo {author} {\bibfnamefont {V.~J.}\ \bibnamefont {Wedeen}}, \bibinfo {author} {\bibfnamefont {P.}~\bibnamefont {Hagmann}}, \bibinfo {author} {\bibfnamefont {W.-Y.~I.}\ \bibnamefont {Tseng}}, \bibinfo {author} {\bibfnamefont {T.~G.}\ \bibnamefont {Reese}},\ and\ \bibinfo {author} {\bibfnamefont {R.~M.}\ \bibnamefont {Weisskoff}},\ }\bibfield  {title} {\bibinfo {title} {Mapping complex tissue architecture with diffusion spectrum magnetic resonance imaging},\ }\href {https://doi.org/10.1002/mrm.20642} {\bibfield  {journal} {\bibinfo  {journal} {Magnetic Resonance in Medicine}\ }\textbf {\bibinfo {volume} {54}},\ \bibinfo {pages} {1377} (\bibinfo {year} {2005})}\BibitemShut {NoStop}%
\bibitem [{\citenamefont {Novikov}\ \emph {et~al.}(2014)\citenamefont {Novikov}, \citenamefont {Jensen}, \citenamefont {Helpern},\ and\ \citenamefont {Fieremans}}]{novikovRevealingMesoscopicStructural2014a}%
  \BibitemOpen
  \bibfield  {author} {\bibinfo {author} {\bibfnamefont {D.~S.}\ \bibnamefont {Novikov}}, \bibinfo {author} {\bibfnamefont {J.~H.}\ \bibnamefont {Jensen}}, \bibinfo {author} {\bibfnamefont {J.~A.}\ \bibnamefont {Helpern}},\ and\ \bibinfo {author} {\bibfnamefont {E.}~\bibnamefont {Fieremans}},\ }\bibfield  {title} {\bibinfo {title} {Revealing mesoscopic structural universality with diffusion},\ }\href {https://doi.org/10.1073/pnas.1316944111} {\bibfield  {journal} {\bibinfo  {journal} {Proceedings of the National Academy of Sciences}\ }\textbf {\bibinfo {volume} {111}},\ \bibinfo {pages} {5088} (\bibinfo {year} {2014})}\BibitemShut {NoStop}%
\bibitem [{\citenamefont {Lee}\ \emph {et~al.}(2020)\citenamefont {Lee}, \citenamefont {Papaioannou}, \citenamefont {Novikov},\ and\ \citenamefont {Fieremans}}]{leeVivoObservationBiophysical2020}%
  \BibitemOpen
  \bibfield  {author} {\bibinfo {author} {\bibfnamefont {H.-H.}\ \bibnamefont {Lee}}, \bibinfo {author} {\bibfnamefont {A.}~\bibnamefont {Papaioannou}}, \bibinfo {author} {\bibfnamefont {D.~S.}\ \bibnamefont {Novikov}},\ and\ \bibinfo {author} {\bibfnamefont {E.}~\bibnamefont {Fieremans}},\ }\bibfield  {title} {\bibinfo {title} {In vivo observation and biophysical interpretation of time-dependent diffusion in human cortical gray matter},\ }\href {https://doi.org/10.1016/j.neuroimage.2020.117054} {\bibfield  {journal} {\bibinfo  {journal} {NeuroImage}\ }\textbf {\bibinfo {volume} {222}},\ \bibinfo {pages} {117054} (\bibinfo {year} {2020})}\BibitemShut {NoStop}%
\bibitem [{\citenamefont {Novikov}\ \emph {et~al.}(2019)\citenamefont {Novikov}, \citenamefont {Fieremans}, \citenamefont {Jespersen},\ and\ \citenamefont {Kiselev}}]{novikovQuantifyingBrainMicrostructure2019}%
  \BibitemOpen
  \bibfield  {author} {\bibinfo {author} {\bibfnamefont {D.~S.}\ \bibnamefont {Novikov}}, \bibinfo {author} {\bibfnamefont {E.}~\bibnamefont {Fieremans}}, \bibinfo {author} {\bibfnamefont {S.~N.}\ \bibnamefont {Jespersen}},\ and\ \bibinfo {author} {\bibfnamefont {V.~G.}\ \bibnamefont {Kiselev}},\ }\bibfield  {title} {\bibinfo {title} {Quantifying brain microstructure with diffusion {{MRI}}: {{Theory}} and parameter estimation},\ }\href {https://doi.org/10.1002/nbm.3998} {\bibfield  {journal} {\bibinfo  {journal} {NMR in Biomedicine}\ }\textbf {\bibinfo {volume} {32}},\ \bibinfo {pages} {e3998} (\bibinfo {year} {2019})}\BibitemShut {NoStop}%
\bibitem [{\citenamefont {Wu}\ \emph {et~al.}(2022)\citenamefont {Wu}, \citenamefont {Jiang}, \citenamefont {Li}, \citenamefont {Zhang}, \citenamefont {Ba}, \citenamefont {Zhang}, \citenamefont {Hsu}, \citenamefont {Sun},\ and\ \citenamefont {Zhang}}]{wuTimeDependentDiffusionMRI2022}%
  \BibitemOpen
  \bibfield  {author} {\bibinfo {author} {\bibfnamefont {D.}~\bibnamefont {Wu}}, \bibinfo {author} {\bibfnamefont {K.}~\bibnamefont {Jiang}}, \bibinfo {author} {\bibfnamefont {H.}~\bibnamefont {Li}}, \bibinfo {author} {\bibfnamefont {Z.}~\bibnamefont {Zhang}}, \bibinfo {author} {\bibfnamefont {R.}~\bibnamefont {Ba}}, \bibinfo {author} {\bibfnamefont {Y.}~\bibnamefont {Zhang}}, \bibinfo {author} {\bibfnamefont {Y.-C.}\ \bibnamefont {Hsu}}, \bibinfo {author} {\bibfnamefont {Y.}~\bibnamefont {Sun}},\ and\ \bibinfo {author} {\bibfnamefont {Y.-D.}\ \bibnamefont {Zhang}},\ }\bibfield  {title} {\bibinfo {title} {Time-{{Dependent Diffusion MRI}} for {{Quantitative Microstructural Mapping}} of {{Prostate Cancer}}},\ }\href {https://doi.org/10.1148/radiol.211180} {\bibfield  {journal} {\bibinfo  {journal} {Radiology}\ }\textbf {\bibinfo {volume} {303}},\ \bibinfo {pages} {578} (\bibinfo {year} {2022})}\BibitemShut {NoStop}%
\bibitem [{\citenamefont {Torquato}(2002)}]{torquatoRandomHeterogeneousMaterials2002}%
  \BibitemOpen
  \bibfield  {author} {\bibinfo {author} {\bibfnamefont {S.}~\bibnamefont {Torquato}},\ }\href@noop {} {\emph {\bibinfo {title} {Random Heterogeneous Materials: Microstructure and Macroscopic Properties}}},\ \bibinfo {series} {Interdisciplinary {{Applied Mathematics}}}\ No.\ \bibinfo {number} {Volume 16}\ (\bibinfo  {publisher} {Springer},\ \bibinfo {address} {New York},\ \bibinfo {year} {2002})\BibitemShut {NoStop}%
\bibitem [{\citenamefont {Brownstein}\ and\ \citenamefont {Tarr}(1979)}]{brownsteinImportanceClassicalDiffusion1979}%
  \BibitemOpen
  \bibfield  {author} {\bibinfo {author} {\bibfnamefont {K.~R.}\ \bibnamefont {Brownstein}}\ and\ \bibinfo {author} {\bibfnamefont {C.~E.}\ \bibnamefont {Tarr}},\ }\bibfield  {title} {\bibinfo {title} {Importance of classical diffusion in {{NMR}} studies of water in biological cells},\ }\href {https://doi.org/10.1103/PhysRevA.19.2446} {\bibfield  {journal} {\bibinfo  {journal} {Physical Review A}\ }\textbf {\bibinfo {volume} {19}},\ \bibinfo {pages} {2446} (\bibinfo {year} {1979})}\BibitemShut {NoStop}%
\bibitem [{\citenamefont {Sosna}\ \emph {et~al.}(2020)\citenamefont {Sosna}, \citenamefont {Korup},\ and\ \citenamefont {Horn}}]{sosnaProbingLocalDiffusion2020}%
  \BibitemOpen
  \bibfield  {author} {\bibinfo {author} {\bibfnamefont {B.}~\bibnamefont {Sosna}}, \bibinfo {author} {\bibfnamefont {O.}~\bibnamefont {Korup}},\ and\ \bibinfo {author} {\bibfnamefont {R.}~\bibnamefont {Horn}},\ }\bibfield  {title} {\bibinfo {title} {Probing local diffusion and reaction in a porous catalyst pellet},\ }\href {https://doi.org/10.1016/j.jcat.2019.11.005} {\bibfield  {journal} {\bibinfo  {journal} {Journal of Catalysis}\ }\textbf {\bibinfo {volume} {381}},\ \bibinfo {pages} {285} (\bibinfo {year} {2020})}\BibitemShut {NoStop}%
\bibitem [{\citenamefont {Tahmasebi}(2018)}]{tahmasebiAccurateModelingEvaluation2018}%
  \BibitemOpen
  \bibfield  {author} {\bibinfo {author} {\bibfnamefont {P.}~\bibnamefont {Tahmasebi}},\ }\bibfield  {title} {\bibinfo {title} {Accurate modeling and evaluation of microstructures in complex materials},\ }\href {https://doi.org/10.1103/PhysRevE.97.023307} {\bibfield  {journal} {\bibinfo  {journal} {Physical Review E}\ }\textbf {\bibinfo {volume} {97}},\ \bibinfo {pages} {023307} (\bibinfo {year} {2018})}\BibitemShut {NoStop}%
\bibitem [{\citenamefont {Langer}\ and\ \citenamefont {Peppas}(1981)}]{langerPresentFutureApplications1981}%
  \BibitemOpen
  \bibfield  {author} {\bibinfo {author} {\bibfnamefont {R.~S.}\ \bibnamefont {Langer}}\ and\ \bibinfo {author} {\bibfnamefont {N.~A.}\ \bibnamefont {Peppas}},\ }\bibfield  {title} {\bibinfo {title} {Present and future applications of biomaterials in controlled drug delivery systems},\ }\href {https://doi.org/10.1016/0142-9612(81)90059-4} {\bibfield  {journal} {\bibinfo  {journal} {Biomaterials}\ }\textbf {\bibinfo {volume} {2}},\ \bibinfo {pages} {201} (\bibinfo {year} {1981})}\BibitemShut {NoStop}%
\bibitem [{\citenamefont {Torquato}(1998)}]{torquatoMorphologyEffectiveProperties1998}%
  \BibitemOpen
  \bibfield  {author} {\bibinfo {author} {\bibfnamefont {S.}~\bibnamefont {Torquato}},\ }\bibfield  {title} {\bibinfo {title} {Morphology and effective properties of disordered heterogeneous media},\ }\href {https://doi.org/10.1016/S0020-7683(97)00142-X} {\bibfield  {journal} {\bibinfo  {journal} {International Journal of Solids and Structures}\ }\textbf {\bibinfo {volume} {35}},\ \bibinfo {pages} {2385} (\bibinfo {year} {1998})}\BibitemShut {NoStop}%
\bibitem [{\citenamefont {Klatt}\ \emph {et~al.}(2019)\citenamefont {Klatt}, \citenamefont {Steinhardt},\ and\ \citenamefont {Torquato}}]{klattPhoamtonicDesignsYield2019}%
  \BibitemOpen
  \bibfield  {author} {\bibinfo {author} {\bibfnamefont {M.~A.}\ \bibnamefont {Klatt}}, \bibinfo {author} {\bibfnamefont {P.~J.}\ \bibnamefont {Steinhardt}},\ and\ \bibinfo {author} {\bibfnamefont {S.}~\bibnamefont {Torquato}},\ }\bibfield  {title} {\bibinfo {title} {Phoamtonic designs yield sizeable {{3D}} photonic band gaps},\ }\href {https://doi.org/10.1073/pnas.1912730116} {\bibfield  {journal} {\bibinfo  {journal} {Proceedings of the National Academy of Sciences}\ }\textbf {\bibinfo {volume} {116}},\ \bibinfo {pages} {23480} (\bibinfo {year} {2019})}\BibitemShut {NoStop}%
\bibitem [{\citenamefont {Huang}\ \emph {et~al.}(2021)\citenamefont {Huang}, \citenamefont {Hu}, \citenamefont {Yang}, \citenamefont {Liu},\ and\ \citenamefont {Zhang}}]{huangCircularSwimmingMotility2021}%
  \BibitemOpen
  \bibfield  {author} {\bibinfo {author} {\bibfnamefont {M.}~\bibnamefont {Huang}}, \bibinfo {author} {\bibfnamefont {W.}~\bibnamefont {Hu}}, \bibinfo {author} {\bibfnamefont {S.}~\bibnamefont {Yang}}, \bibinfo {author} {\bibfnamefont {Q.-X.}\ \bibnamefont {Liu}},\ and\ \bibinfo {author} {\bibfnamefont {H.~P.}\ \bibnamefont {Zhang}},\ }\bibfield  {title} {\bibinfo {title} {Circular swimming motility and disordered hyperuniform state in an algae system},\ }\href {https://doi.org/10.1073/pnas.2100493118} {\bibfield  {journal} {\bibinfo  {journal} {Proceedings of the National Academy of Sciences}\ }\textbf {\bibinfo {volume} {118}},\ \bibinfo {pages} {e2100493118} (\bibinfo {year} {2021})}\BibitemShut {NoStop}%
\bibitem [{\citenamefont {Sahimi}(2003)}]{Sa03}%
  \BibitemOpen
  \bibfield  {author} {\bibinfo {author} {\bibfnamefont {M.}~\bibnamefont {Sahimi}},\ }\href@noop {} {\emph {\bibinfo {title} {Heterogeneous Materials {I}: {L}inear Transport and Optical Properties}}}\ (\bibinfo  {publisher} {Springer-Verlag},\ \bibinfo {address} {New York},\ \bibinfo {year} {2003})\BibitemShut {NoStop}%
\bibitem [{\citenamefont {Sen}\ and\ \citenamefont {Torquato}(1989)}]{senEffectiveConductivityAnisotropic1989}%
  \BibitemOpen
  \bibfield  {author} {\bibinfo {author} {\bibfnamefont {A.~K.}\ \bibnamefont {Sen}}\ and\ \bibinfo {author} {\bibfnamefont {S.}~\bibnamefont {Torquato}},\ }\bibfield  {title} {\bibinfo {title} {Effective conductivity of anisotropic two-phase composite media},\ }\href {https://doi.org/10.1103/PhysRevB.39.4504} {\bibfield  {journal} {\bibinfo  {journal} {Physical Review B}\ }\textbf {\bibinfo {volume} {39}},\ \bibinfo {pages} {4504} (\bibinfo {year} {1989})}\BibitemShut {NoStop}%
\bibitem [{\citenamefont {Torquato}\ and\ \citenamefont {Kim}(2021)}]{torquatoNonlocalEffectiveElectromagnetic2021}%
  \BibitemOpen
  \bibfield  {author} {\bibinfo {author} {\bibfnamefont {S.}~\bibnamefont {Torquato}}\ and\ \bibinfo {author} {\bibfnamefont {J.}~\bibnamefont {Kim}},\ }\bibfield  {title} {\bibinfo {title} {Nonlocal {{Effective Electromagnetic Wave Characteristics}} of {{Composite Media}}: {{Beyond}} the {{Quasistatic Regime}}},\ }\href {https://doi.org/10.1103/PhysRevX.11.021002} {\bibfield  {journal} {\bibinfo  {journal} {Physical Review X}\ }\textbf {\bibinfo {volume} {11}},\ \bibinfo {pages} {021002} (\bibinfo {year} {2021})}\BibitemShut {NoStop}%
\bibitem [{\citenamefont {Torquato}(2021{\natexlab{a}})}]{torquatoDiffusionSpreadabilityProbe2021}%
  \BibitemOpen
  \bibfield  {author} {\bibinfo {author} {\bibfnamefont {S.}~\bibnamefont {Torquato}},\ }\bibfield  {title} {\bibinfo {title} {Diffusion spreadability as a probe of the microstructure of complex media across length scales},\ }\href {https://doi.org/10.1103/PhysRevE.104.054102} {\bibfield  {journal} {\bibinfo  {journal} {Physical Review E}\ }\textbf {\bibinfo {volume} {104}},\ \bibinfo {pages} {054102} (\bibinfo {year} {2021}{\natexlab{a}})}\BibitemShut {NoStop}%
\bibitem [{\citenamefont {Prager}(1963)}]{pragerInterphaseTransferStationary1963}%
  \BibitemOpen
  \bibfield  {author} {\bibinfo {author} {\bibfnamefont {S.}~\bibnamefont {Prager}},\ }\bibfield  {title} {\bibinfo {title} {Interphase transfer in stationary two-phase media},\ }\href {https://doi.org/10.1016/0009-2509(63)87003-7} {\bibfield  {journal} {\bibinfo  {journal} {Chemical Engineering Science}\ }\textbf {\bibinfo {volume} {18}},\ \bibinfo {pages} {227} (\bibinfo {year} {1963})}\BibitemShut {NoStop}%
\bibitem [{\citenamefont {Wang}\ and\ \citenamefont {Torquato}(2022)}]{wangDynamicMeasureHyperuniformity2022}%
  \BibitemOpen
  \bibfield  {author} {\bibinfo {author} {\bibfnamefont {H.}~\bibnamefont {Wang}}\ and\ \bibinfo {author} {\bibfnamefont {S.}~\bibnamefont {Torquato}},\ }\bibfield  {title} {\bibinfo {title} {Dynamic {{Measure}} of {{Hyperuniformity}} and {{Nonhyperuniformity}} in {{Heterogeneous Media}} via the {{Diffusion Spreadability}}},\ }\href {https://doi.org/10.1103/PhysRevApplied.17.034022} {\bibfield  {journal} {\bibinfo  {journal} {Physical Review Applied}\ }\textbf {\bibinfo {volume} {17}},\ \bibinfo {pages} {034022} (\bibinfo {year} {2022})}\BibitemShut {NoStop}%
\bibitem [{\citenamefont {Torquato}\ and\ \citenamefont {Stillinger}(2003)}]{torquatoLocalDensityFluctuations2003a}%
  \BibitemOpen
  \bibfield  {author} {\bibinfo {author} {\bibfnamefont {S.}~\bibnamefont {Torquato}}\ and\ \bibinfo {author} {\bibfnamefont {F.~H.}\ \bibnamefont {Stillinger}},\ }\bibfield  {title} {\bibinfo {title} {Local density fluctuations, hyperuniformity, and order metrics},\ }\href {https://doi.org/10.1103/PhysRevE.68.041113} {\bibfield  {journal} {\bibinfo  {journal} {Physical Review E}\ }\textbf {\bibinfo {volume} {68}},\ \bibinfo {pages} {041113} (\bibinfo {year} {2003})}\BibitemShut {NoStop}%
\bibitem [{\citenamefont {Zachary}\ and\ \citenamefont {Torquato}(2009)}]{zacharyHyperuniformityPointPatterns2009}%
  \BibitemOpen
  \bibfield  {author} {\bibinfo {author} {\bibfnamefont {C.~E.}\ \bibnamefont {Zachary}}\ and\ \bibinfo {author} {\bibfnamefont {S.}~\bibnamefont {Torquato}},\ }\bibfield  {title} {\bibinfo {title} {Hyperuniformity in point patterns and two-phase random heterogeneous media},\ }\href {https://doi.org/10.1088/1742-5468/2009/12/P12015} {\bibfield  {journal} {\bibinfo  {journal} {Journal of Statistical Mechanics: Theory and Experiment}\ }\textbf {\bibinfo {volume} {2009}},\ \bibinfo {pages} {P12015} (\bibinfo {year} {2009})}\BibitemShut {NoStop}%
\bibitem [{\citenamefont {Torquato}(2018)}]{torquatoHyperuniformStatesMatter2018a}%
  \BibitemOpen
  \bibfield  {author} {\bibinfo {author} {\bibfnamefont {S.}~\bibnamefont {Torquato}},\ }\bibfield  {title} {\bibinfo {title} {Hyperuniform states of matter},\ }\href {https://doi.org/10.1016/j.physrep.2018.03.001} {\bibfield  {journal} {\bibinfo  {journal} {Physics Reports}\ }\bibinfo {series} {Hyperuniform {{States}} of {{Matter}}},\ \textbf {\bibinfo {volume} {745}},\ \bibinfo {pages} {1} (\bibinfo {year} {2018})}\BibitemShut {NoStop}%
\bibitem [{\citenamefont {Marcotte}\ \emph {et~al.}(2013)\citenamefont {Marcotte}, \citenamefont {Stillinger},\ and\ \citenamefont {Torquato}}]{marcotteNonequilibriumStaticGrowing2013}%
  \BibitemOpen
  \bibfield  {author} {\bibinfo {author} {\bibfnamefont {{\'E}.}~\bibnamefont {Marcotte}}, \bibinfo {author} {\bibfnamefont {F.~H.}\ \bibnamefont {Stillinger}},\ and\ \bibinfo {author} {\bibfnamefont {S.}~\bibnamefont {Torquato}},\ }\bibfield  {title} {\bibinfo {title} {Nonequilibrium static growing length scales in supercooled liquids on approaching the glass transition},\ }\href {https://doi.org/10.1063/1.4769422} {\bibfield  {journal} {\bibinfo  {journal} {The Journal of Chemical Physics}\ }\textbf {\bibinfo {volume} {138}},\ \bibinfo {pages} {12A508} (\bibinfo {year} {2013})}\BibitemShut {NoStop}%
\bibitem [{\citenamefont {Coniglio}\ \emph {et~al.}(2017)\citenamefont {Coniglio}, \citenamefont {Ciamarra},\ and\ \citenamefont {Aste}}]{coniglioUniversalBehaviourGlass2017}%
  \BibitemOpen
  \bibfield  {author} {\bibinfo {author} {\bibfnamefont {A.}~\bibnamefont {Coniglio}}, \bibinfo {author} {\bibfnamefont {M.~P.}\ \bibnamefont {Ciamarra}},\ and\ \bibinfo {author} {\bibfnamefont {T.}~\bibnamefont {Aste}},\ }\bibfield  {title} {\bibinfo {title} {Universal behaviour of the glass and the jamming transitions in finite dimensions for hard spheres},\ }\href {https://doi.org/10.1039/C7SM01481C} {\bibfield  {journal} {\bibinfo  {journal} {Soft Matter}\ }\textbf {\bibinfo {volume} {13}},\ \bibinfo {pages} {8766} (\bibinfo {year} {2017})}\BibitemShut {NoStop}%
\bibitem [{\citenamefont {Zhang}\ \emph {et~al.}(2023)\citenamefont {Zhang}, \citenamefont {Wang}, \citenamefont {Zhang}, \citenamefont {Yu},\ and\ \citenamefont {Douglas}}]{zhangApproachHyperuniformityMetallic2023}%
  \BibitemOpen
  \bibfield  {author} {\bibinfo {author} {\bibfnamefont {H.}~\bibnamefont {Zhang}}, \bibinfo {author} {\bibfnamefont {X.}~\bibnamefont {Wang}}, \bibinfo {author} {\bibfnamefont {J.}~\bibnamefont {Zhang}}, \bibinfo {author} {\bibfnamefont {H.-B.}\ \bibnamefont {Yu}},\ and\ \bibinfo {author} {\bibfnamefont {J.~F.}\ \bibnamefont {Douglas}},\ }\bibfield  {title} {\bibinfo {title} {Approach to hyperuniformity in a metallic glass-forming material exhibiting a fragile to strong glass transition},\ }\href {https://doi.org/10.1140/epje/s10189-023-00308-4} {\bibfield  {journal} {\bibinfo  {journal} {The European Physical Journal E}\ }\textbf {\bibinfo {volume} {46}},\ \bibinfo {pages} {50} (\bibinfo {year} {2023})}\BibitemShut {NoStop}%
\bibitem [{\citenamefont {Wang}\ \emph {et~al.}(2025)\citenamefont {Wang}, \citenamefont {Qian}, \citenamefont {Tong},\ and\ \citenamefont {Tanaka}}]{wangHyperuniformDisorderedSolids2025}%
  \BibitemOpen
  \bibfield  {author} {\bibinfo {author} {\bibfnamefont {Y.}~\bibnamefont {Wang}}, \bibinfo {author} {\bibfnamefont {Z.}~\bibnamefont {Qian}}, \bibinfo {author} {\bibfnamefont {H.}~\bibnamefont {Tong}},\ and\ \bibinfo {author} {\bibfnamefont {H.}~\bibnamefont {Tanaka}},\ }\bibfield  {title} {\bibinfo {title} {Hyperuniform disordered solids with crystal-like stability},\ }\href {https://doi.org/10.1038/s41467-025-56283-1} {\bibfield  {journal} {\bibinfo  {journal} {Nature Communications}\ }\textbf {\bibinfo {volume} {16}},\ \bibinfo {pages} {1398} (\bibinfo {year} {2025})}\BibitemShut {NoStop}%
\bibitem [{\citenamefont {Mitra}\ \emph {et~al.}(2021)\citenamefont {Mitra}, \citenamefont {Parmar}, \citenamefont {Leishangthem}, \citenamefont {Sastry},\ and\ \citenamefont {Foffi}}]{mitraHyperuniformityCyclicallyDriven2021}%
  \BibitemOpen
  \bibfield  {author} {\bibinfo {author} {\bibfnamefont {S.}~\bibnamefont {Mitra}}, \bibinfo {author} {\bibfnamefont {A.~D.~S.}\ \bibnamefont {Parmar}}, \bibinfo {author} {\bibfnamefont {P.}~\bibnamefont {Leishangthem}}, \bibinfo {author} {\bibfnamefont {S.}~\bibnamefont {Sastry}},\ and\ \bibinfo {author} {\bibfnamefont {G.}~\bibnamefont {Foffi}},\ }\bibfield  {title} {\bibinfo {title} {Hyperuniformity in cyclically driven glasses},\ }\href {https://doi.org/10.1088/1742-5468/abdeb0} {\bibfield  {journal} {\bibinfo  {journal} {Journal of Statistical Mechanics: Theory and Experiment}\ }\textbf {\bibinfo {volume} {2021}},\ \bibinfo {pages} {033203} (\bibinfo {year} {2021})}\BibitemShut {NoStop}%
\bibitem [{\citenamefont {Atkinson}\ \emph {et~al.}(2016)\citenamefont {Atkinson}, \citenamefont {Zhang}, \citenamefont {Hopkins},\ and\ \citenamefont {Torquato}}]{atkinsonCriticalSlowingHyperuniformity2016}%
  \BibitemOpen
  \bibfield  {author} {\bibinfo {author} {\bibfnamefont {S.}~\bibnamefont {Atkinson}}, \bibinfo {author} {\bibfnamefont {G.}~\bibnamefont {Zhang}}, \bibinfo {author} {\bibfnamefont {A.~B.}\ \bibnamefont {Hopkins}},\ and\ \bibinfo {author} {\bibfnamefont {S.}~\bibnamefont {Torquato}},\ }\bibfield  {title} {\bibinfo {title} {Critical slowing down and hyperuniformity on approach to jamming},\ }\href {https://doi.org/10.1103/PhysRevE.94.012902} {\bibfield  {journal} {\bibinfo  {journal} {Physical Review E}\ }\textbf {\bibinfo {volume} {94}},\ \bibinfo {pages} {012902} (\bibinfo {year} {2016})}\BibitemShut {NoStop}%
\bibitem [{\citenamefont {Rissone}\ \emph {et~al.}(2021)\citenamefont {Rissone}, \citenamefont {Corwin},\ and\ \citenamefont {Parisi}}]{rissoneLongRangeAnomalousDecay2021}%
  \BibitemOpen
  \bibfield  {author} {\bibinfo {author} {\bibfnamefont {P.}~\bibnamefont {Rissone}}, \bibinfo {author} {\bibfnamefont {E.~I.}\ \bibnamefont {Corwin}},\ and\ \bibinfo {author} {\bibfnamefont {G.}~\bibnamefont {Parisi}},\ }\bibfield  {title} {\bibinfo {title} {Long-{{Range Anomalous Decay}} of the {{Correlation}} in {{Jammed Packings}}},\ }\href {https://doi.org/10.1103/PhysRevLett.127.038001} {\bibfield  {journal} {\bibinfo  {journal} {Physical Review Letters}\ }\textbf {\bibinfo {volume} {127}},\ \bibinfo {pages} {038001} (\bibinfo {year} {2021})}\BibitemShut {NoStop}%
\bibitem [{\citenamefont {Torquato}(2021{\natexlab{b}})}]{torquatoStructuralCharacterizationManyparticle2021}%
  \BibitemOpen
  \bibfield  {author} {\bibinfo {author} {\bibfnamefont {S.}~\bibnamefont {Torquato}},\ }\bibfield  {title} {\bibinfo {title} {Structural characterization of many-particle systems on approach to hyperuniform states},\ }\href {https://doi.org/10.1103/PhysRevE.103.052126} {\bibfield  {journal} {\bibinfo  {journal} {Physical Review E}\ }\textbf {\bibinfo {volume} {103}},\ \bibinfo {pages} {052126} (\bibinfo {year} {2021}{\natexlab{b}})}\BibitemShut {NoStop}%
\bibitem [{\citenamefont {Torquato}\ and\ \citenamefont {Stillinger}(2007)}]{torquatoJammingThresholdSphere2007}%
  \BibitemOpen
  \bibfield  {author} {\bibinfo {author} {\bibfnamefont {S.}~\bibnamefont {Torquato}}\ and\ \bibinfo {author} {\bibfnamefont {F.~H.}\ \bibnamefont {Stillinger}},\ }\bibfield  {title} {\bibinfo {title} {Toward the jamming threshold of sphere packings: {{Tunneled}} crystals},\ }\href {https://doi.org/10.1063/1.2802184} {\bibfield  {journal} {\bibinfo  {journal} {Journal of Applied Physics}\ }\textbf {\bibinfo {volume} {102}},\ \bibinfo {pages} {093511} (\bibinfo {year} {2007})}\BibitemShut {NoStop}%
\bibitem [{\citenamefont {Donev}\ \emph {et~al.}(2005)\citenamefont {Donev}, \citenamefont {Stillinger},\ and\ \citenamefont {Torquato}}]{donevUnexpectedDensityFluctuations2005}%
  \BibitemOpen
  \bibfield  {author} {\bibinfo {author} {\bibfnamefont {A.}~\bibnamefont {Donev}}, \bibinfo {author} {\bibfnamefont {F.~H.}\ \bibnamefont {Stillinger}},\ and\ \bibinfo {author} {\bibfnamefont {S.}~\bibnamefont {Torquato}},\ }\bibfield  {title} {\bibinfo {title} {Unexpected {{Density Fluctuations}} in {{Jammed Disordered Sphere Packings}}},\ }\href {https://doi.org/10.1103/PhysRevLett.95.090604} {\bibfield  {journal} {\bibinfo  {journal} {Physical Review Letters}\ }\textbf {\bibinfo {volume} {95}},\ \bibinfo {pages} {090604} (\bibinfo {year} {2005})}\BibitemShut {NoStop}%
\bibitem [{\citenamefont {Dreyfus}\ \emph {et~al.}(2015)\citenamefont {Dreyfus}, \citenamefont {Xu}, \citenamefont {Still}, \citenamefont {Hough}, \citenamefont {Yodh},\ and\ \citenamefont {Torquato}}]{dreyfusDiagnosingHyperuniformityTwodimensional2015}%
  \BibitemOpen
  \bibfield  {author} {\bibinfo {author} {\bibfnamefont {R.}~\bibnamefont {Dreyfus}}, \bibinfo {author} {\bibfnamefont {Y.}~\bibnamefont {Xu}}, \bibinfo {author} {\bibfnamefont {T.}~\bibnamefont {Still}}, \bibinfo {author} {\bibfnamefont {L.~A.}\ \bibnamefont {Hough}}, \bibinfo {author} {\bibfnamefont {A.~G.}\ \bibnamefont {Yodh}},\ and\ \bibinfo {author} {\bibfnamefont {S.}~\bibnamefont {Torquato}},\ }\bibfield  {title} {\bibinfo {title} {Diagnosing hyperuniformity in two-dimensional, disordered, jammed packings of soft spheres},\ }\href {https://doi.org/10.1103/PhysRevE.91.012302} {\bibfield  {journal} {\bibinfo  {journal} {Physical Review E}\ }\textbf {\bibinfo {volume} {91}},\ \bibinfo {pages} {012302} (\bibinfo {year} {2015})}\BibitemShut {NoStop}%
\bibitem [{\citenamefont {Dale}\ \emph {et~al.}(2022)\citenamefont {Dale}, \citenamefont {Sartor}, \citenamefont {Dennis},\ and\ \citenamefont {Corwin}}]{daleHyperuniformJammedSphere2022}%
  \BibitemOpen
  \bibfield  {author} {\bibinfo {author} {\bibfnamefont {J.~R.}\ \bibnamefont {Dale}}, \bibinfo {author} {\bibfnamefont {J.~D.}\ \bibnamefont {Sartor}}, \bibinfo {author} {\bibfnamefont {R.~C.}\ \bibnamefont {Dennis}},\ and\ \bibinfo {author} {\bibfnamefont {E.~I.}\ \bibnamefont {Corwin}},\ }\bibfield  {title} {\bibinfo {title} {Hyperuniform jammed sphere packings have anomalous material properties},\ }\href {https://doi.org/10.1103/PhysRevE.106.024903} {\bibfield  {journal} {\bibinfo  {journal} {Physical Review E}\ }\textbf {\bibinfo {volume} {106}},\ \bibinfo {pages} {024903} (\bibinfo {year} {2022})}\BibitemShut {NoStop}%
\bibitem [{\citenamefont {Florescu}\ \emph {et~al.}(2009)\citenamefont {Florescu}, \citenamefont {Torquato},\ and\ \citenamefont {Steinhardt}}]{florescuDesignerDisorderedMaterials2009a}%
  \BibitemOpen
  \bibfield  {author} {\bibinfo {author} {\bibfnamefont {M.}~\bibnamefont {Florescu}}, \bibinfo {author} {\bibfnamefont {S.}~\bibnamefont {Torquato}},\ and\ \bibinfo {author} {\bibfnamefont {P.~J.}\ \bibnamefont {Steinhardt}},\ }\bibfield  {title} {\bibinfo {title} {Designer disordered materials with large, complete photonic band gaps},\ }\href {https://doi.org/10.1073/pnas.0907744106} {\bibfield  {journal} {\bibinfo  {journal} {Proceedings of the National Academy of Sciences}\ }\textbf {\bibinfo {volume} {106}},\ \bibinfo {pages} {20658} (\bibinfo {year} {2009})}\BibitemShut {NoStop}%
\bibitem [{\citenamefont {{Froufe-P{\'e}rez}}\ \emph {et~al.}(2017)\citenamefont {{Froufe-P{\'e}rez}}, \citenamefont {Engel}, \citenamefont {S{\'a}enz},\ and\ \citenamefont {Scheffold}}]{froufe-perezBandGapFormation2017}%
  \BibitemOpen
  \bibfield  {author} {\bibinfo {author} {\bibfnamefont {L.~S.}\ \bibnamefont {{Froufe-P{\'e}rez}}}, \bibinfo {author} {\bibfnamefont {M.}~\bibnamefont {Engel}}, \bibinfo {author} {\bibfnamefont {J.~J.}\ \bibnamefont {S{\'a}enz}},\ and\ \bibinfo {author} {\bibfnamefont {F.}~\bibnamefont {Scheffold}},\ }\bibfield  {title} {\bibinfo {title} {Band gap formation and {{Anderson}} localization in disordered photonic materials with structural correlations},\ }\href {https://doi.org/10.1073/pnas.1705130114} {\bibfield  {journal} {\bibinfo  {journal} {Proceedings of the National Academy of Sciences}\ }\textbf {\bibinfo {volume} {114}},\ \bibinfo {pages} {9570} (\bibinfo {year} {2017})}\BibitemShut {NoStop}%
\bibitem [{\citenamefont {Man}\ \emph {et~al.}(2013)\citenamefont {Man}, \citenamefont {Florescu}, \citenamefont {Matsuyama}, \citenamefont {Yadak}, \citenamefont {Nahal}, \citenamefont {Hashemizad}, \citenamefont {Williamson}, \citenamefont {Steinhardt}, \citenamefont {Torquato},\ and\ \citenamefont {Chaikin}}]{manPhotonicBandGap2013}%
  \BibitemOpen
  \bibfield  {author} {\bibinfo {author} {\bibfnamefont {W.}~\bibnamefont {Man}}, \bibinfo {author} {\bibfnamefont {M.}~\bibnamefont {Florescu}}, \bibinfo {author} {\bibfnamefont {K.}~\bibnamefont {Matsuyama}}, \bibinfo {author} {\bibfnamefont {P.}~\bibnamefont {Yadak}}, \bibinfo {author} {\bibfnamefont {G.}~\bibnamefont {Nahal}}, \bibinfo {author} {\bibfnamefont {S.}~\bibnamefont {Hashemizad}}, \bibinfo {author} {\bibfnamefont {E.}~\bibnamefont {Williamson}}, \bibinfo {author} {\bibfnamefont {P.}~\bibnamefont {Steinhardt}}, \bibinfo {author} {\bibfnamefont {S.}~\bibnamefont {Torquato}},\ and\ \bibinfo {author} {\bibfnamefont {P.}~\bibnamefont {Chaikin}},\ }\bibfield  {title} {\bibinfo {title} {Photonic band gap in isotropic hyperuniform disordered solids with low dielectric contrast},\ }\href {https://doi.org/10.1364/OE.21.019972} {\bibfield  {journal} {\bibinfo  {journal} {Optics Express}\ }\textbf {\bibinfo {volume} {21}},\ \bibinfo {pages} {19972} (\bibinfo {year} {2013})}\BibitemShut {NoStop}%
\bibitem [{\citenamefont {Klatt}\ \emph {et~al.}(2022)\citenamefont {Klatt}, \citenamefont {Steinhardt},\ and\ \citenamefont {Torquato}}]{klattWavePropagationBand2022}%
  \BibitemOpen
  \bibfield  {author} {\bibinfo {author} {\bibfnamefont {M.~A.}\ \bibnamefont {Klatt}}, \bibinfo {author} {\bibfnamefont {P.~J.}\ \bibnamefont {Steinhardt}},\ and\ \bibinfo {author} {\bibfnamefont {S.}~\bibnamefont {Torquato}},\ }\bibfield  {title} {\bibinfo {title} {Wave propagation and band tails of two-dimensional disordered systems in the thermodynamic limit},\ }\href {https://doi.org/10.1073/pnas.2213633119} {\bibfield  {journal} {\bibinfo  {journal} {Proceedings of the National Academy of Sciences}\ }\textbf {\bibinfo {volume} {119}},\ \bibinfo {pages} {e2213633119} (\bibinfo {year} {2022})}\BibitemShut {NoStop}%
\bibitem [{\citenamefont {Christogeorgos}\ \emph {et~al.}(2021)\citenamefont {Christogeorgos}, \citenamefont {Zhang}, \citenamefont {Cheng},\ and\ \citenamefont {Hao}}]{christogeorgosExtraordinaryDirectiveEmission2021}%
  \BibitemOpen
  \bibfield  {author} {\bibinfo {author} {\bibfnamefont {O.}~\bibnamefont {Christogeorgos}}, \bibinfo {author} {\bibfnamefont {H.}~\bibnamefont {Zhang}}, \bibinfo {author} {\bibfnamefont {Q.}~\bibnamefont {Cheng}},\ and\ \bibinfo {author} {\bibfnamefont {Y.}~\bibnamefont {Hao}},\ }\bibfield  {title} {\bibinfo {title} {Extraordinary {{Directive Emission}} and {{Scanning}} from an {{Array}} of {{Radiation Sources}} with {{Hyperuniform Disorder}}},\ }\href {https://doi.org/10.1103/PhysRevApplied.15.014062} {\bibfield  {journal} {\bibinfo  {journal} {Physical Review Applied}\ }\textbf {\bibinfo {volume} {15}},\ \bibinfo {pages} {014062} (\bibinfo {year} {2021})}\BibitemShut {NoStop}%
\bibitem [{\citenamefont {Aubry}\ \emph {et~al.}(2020)\citenamefont {Aubry}, \citenamefont {{Froufe-P{\'e}rez}}, \citenamefont {Kuhl}, \citenamefont {Legrand}, \citenamefont {Scheffold},\ and\ \citenamefont {Mortessagne}}]{aubryExperimentalTuningTransport2020}%
  \BibitemOpen
  \bibfield  {author} {\bibinfo {author} {\bibfnamefont {G.~J.}\ \bibnamefont {Aubry}}, \bibinfo {author} {\bibfnamefont {L.~S.}\ \bibnamefont {{Froufe-P{\'e}rez}}}, \bibinfo {author} {\bibfnamefont {U.}~\bibnamefont {Kuhl}}, \bibinfo {author} {\bibfnamefont {O.}~\bibnamefont {Legrand}}, \bibinfo {author} {\bibfnamefont {F.}~\bibnamefont {Scheffold}},\ and\ \bibinfo {author} {\bibfnamefont {F.}~\bibnamefont {Mortessagne}},\ }\bibfield  {title} {\bibinfo {title} {Experimental {{Tuning}} of {{Transport Regimes}} in {{Hyperuniform Disordered Photonic Materials}}},\ }\href {https://doi.org/10.1103/PhysRevLett.125.127402} {\bibfield  {journal} {\bibinfo  {journal} {Physical Review Letters}\ }\textbf {\bibinfo {volume} {125}},\ \bibinfo {pages} {127402} (\bibinfo {year} {2020})}\BibitemShut {NoStop}%
\bibitem [{\citenamefont {Gkantzounis}\ \emph {et~al.}(2017)\citenamefont {Gkantzounis}, \citenamefont {Amoah},\ and\ \citenamefont {Florescu}}]{gkantzounisHyperuniformDisorderedPhononic2017}%
  \BibitemOpen
  \bibfield  {author} {\bibinfo {author} {\bibfnamefont {G.}~\bibnamefont {Gkantzounis}}, \bibinfo {author} {\bibfnamefont {T.}~\bibnamefont {Amoah}},\ and\ \bibinfo {author} {\bibfnamefont {M.}~\bibnamefont {Florescu}},\ }\bibfield  {title} {\bibinfo {title} {Hyperuniform disordered phononic structures},\ }\href {https://doi.org/10.1103/PhysRevB.95.094120} {\bibfield  {journal} {\bibinfo  {journal} {Physical Review B}\ }\textbf {\bibinfo {volume} {95}},\ \bibinfo {pages} {094120} (\bibinfo {year} {2017})}\BibitemShut {NoStop}%
\bibitem [{\citenamefont {Torquato}\ \emph {et~al.}(2019)\citenamefont {Torquato}, \citenamefont {Zhang},\ and\ \citenamefont {{De Courcy-Ireland}}}]{torquatoHiddenMultiscaleOrder2019}%
  \BibitemOpen
  \bibfield  {author} {\bibinfo {author} {\bibfnamefont {S.}~\bibnamefont {Torquato}}, \bibinfo {author} {\bibfnamefont {G.}~\bibnamefont {Zhang}},\ and\ \bibinfo {author} {\bibfnamefont {M.}~\bibnamefont {{De Courcy-Ireland}}},\ }\bibfield  {title} {\bibinfo {title} {Hidden multiscale order in the primes},\ }\href {https://doi.org/10.1088/1751-8121/ab0588} {\bibfield  {journal} {\bibinfo  {journal} {Journal of Physics A: Mathematical and Theoretical}\ }\textbf {\bibinfo {volume} {52}},\ \bibinfo {pages} {135002} (\bibinfo {year} {2019})}\BibitemShut {NoStop}%
\bibitem [{\citenamefont {Brauchart}\ \emph {et~al.}(2019)\citenamefont {Brauchart}, \citenamefont {Grabner},\ and\ \citenamefont {Kusner}}]{brauchartHyperuniformPointSets2019}%
  \BibitemOpen
  \bibfield  {author} {\bibinfo {author} {\bibfnamefont {J.~S.}\ \bibnamefont {Brauchart}}, \bibinfo {author} {\bibfnamefont {P.~J.}\ \bibnamefont {Grabner}},\ and\ \bibinfo {author} {\bibfnamefont {W.}~\bibnamefont {Kusner}},\ }\bibfield  {title} {\bibinfo {title} {Hyperuniform {{Point Sets}} on the {{Sphere}}: {{Deterministic Aspects}}},\ }\href {https://doi.org/10.1007/s00365-018-9432-8} {\bibfield  {journal} {\bibinfo  {journal} {Constructive Approximation}\ }\textbf {\bibinfo {volume} {50}},\ \bibinfo {pages} {45} (\bibinfo {year} {2019})}\BibitemShut {NoStop}%
\bibitem [{\citenamefont {Brauchart}\ \emph {et~al.}(2020)\citenamefont {Brauchart}, \citenamefont {Grabner}, \citenamefont {Kusner},\ and\ \citenamefont {Ziefle}}]{brauchartHyperuniformPointSets2020}%
  \BibitemOpen
  \bibfield  {author} {\bibinfo {author} {\bibfnamefont {J.~S.}\ \bibnamefont {Brauchart}}, \bibinfo {author} {\bibfnamefont {P.~J.}\ \bibnamefont {Grabner}}, \bibinfo {author} {\bibfnamefont {W.}~\bibnamefont {Kusner}},\ and\ \bibinfo {author} {\bibfnamefont {J.}~\bibnamefont {Ziefle}},\ }\bibfield  {title} {\bibinfo {title} {Hyperuniform point sets on the sphere: Probabilistic aspects},\ }\href {https://doi.org/10.1007/s00605-020-01439-y} {\bibfield  {journal} {\bibinfo  {journal} {Monatshefte f{\"u}r Mathematik}\ }\textbf {\bibinfo {volume} {192}},\ \bibinfo {pages} {763} (\bibinfo {year} {2020})}\BibitemShut {NoStop}%
\bibitem [{\citenamefont {Torquato}\ \emph {et~al.}(2018)\citenamefont {Torquato}, \citenamefont {Zhang},\ and\ \citenamefont {{de Courcy-Ireland}}}]{torquatoUncoveringMultiscaleOrder2018}%
  \BibitemOpen
  \bibfield  {author} {\bibinfo {author} {\bibfnamefont {S.}~\bibnamefont {Torquato}}, \bibinfo {author} {\bibfnamefont {G.}~\bibnamefont {Zhang}},\ and\ \bibinfo {author} {\bibfnamefont {M.}~\bibnamefont {{de Courcy-Ireland}}},\ }\bibfield  {title} {\bibinfo {title} {Uncovering multiscale order in the prime numbers via scattering},\ }\href {https://doi.org/10.1088/1742-5468/aad6be} {\bibfield  {journal} {\bibinfo  {journal} {Journal of Statistical Mechanics: Theory and Experiment}\ }\textbf {\bibinfo {volume} {2018}},\ \bibinfo {pages} {093401} (\bibinfo {year} {2018})}\BibitemShut {NoStop}%
\bibitem [{\citenamefont {Bj{\"o}rklund}\ and\ \citenamefont {Byl{\'e}hn}(2026)}]{bjorklundHyperuniformityRandomMeasures2026}%
  \BibitemOpen
  \bibfield  {author} {\bibinfo {author} {\bibfnamefont {M.}~\bibnamefont {Bj{\"o}rklund}}\ and\ \bibinfo {author} {\bibfnamefont {M.}~\bibnamefont {Byl{\'e}hn}},\ }\bibfield  {title} {\bibinfo {title} {Hyperuniformity of random measures on {{Euclidean}} and hyperbolic spaces},\ }\href {https://doi.org/10.1007/s00208-026-03349-0} {\bibfield  {journal} {\bibinfo  {journal} {Mathematische Annalen}\ }\textbf {\bibinfo {volume} {394}},\ \bibinfo {pages} {46} (\bibinfo {year} {2026})}\BibitemShut {NoStop}%
\bibitem [{\citenamefont {Byl{\'e}hn}(2025)}]{bylehnHyperuniformityRegularTrees2025}%
  \BibitemOpen
  \bibfield  {author} {\bibinfo {author} {\bibfnamefont {M.}~\bibnamefont {Byl{\'e}hn}},\ }\bibfield  {title} {\bibinfo {title} {Hyperuniformity in {{Regular Trees}}},\ }\href {https://doi.org/10.1007/s00041-025-10198-z} {\bibfield  {journal} {\bibinfo  {journal} {Journal of Fourier Analysis and Applications}\ }\textbf {\bibinfo {volume} {31}},\ \bibinfo {pages} {66} (\bibinfo {year} {2025})}\BibitemShut {NoStop}%
\bibitem [{\citenamefont {Ghosh}\ and\ \citenamefont {Lebowitz}(2018)}]{ghoshGeneralizedStealthyHyperuniform2018}%
  \BibitemOpen
  \bibfield  {author} {\bibinfo {author} {\bibfnamefont {S.}~\bibnamefont {Ghosh}}\ and\ \bibinfo {author} {\bibfnamefont {J.~L.}\ \bibnamefont {Lebowitz}},\ }\bibfield  {title} {\bibinfo {title} {Generalized {{Stealthy Hyperuniform Processes}}: {{Maximal Rigidity}} and the {{Bounded Holes Conjecture}}},\ }\href {https://doi.org/10.1007/s00220-018-3226-5} {\bibfield  {journal} {\bibinfo  {journal} {Communications in Mathematical Physics}\ }\textbf {\bibinfo {volume} {363}},\ \bibinfo {pages} {97} (\bibinfo {year} {2018})}\BibitemShut {NoStop}%
\bibitem [{\citenamefont {Jiao}\ \emph {et~al.}(2014)\citenamefont {Jiao}, \citenamefont {Lau}, \citenamefont {Hatzikirou}, \citenamefont {{Meyer-Hermann}}, \citenamefont {{Joseph C. Corbo}},\ and\ \citenamefont {Torquato}}]{jiaoAvianPhotoreceptorPatterns2014}%
  \BibitemOpen
  \bibfield  {author} {\bibinfo {author} {\bibfnamefont {Y.}~\bibnamefont {Jiao}}, \bibinfo {author} {\bibfnamefont {T.}~\bibnamefont {Lau}}, \bibinfo {author} {\bibfnamefont {H.}~\bibnamefont {Hatzikirou}}, \bibinfo {author} {\bibfnamefont {M.}~\bibnamefont {{Meyer-Hermann}}}, \bibinfo {author} {\bibnamefont {{Joseph C. Corbo}}},\ and\ \bibinfo {author} {\bibfnamefont {S.}~\bibnamefont {Torquato}},\ }\bibfield  {title} {\bibinfo {title} {Avian photoreceptor patterns represent a disordered hyperuniform solution to a multiscale packing problem},\ }\href {https://doi.org/10.1103/PhysRevE.89.022721} {\bibfield  {journal} {\bibinfo  {journal} {Physical Review E}\ }\textbf {\bibinfo {volume} {89}},\ \bibinfo {pages} {022721} (\bibinfo {year} {2014})}\BibitemShut {NoStop}%
\bibitem [{\citenamefont {Mayer}\ \emph {et~al.}(2015)\citenamefont {Mayer}, \citenamefont {Balasubramanian}, \citenamefont {Mora},\ and\ \citenamefont {Walczak}}]{mayerHowWelladaptedImmune2015}%
  \BibitemOpen
  \bibfield  {author} {\bibinfo {author} {\bibfnamefont {A.}~\bibnamefont {Mayer}}, \bibinfo {author} {\bibfnamefont {V.}~\bibnamefont {Balasubramanian}}, \bibinfo {author} {\bibfnamefont {T.}~\bibnamefont {Mora}},\ and\ \bibinfo {author} {\bibfnamefont {A.~M.}\ \bibnamefont {Walczak}},\ }\bibfield  {title} {\bibinfo {title} {How a well-adapted immune system is organized},\ }\href {https://doi.org/10.1073/pnas.1421827112} {\bibfield  {journal} {\bibinfo  {journal} {Proceedings of the National Academy of Sciences}\ }\textbf {\bibinfo {volume} {112}},\ \bibinfo {pages} {5950} (\bibinfo {year} {2015})}\BibitemShut {NoStop}%
\bibitem [{\citenamefont {Zheng}\ \emph {et~al.}(2020)\citenamefont {Zheng}, \citenamefont {Li},\ and\ \citenamefont {Pica~Ciamarra}}]{zhengHyperuniformityDensityFluctuations2020}%
  \BibitemOpen
  \bibfield  {author} {\bibinfo {author} {\bibfnamefont {Y.}~\bibnamefont {Zheng}}, \bibinfo {author} {\bibfnamefont {Y.-W.}\ \bibnamefont {Li}},\ and\ \bibinfo {author} {\bibfnamefont {M.}~\bibnamefont {Pica~Ciamarra}},\ }\bibfield  {title} {\bibinfo {title} {Hyperuniformity and density fluctuations at a rigidity transition in a model of biological tissues},\ }\href {https://doi.org/10.1039/D0SM00776E} {\bibfield  {journal} {\bibinfo  {journal} {Soft Matter}\ }\textbf {\bibinfo {volume} {16}},\ \bibinfo {pages} {5942} (\bibinfo {year} {2020})}\BibitemShut {NoStop}%
\bibitem [{\citenamefont {Hu}\ \emph {et~al.}(2025)\citenamefont {Hu}, \citenamefont {Cui}, \citenamefont {{Delgado-Baquerizo}}, \citenamefont {Sol{\'e}}, \citenamefont {K{\'e}fi}, \citenamefont {Berdugo}, \citenamefont {Xu}, \citenamefont {Wang}, \citenamefont {Liu},\ and\ \citenamefont {Xu}}]{huCausesConsequencesDisordered2025}%
  \BibitemOpen
  \bibfield  {author} {\bibinfo {author} {\bibfnamefont {W.}~\bibnamefont {Hu}}, \bibinfo {author} {\bibfnamefont {L.}~\bibnamefont {Cui}}, \bibinfo {author} {\bibfnamefont {M.}~\bibnamefont {{Delgado-Baquerizo}}}, \bibinfo {author} {\bibfnamefont {R.}~\bibnamefont {Sol{\'e}}}, \bibinfo {author} {\bibfnamefont {S.}~\bibnamefont {K{\'e}fi}}, \bibinfo {author} {\bibfnamefont {M.}~\bibnamefont {Berdugo}}, \bibinfo {author} {\bibfnamefont {N.}~\bibnamefont {Xu}}, \bibinfo {author} {\bibfnamefont {B.}~\bibnamefont {Wang}}, \bibinfo {author} {\bibfnamefont {Q.-X.}\ \bibnamefont {Liu}},\ and\ \bibinfo {author} {\bibfnamefont {C.}~\bibnamefont {Xu}},\ }\bibfield  {title} {\bibinfo {title} {Causes and consequences of disordered hyperuniformity in global drylands},\ }\href {https://doi.org/10.1073/pnas.2504496122} {\bibfield  {journal} {\bibinfo  {journal} {Proceedings of the National Academy of Sciences}\ }\textbf {\bibinfo {volume} {122}},\ \bibinfo {pages} {e2504496122} (\bibinfo {year} {2025})}\BibitemShut {NoStop}%
\bibitem [{\citenamefont {Hou}\ \emph {et~al.}(2026)\citenamefont {Hou}, \citenamefont {Zhang}, \citenamefont {Sun},\ and\ \citenamefont {Jin}}]{houVegetationPatternsStructures2026}%
  \BibitemOpen
  \bibfield  {author} {\bibinfo {author} {\bibfnamefont {L.-F.}\ \bibnamefont {Hou}}, \bibinfo {author} {\bibfnamefont {J.}~\bibnamefont {Zhang}}, \bibinfo {author} {\bibfnamefont {G.-Q.}\ \bibnamefont {Sun}},\ and\ \bibinfo {author} {\bibfnamefont {Z.}~\bibnamefont {Jin}},\ }\bibfield  {title} {\bibinfo {title} {Vegetation {{Patterns}}: {{Structures}} and {{Dynamics}}},\ }\href {https://doi.org/10.4208/csiam-ls.SO-2025-0024} {\bibfield  {journal} {\bibinfo  {journal} {CSIAM Transactions on Life Sciences}\ }\textbf {\bibinfo {volume} {2}},\ \bibinfo {pages} {91} (\bibinfo {year} {2026})}\BibitemShut {NoStop}%
\bibitem [{\citenamefont {Hexner}\ \emph {et~al.}(2017)\citenamefont {Hexner}, \citenamefont {Chaikin},\ and\ \citenamefont {Levine}}]{hexnerEnhancedHyperuniformityRandom2017}%
  \BibitemOpen
  \bibfield  {author} {\bibinfo {author} {\bibfnamefont {D.}~\bibnamefont {Hexner}}, \bibinfo {author} {\bibfnamefont {P.~M.}\ \bibnamefont {Chaikin}},\ and\ \bibinfo {author} {\bibfnamefont {D.}~\bibnamefont {Levine}},\ }\bibfield  {title} {\bibinfo {title} {Enhanced hyperuniformity from random reorganization},\ }\href {https://doi.org/10.1073/pnas.1619260114} {\bibfield  {journal} {\bibinfo  {journal} {Proceedings of the National Academy of Sciences}\ }\textbf {\bibinfo {volume} {114}},\ \bibinfo {pages} {4294} (\bibinfo {year} {2017})}\BibitemShut {NoStop}%
\bibitem [{\citenamefont {Hexner}\ and\ \citenamefont {Levine}(2015)}]{hexnerHyperuniformityCriticalAbsorbing2015a}%
  \BibitemOpen
  \bibfield  {author} {\bibinfo {author} {\bibfnamefont {D.}~\bibnamefont {Hexner}}\ and\ \bibinfo {author} {\bibfnamefont {D.}~\bibnamefont {Levine}},\ }\bibfield  {title} {\bibinfo {title} {Hyperuniformity of {{Critical Absorbing States}}},\ }\href {https://doi.org/10.1103/PhysRevLett.114.110602} {\bibfield  {journal} {\bibinfo  {journal} {Physical Review Letters}\ }\textbf {\bibinfo {volume} {114}},\ \bibinfo {pages} {110602} (\bibinfo {year} {2015})}\BibitemShut {NoStop}%
\bibitem [{\citenamefont {Ma}\ and\ \citenamefont {Torquato}(2019)}]{maHyperuniformityGeneralizedRandom2019}%
  \BibitemOpen
  \bibfield  {author} {\bibinfo {author} {\bibfnamefont {Z.}~\bibnamefont {Ma}}\ and\ \bibinfo {author} {\bibfnamefont {S.}~\bibnamefont {Torquato}},\ }\bibfield  {title} {\bibinfo {title} {Hyperuniformity of generalized random organization models},\ }\href {https://doi.org/10.1103/PhysRevE.99.022115} {\bibfield  {journal} {\bibinfo  {journal} {Physical Review E}\ }\textbf {\bibinfo {volume} {99}},\ \bibinfo {pages} {022115} (\bibinfo {year} {2019})}\BibitemShut {NoStop}%
\bibitem [{\citenamefont {Torquato}\ \emph {et~al.}(2015)\citenamefont {Torquato}, \citenamefont {Zhang},\ and\ \citenamefont {Stillinger}}]{torquatoEnsembleTheoryStealthy2015}%
  \BibitemOpen
  \bibfield  {author} {\bibinfo {author} {\bibfnamefont {S.}~\bibnamefont {Torquato}}, \bibinfo {author} {\bibfnamefont {G.}~\bibnamefont {Zhang}},\ and\ \bibinfo {author} {\bibfnamefont {F.~H.}\ \bibnamefont {Stillinger}},\ }\bibfield  {title} {\bibinfo {title} {Ensemble {{Theory}} for {{Stealthy Hyperuniform Disordered Ground States}}},\ }\href {https://doi.org/10.1103/PhysRevX.5.021020} {\bibfield  {journal} {\bibinfo  {journal} {Physical Review X}\ }\textbf {\bibinfo {volume} {5}},\ \bibinfo {pages} {021020} (\bibinfo {year} {2015})}\BibitemShut {NoStop}%
\bibitem [{\citenamefont {Zachary}\ and\ \citenamefont {Torquato}(2011)}]{zacharyAnomalousLocalCoordination2011}%
  \BibitemOpen
  \bibfield  {author} {\bibinfo {author} {\bibfnamefont {C.~E.}\ \bibnamefont {Zachary}}\ and\ \bibinfo {author} {\bibfnamefont {S.}~\bibnamefont {Torquato}},\ }\bibfield  {title} {\bibinfo {title} {Anomalous local coordination, density fluctuations, and void statistics in disordered hyperuniform many-particle ground states},\ }\href {https://doi.org/10.1103/PhysRevE.83.051133} {\bibfield  {journal} {\bibinfo  {journal} {Physical Review E}\ }\textbf {\bibinfo {volume} {83}},\ \bibinfo {pages} {051133} (\bibinfo {year} {2011})}\BibitemShut {NoStop}%
\bibitem [{\citenamefont {Ganguly}\ and\ \citenamefont {Sarkar}(2020)}]{gangulyGroundStatesHyperuniformity2020}%
  \BibitemOpen
  \bibfield  {author} {\bibinfo {author} {\bibfnamefont {S.}~\bibnamefont {Ganguly}}\ and\ \bibinfo {author} {\bibfnamefont {S.}~\bibnamefont {Sarkar}},\ }\bibfield  {title} {\bibinfo {title} {Ground states and hyperuniformity of the hierarchical {{Coulomb}} gas in all dimensions},\ }\href {https://doi.org/10.1007/s00440-019-00955-9} {\bibfield  {journal} {\bibinfo  {journal} {Probability Theory and Related Fields}\ }\textbf {\bibinfo {volume} {177}},\ \bibinfo {pages} {621} (\bibinfo {year} {2020})}\BibitemShut {NoStop}%
\bibitem [{\citenamefont {Torquato}\ and\ \citenamefont {Kim}(2025)}]{torquatoExistenceNonequilibriumGlasses2025}%
  \BibitemOpen
  \bibfield  {author} {\bibinfo {author} {\bibfnamefont {S.}~\bibnamefont {Torquato}}\ and\ \bibinfo {author} {\bibfnamefont {J.}~\bibnamefont {Kim}},\ }\bibfield  {title} {\bibinfo {title} {Existence of nonequilibrium glasses in the degenerate stealthy hyperuniform ground-state manifold},\ }\href {https://doi.org/10.1039/D5SM00379B} {\bibfield  {journal} {\bibinfo  {journal} {Soft Matter}\ }\textbf {\bibinfo {volume} {21}},\ \bibinfo {pages} {4898} (\bibinfo {year} {2025})}\BibitemShut {NoStop}%
\bibitem [{\citenamefont {Dawley}\ and\ \citenamefont {Torquato}(2025)}]{dawleyEvolutionVariousInitial2025}%
  \BibitemOpen
  \bibfield  {author} {\bibinfo {author} {\bibfnamefont {S.~J.}\ \bibnamefont {Dawley}}\ and\ \bibinfo {author} {\bibfnamefont {S.}~\bibnamefont {Torquato}},\ }\bibfield  {title} {\bibinfo {title} {Evolution of various initial many-particle configurations to disordered stealthy hyperuniform ground states},\ }\href {https://doi.org/10.1103/9wbz-h7t4} {\bibfield  {journal} {\bibinfo  {journal} {Physical Review E}\ }\textbf {\bibinfo {volume} {112}},\ \bibinfo {pages} {045310} (\bibinfo {year} {2025})}\BibitemShut {NoStop}%
\bibitem [{\citenamefont {Morse}\ \emph {et~al.}(2023)\citenamefont {Morse}, \citenamefont {Kim}, \citenamefont {Steinhardt},\ and\ \citenamefont {Torquato}}]{morseGeneratingLargeDisordered2023}%
  \BibitemOpen
  \bibfield  {author} {\bibinfo {author} {\bibfnamefont {P.~K.}\ \bibnamefont {Morse}}, \bibinfo {author} {\bibfnamefont {J.}~\bibnamefont {Kim}}, \bibinfo {author} {\bibfnamefont {P.~J.}\ \bibnamefont {Steinhardt}},\ and\ \bibinfo {author} {\bibfnamefont {S.}~\bibnamefont {Torquato}},\ }\bibfield  {title} {\bibinfo {title} {Generating large disordered stealthy hyperuniform systems with ultrahigh accuracy to determine their physical properties},\ }\href {https://doi.org/10.1103/PhysRevResearch.5.033190} {\bibfield  {journal} {\bibinfo  {journal} {Physical Review Research}\ }\textbf {\bibinfo {volume} {5}},\ \bibinfo {pages} {033190} (\bibinfo {year} {2023})}\BibitemShut {NoStop}%
\bibitem [{\citenamefont {Chen}\ \emph {et~al.}(2023)\citenamefont {Chen}, \citenamefont {Zhuang}, \citenamefont {Chen}, \citenamefont {Huang}, \citenamefont {Vlcek},\ and\ \citenamefont {Jiao}}]{chenDisorderedHyperuniformSolid2023}%
  \BibitemOpen
  \bibfield  {author} {\bibinfo {author} {\bibfnamefont {D.}~\bibnamefont {Chen}}, \bibinfo {author} {\bibfnamefont {H.}~\bibnamefont {Zhuang}}, \bibinfo {author} {\bibfnamefont {M.}~\bibnamefont {Chen}}, \bibinfo {author} {\bibfnamefont {P.~Y.}\ \bibnamefont {Huang}}, \bibinfo {author} {\bibfnamefont {V.}~\bibnamefont {Vlcek}},\ and\ \bibinfo {author} {\bibfnamefont {Y.}~\bibnamefont {Jiao}},\ }\bibfield  {title} {\bibinfo {title} {Disordered hyperuniform solid state materials},\ }\href {https://doi.org/10.1063/5.0137187} {\bibfield  {journal} {\bibinfo  {journal} {Applied Physics Reviews}\ }\textbf {\bibinfo {volume} {10}},\ \bibinfo {pages} {021310} (\bibinfo {year} {2023})}\BibitemShut {NoStop}%
\bibitem [{\citenamefont {Torquato}\ \emph {et~al.}(2008)\citenamefont {Torquato}, \citenamefont {Scardicchio},\ and\ \citenamefont {Zachary}}]{torquatoPointProcessesArbitrary2008a}%
  \BibitemOpen
  \bibfield  {author} {\bibinfo {author} {\bibfnamefont {S.}~\bibnamefont {Torquato}}, \bibinfo {author} {\bibfnamefont {A.}~\bibnamefont {Scardicchio}},\ and\ \bibinfo {author} {\bibfnamefont {C.~E.}\ \bibnamefont {Zachary}},\ }\bibfield  {title} {\bibinfo {title} {Point processes in arbitrary dimension from fermionic gases, random matrix theory, and number theory},\ }\href {https://doi.org/10.1088/1742-5468/2008/11/P11019} {\bibfield  {journal} {\bibinfo  {journal} {Journal of Statistical Mechanics: Theory and Experiment}\ }\textbf {\bibinfo {volume} {2008}},\ \bibinfo {pages} {P11019} (\bibinfo {year} {2008})}\BibitemShut {NoStop}%
\bibitem [{\citenamefont {Crowley}\ \emph {et~al.}(2019)\citenamefont {Crowley}, \citenamefont {Laumann},\ and\ \citenamefont {Gopalakrishnan}}]{crowleyQuantumCriticalityIsing2019}%
  \BibitemOpen
  \bibfield  {author} {\bibinfo {author} {\bibfnamefont {P.~J.~D.}\ \bibnamefont {Crowley}}, \bibinfo {author} {\bibfnamefont {C.~R.}\ \bibnamefont {Laumann}},\ and\ \bibinfo {author} {\bibfnamefont {S.}~\bibnamefont {Gopalakrishnan}},\ }\bibfield  {title} {\bibinfo {title} {Quantum criticality in {{Ising}} chains with random hyperuniform couplings},\ }\href {https://doi.org/10.1103/PhysRevB.100.134206} {\bibfield  {journal} {\bibinfo  {journal} {Physical Review B}\ }\textbf {\bibinfo {volume} {100}},\ \bibinfo {pages} {134206} (\bibinfo {year} {2019})}\BibitemShut {NoStop}%
\bibitem [{\citenamefont {Abreu}\ \emph {et~al.}(2017)\citenamefont {Abreu}, \citenamefont {Pereira}, \citenamefont {Romero},\ and\ \citenamefont {Torquato}}]{abreuWeylHeisenbergEnsemble2017}%
  \BibitemOpen
  \bibfield  {author} {\bibinfo {author} {\bibfnamefont {L.~D.}\ \bibnamefont {Abreu}}, \bibinfo {author} {\bibfnamefont {J.~M.}\ \bibnamefont {Pereira}}, \bibinfo {author} {\bibfnamefont {J.~L.}\ \bibnamefont {Romero}},\ and\ \bibinfo {author} {\bibfnamefont {S.}~\bibnamefont {Torquato}},\ }\bibfield  {title} {\bibinfo {title} {The {{Weyl}}--{{Heisenberg}} ensemble: Hyperuniformity and higher {{Landau}} levels},\ }\href {https://doi.org/10.1088/1742-5468/aa68a7} {\bibfield  {journal} {\bibinfo  {journal} {Journal of Statistical Mechanics: Theory and Experiment}\ }\textbf {\bibinfo {volume} {2017}},\ \bibinfo {pages} {043103} (\bibinfo {year} {2017})}\BibitemShut {NoStop}%
\bibitem [{\citenamefont {Wang}\ \emph {et~al.}(2024)\citenamefont {Wang}, \citenamefont {Samajdar},\ and\ \citenamefont {Torquato}}]{wangCorrelationsInteractingElectron2024}%
  \BibitemOpen
  \bibfield  {author} {\bibinfo {author} {\bibfnamefont {H.}~\bibnamefont {Wang}}, \bibinfo {author} {\bibfnamefont {R.}~\bibnamefont {Samajdar}},\ and\ \bibinfo {author} {\bibfnamefont {S.}~\bibnamefont {Torquato}},\ }\bibfield  {title} {\bibinfo {title} {Correlations in interacting electron liquids: {{Many-body}} statistics and hyperuniformity},\ }\href {https://doi.org/10.1103/PhysRevB.110.104201} {\bibfield  {journal} {\bibinfo  {journal} {Physical Review B}\ }\textbf {\bibinfo {volume} {110}},\ \bibinfo {pages} {104201} (\bibinfo {year} {2024})}\BibitemShut {NoStop}%
\bibitem [{\citenamefont {Chen}\ \emph {et~al.}(2025)\citenamefont {Chen}, \citenamefont {Samajdar}, \citenamefont {Jiao},\ and\ \citenamefont {Torquato}}]{chenAnomalousSuppressionLargescale2025}%
  \BibitemOpen
  \bibfield  {author} {\bibinfo {author} {\bibfnamefont {D.}~\bibnamefont {Chen}}, \bibinfo {author} {\bibfnamefont {R.}~\bibnamefont {Samajdar}}, \bibinfo {author} {\bibfnamefont {Y.}~\bibnamefont {Jiao}},\ and\ \bibinfo {author} {\bibfnamefont {S.}~\bibnamefont {Torquato}},\ }\bibfield  {title} {\bibinfo {title} {Anomalous suppression of large-scale density fluctuations in classical and quantum spin liquids},\ }\href {https://doi.org/10.1073/pnas.2416111122} {\bibfield  {journal} {\bibinfo  {journal} {Proceedings of the National Academy of Sciences}\ }\textbf {\bibinfo {volume} {122}},\ \bibinfo {pages} {e2416111122} (\bibinfo {year} {2025})}\BibitemShut {NoStop}%
\bibitem [{\citenamefont {Degl'Innocenti}\ \emph {et~al.}(2016)\citenamefont {Degl'Innocenti}, \citenamefont {Shah}, \citenamefont {Masini}, \citenamefont {Ronzani}, \citenamefont {Pitanti}, \citenamefont {Ren}, \citenamefont {Jessop}, \citenamefont {Tredicucci}, \citenamefont {Beere},\ and\ \citenamefont {Ritchie}}]{deglinnocentiHyperuniformDisorderedTerahertz2016}%
  \BibitemOpen
  \bibfield  {author} {\bibinfo {author} {\bibfnamefont {R.}~\bibnamefont {Degl'Innocenti}}, \bibinfo {author} {\bibfnamefont {Y.~D.}\ \bibnamefont {Shah}}, \bibinfo {author} {\bibfnamefont {L.}~\bibnamefont {Masini}}, \bibinfo {author} {\bibfnamefont {A.}~\bibnamefont {Ronzani}}, \bibinfo {author} {\bibfnamefont {A.}~\bibnamefont {Pitanti}}, \bibinfo {author} {\bibfnamefont {Y.}~\bibnamefont {Ren}}, \bibinfo {author} {\bibfnamefont {D.~S.}\ \bibnamefont {Jessop}}, \bibinfo {author} {\bibfnamefont {A.}~\bibnamefont {Tredicucci}}, \bibinfo {author} {\bibfnamefont {H.~E.}\ \bibnamefont {Beere}},\ and\ \bibinfo {author} {\bibfnamefont {D.~A.}\ \bibnamefont {Ritchie}},\ }\bibfield  {title} {\bibinfo {title} {Hyperuniform disordered terahertz quantum cascade laser},\ }\href {https://doi.org/10.1038/srep19325} {\bibfield  {journal} {\bibinfo  {journal} {Scientific Reports}\ }\textbf {\bibinfo {volume} {6}},\ \bibinfo {pages} {19325} (\bibinfo {year} {2016})}\BibitemShut {NoStop}%
\bibitem [{\citenamefont {Sakai}\ \emph {et~al.}(2022)\citenamefont {Sakai}, \citenamefont {Arita},\ and\ \citenamefont {Ohtsuki}}]{sakaiQuantumPhaseTransition2022}%
  \BibitemOpen
  \bibfield  {author} {\bibinfo {author} {\bibfnamefont {S.}~\bibnamefont {Sakai}}, \bibinfo {author} {\bibfnamefont {R.}~\bibnamefont {Arita}},\ and\ \bibinfo {author} {\bibfnamefont {T.}~\bibnamefont {Ohtsuki}},\ }\bibfield  {title} {\bibinfo {title} {Quantum phase transition between hyperuniform density distributions},\ }\href {https://doi.org/10.1103/PhysRevResearch.4.033241} {\bibfield  {journal} {\bibinfo  {journal} {Physical Review Research}\ }\textbf {\bibinfo {volume} {4}},\ \bibinfo {pages} {033241} (\bibinfo {year} {2022})}\BibitemShut {NoStop}%
\bibitem [{\citenamefont {Vanoni}\ \emph {et~al.}(2025)\citenamefont {Vanoni}, \citenamefont {Steinhardt},\ and\ \citenamefont {Torquato}}]{vanoniQuantifyingWhenHyperuniformity2025}%
  \BibitemOpen
  \bibfield  {author} {\bibinfo {author} {\bibfnamefont {C.}~\bibnamefont {Vanoni}}, \bibinfo {author} {\bibfnamefont {P.~J.}\ \bibnamefont {Steinhardt}},\ and\ \bibinfo {author} {\bibfnamefont {S.}~\bibnamefont {Torquato}},\ }\bibfield  {title} {\bibinfo {title} {Quantifying when hyperuniformity of a many-particle system leads to uniformity across length scales},\ }\href {https://doi.org/10.1103/511q-ltf1} {\bibfield  {journal} {\bibinfo  {journal} {Physical Review E}\ }\textbf {\bibinfo {volume} {112}},\ \bibinfo {pages} {044142} (\bibinfo {year} {2025})}\BibitemShut {NoStop}%
\bibitem [{\citenamefont {Mitra}\ \emph {et~al.}(1992)\citenamefont {Mitra}, \citenamefont {Sen}, \citenamefont {Schwartz},\ and\ \citenamefont {Le~Doussal}}]{mitraDiffusionPropagatorProbe1992a}%
  \BibitemOpen
  \bibfield  {author} {\bibinfo {author} {\bibfnamefont {P.~P.}\ \bibnamefont {Mitra}}, \bibinfo {author} {\bibfnamefont {P.~N.}\ \bibnamefont {Sen}}, \bibinfo {author} {\bibfnamefont {L.~M.}\ \bibnamefont {Schwartz}},\ and\ \bibinfo {author} {\bibfnamefont {P.}~\bibnamefont {Le~Doussal}},\ }\bibfield  {title} {\bibinfo {title} {Diffusion propagator as a probe of the structure of porous media},\ }\href {https://doi.org/10.1103/PhysRevLett.68.3555} {\bibfield  {journal} {\bibinfo  {journal} {Physical Review Letters}\ }\textbf {\bibinfo {volume} {68}},\ \bibinfo {pages} {3555} (\bibinfo {year} {1992})}\BibitemShut {NoStop}%
\bibitem [{\citenamefont {Wilken}\ \emph {et~al.}(2023)\citenamefont {Wilken}, \citenamefont {Chaderjian},\ and\ \citenamefont {Saleh}}]{wilkenSpatialOrganizationPhaseSeparated2023}%
  \BibitemOpen
  \bibfield  {author} {\bibinfo {author} {\bibfnamefont {S.}~\bibnamefont {Wilken}}, \bibinfo {author} {\bibfnamefont {A.}~\bibnamefont {Chaderjian}},\ and\ \bibinfo {author} {\bibfnamefont {O.~A.}\ \bibnamefont {Saleh}},\ }\bibfield  {title} {\bibinfo {title} {Spatial {{Organization}} of {{Phase-Separated DNA Droplets}}},\ }\href {https://doi.org/10.1103/PhysRevX.13.031014} {\bibfield  {journal} {\bibinfo  {journal} {Physical Review X}\ }\textbf {\bibinfo {volume} {13}},\ \bibinfo {pages} {031014} (\bibinfo {year} {2023})}\BibitemShut {NoStop}%
\bibitem [{\citenamefont {Skolnick}\ and\ \citenamefont {Torquato}(2023)}]{skolnickSimulatedDiffusionSpreadability2023}%
  \BibitemOpen
  \bibfield  {author} {\bibinfo {author} {\bibfnamefont {M.}~\bibnamefont {Skolnick}}\ and\ \bibinfo {author} {\bibfnamefont {S.}~\bibnamefont {Torquato}},\ }\bibfield  {title} {\bibinfo {title} {Simulated diffusion spreadability for characterizing the structure and transport properties of two-phase materials},\ }\href {https://doi.org/10.1016/j.actamat.2023.118857} {\bibfield  {journal} {\bibinfo  {journal} {Acta Materialia}\ }\textbf {\bibinfo {volume} {250}},\ \bibinfo {pages} {118857} (\bibinfo {year} {2023})}\BibitemShut {NoStop}%
\bibitem [{\citenamefont {Maher}\ \emph {et~al.}(2022)\citenamefont {Maher}, \citenamefont {Stillinger},\ and\ \citenamefont {Torquato}}]{maherCharacterizationVoidSpace2022}%
  \BibitemOpen
  \bibfield  {author} {\bibinfo {author} {\bibfnamefont {C.~E.}\ \bibnamefont {Maher}}, \bibinfo {author} {\bibfnamefont {F.~H.}\ \bibnamefont {Stillinger}},\ and\ \bibinfo {author} {\bibfnamefont {S.}~\bibnamefont {Torquato}},\ }\bibfield  {title} {\bibinfo {title} {Characterization of void space, large-scale structure, and transport properties of maximally random jammed packings of superballs},\ }\href {https://doi.org/10.1103/PhysRevMaterials.6.025603} {\bibfield  {journal} {\bibinfo  {journal} {Physical Review Materials}\ }\textbf {\bibinfo {volume} {6}},\ \bibinfo {pages} {025603} (\bibinfo {year} {2022})}\BibitemShut {NoStop}%
\bibitem [{\citenamefont {Maher}\ \emph {et~al.}(2023)\citenamefont {Maher}, \citenamefont {Jiao},\ and\ \citenamefont {Torquato}}]{maherHyperuniformityMaximallyRandom2023}%
  \BibitemOpen
  \bibfield  {author} {\bibinfo {author} {\bibfnamefont {C.~E.}\ \bibnamefont {Maher}}, \bibinfo {author} {\bibfnamefont {Y.}~\bibnamefont {Jiao}},\ and\ \bibinfo {author} {\bibfnamefont {S.}~\bibnamefont {Torquato}},\ }\bibfield  {title} {\bibinfo {title} {Hyperuniformity of maximally random jammed packings of hyperspheres across spatial dimensions},\ }\href {https://doi.org/10.1103/PhysRevE.108.064602} {\bibfield  {journal} {\bibinfo  {journal} {Physical Review E}\ }\textbf {\bibinfo {volume} {108}},\ \bibinfo {pages} {064602} (\bibinfo {year} {2023})}\BibitemShut {NoStop}%
\bibitem [{\citenamefont {Maher}\ and\ \citenamefont {Torquato}(2024)}]{maherHyperuniformityScalingMaximally2024a}%
  \BibitemOpen
  \bibfield  {author} {\bibinfo {author} {\bibfnamefont {C.~E.}\ \bibnamefont {Maher}}\ and\ \bibinfo {author} {\bibfnamefont {S.}~\bibnamefont {Torquato}},\ }\bibfield  {title} {\bibinfo {title} {Hyperuniformity scaling of maximally random jammed packings of two-dimensional binary disks},\ }\href {https://doi.org/10.1103/PhysRevE.110.064605} {\bibfield  {journal} {\bibinfo  {journal} {Physical Review E}\ }\textbf {\bibinfo {volume} {110}},\ \bibinfo {pages} {064605} (\bibinfo {year} {2024})}\BibitemShut {NoStop}%
\bibitem [{\citenamefont {Wiese}(2024)}]{wieseHyperuniformityMannaModel2024}%
  \BibitemOpen
  \bibfield  {author} {\bibinfo {author} {\bibfnamefont {K.~J.}\ \bibnamefont {Wiese}},\ }\bibfield  {title} {\bibinfo {title} {Hyperuniformity in the {{Manna Model}}, {{Conserved Directed Percolation}} and {{Depinning}}},\ }\href {https://doi.org/10.1103/PhysRevLett.133.067103} {\bibfield  {journal} {\bibinfo  {journal} {Physical Review Letters}\ }\textbf {\bibinfo {volume} {133}},\ \bibinfo {pages} {067103} (\bibinfo {year} {2024})}\BibitemShut {NoStop}%
\bibitem [{\citenamefont {{Hitin-Bialus}}\ \emph {et~al.}(2024)\citenamefont {{Hitin-Bialus}}, \citenamefont {Maher}, \citenamefont {Steinhardt},\ and\ \citenamefont {Torquato}}]{hitin-bialusHyperuniformityClassesQuasiperiodic2024}%
  \BibitemOpen
  \bibfield  {author} {\bibinfo {author} {\bibfnamefont {A.}~\bibnamefont {{Hitin-Bialus}}}, \bibinfo {author} {\bibfnamefont {C.~E.}\ \bibnamefont {Maher}}, \bibinfo {author} {\bibfnamefont {P.~J.}\ \bibnamefont {Steinhardt}},\ and\ \bibinfo {author} {\bibfnamefont {S.}~\bibnamefont {Torquato}},\ }\bibfield  {title} {\bibinfo {title} {Hyperuniformity classes of quasiperiodic tilings via diffusion spreadability},\ }\href {https://doi.org/10.1103/PhysRevE.109.064108} {\bibfield  {journal} {\bibinfo  {journal} {Physical Review E}\ }\textbf {\bibinfo {volume} {109}},\ \bibinfo {pages} {064108} (\bibinfo {year} {2024})}\BibitemShut {NoStop}%
\bibitem [{\citenamefont {Debye}\ and\ \citenamefont {Bueche}(1949)}]{debyeScatteringInhomogeneousSolid1949}%
  \BibitemOpen
  \bibfield  {author} {\bibinfo {author} {\bibfnamefont {P.}~\bibnamefont {Debye}}\ and\ \bibinfo {author} {\bibfnamefont {A.~M.}\ \bibnamefont {Bueche}},\ }\bibfield  {title} {\bibinfo {title} {Scattering by an {{Inhomogeneous Solid}}},\ }\href {https://doi.org/10.1063/1.1698419} {\bibfield  {journal} {\bibinfo  {journal} {Journal of Applied Physics}\ }\textbf {\bibinfo {volume} {20}},\ \bibinfo {pages} {518} (\bibinfo {year} {1949})}\BibitemShut {NoStop}%
\bibitem [{\citenamefont {Debye}\ \emph {et~al.}(1957)\citenamefont {Debye}, \citenamefont {Anderson},\ and\ \citenamefont {Brumberger}}]{debyeScatteringInhomogeneousSolid1957}%
  \BibitemOpen
  \bibfield  {author} {\bibinfo {author} {\bibfnamefont {P.}~\bibnamefont {Debye}}, \bibinfo {author} {\bibfnamefont {H.~R.}\ \bibnamefont {Anderson}, \bibfnamefont {Jr.}},\ and\ \bibinfo {author} {\bibfnamefont {H.}~\bibnamefont {Brumberger}},\ }\bibfield  {title} {\bibinfo {title} {Scattering by an {{Inhomogeneous Solid}}. {{II}}. {{The Correlation Function}} and {{Its Application}}},\ }\href {https://doi.org/10.1063/1.1722830} {\bibfield  {journal} {\bibinfo  {journal} {Journal of Applied Physics}\ }\textbf {\bibinfo {volume} {28}},\ \bibinfo {pages} {679} (\bibinfo {year} {1957})}\BibitemShut {NoStop}%
\bibitem [{\citenamefont {Torquato}(2016)}]{torquatoDisorderedHyperuniformHeterogeneous2016b}%
  \BibitemOpen
  \bibfield  {author} {\bibinfo {author} {\bibfnamefont {S.}~\bibnamefont {Torquato}},\ }\bibfield  {title} {\bibinfo {title} {Disordered hyperuniform heterogeneous materials},\ }\href {https://doi.org/10.1088/0953-8984/28/41/414012} {\bibfield  {journal} {\bibinfo  {journal} {Journal of Physics: Condensed Matter}\ }\textbf {\bibinfo {volume} {28}},\ \bibinfo {pages} {414012} (\bibinfo {year} {2016})}\BibitemShut {NoStop}%
\bibitem [{\citenamefont {Stanley}(1987)}]{stanleyIntroductionPhaseTransitions1987}%
  \BibitemOpen
  \bibfield  {author} {\bibinfo {author} {\bibfnamefont {H.~E. H.~E.}\ \bibnamefont {Stanley}},\ }\href@noop {} {\emph {\bibinfo {title} {Introduction to Phase Transitions and Critical Phenomena}}}\ (\bibinfo  {publisher} {New York : Oxford University Press},\ \bibinfo {year} {1987})\BibitemShut {NoStop}%
\bibitem [{\citenamefont {Binney}\ \emph {et~al.}(1992)\citenamefont {Binney}, \citenamefont {Dowrick}, \citenamefont {Fisher},\ and\ \citenamefont {Newman}}]{binneyTheoryCriticalPhenomena1992}%
  \BibitemOpen
  \bibfield  {author} {\bibinfo {author} {\bibfnamefont {J.~J.}\ \bibnamefont {Binney}}, \bibinfo {author} {\bibfnamefont {N.~J.}\ \bibnamefont {Dowrick}}, \bibinfo {author} {\bibfnamefont {A.~J.}\ \bibnamefont {Fisher}},\ and\ \bibinfo {author} {\bibfnamefont {M.~E.~J.}\ \bibnamefont {Newman}},\ }\href {https://doi.org/10.1093/oso/9780198513940.001.0001} {\emph {\bibinfo {title} {The {{Theory}} of {{Critical Phenomena}}: {{An Introduction}} to the {{Renormalization Group}}}}}\ (\bibinfo  {publisher} {Oxford University Press},\ \bibinfo {address} {Oxford, England},\ \bibinfo {year} {1992})\BibitemShut {NoStop}%
\bibitem [{\citenamefont {Torquato}\ and\ \citenamefont {Wang}(2022)}]{torquatoPreciseDeterminationPair2022}%
  \BibitemOpen
  \bibfield  {author} {\bibinfo {author} {\bibfnamefont {S.}~\bibnamefont {Torquato}}\ and\ \bibinfo {author} {\bibfnamefont {H.}~\bibnamefont {Wang}},\ }\bibfield  {title} {\bibinfo {title} {Precise determination of pair interactions from pair statistics of many-body systems in and out of equilibrium},\ }\href {https://doi.org/10.1103/PhysRevE.106.044122} {\bibfield  {journal} {\bibinfo  {journal} {Physical Review E}\ }\textbf {\bibinfo {volume} {106}},\ \bibinfo {pages} {044122} (\bibinfo {year} {2022})}\BibitemShut {NoStop}%
\bibitem [{\citenamefont {Hansen}\ and\ \citenamefont {McDonald}(2013)}]{hansenTheorySimpleLiquids2013}%
  \BibitemOpen
  \bibfield  {author} {\bibinfo {author} {\bibfnamefont {J.-P.}\ \bibnamefont {Hansen}}\ and\ \bibinfo {author} {\bibfnamefont {I.~R.}\ \bibnamefont {McDonald}},\ }\href {https://doi.org/10.1016/B978-0-12-387032-2.00013-1} {\emph {\bibinfo {title} {Theory of {{Simple Liquids}}}}}\ (\bibinfo  {publisher} {Academic Press},\ \bibinfo {address} {Oxford},\ \bibinfo {year} {2013})\BibitemShut {NoStop}%
\bibitem [{\citenamefont {H{\"o}rmander}(2003)}]{hormanderAnalysisLinearPartial2003}%
  \BibitemOpen
  \bibfield  {author} {\bibinfo {author} {\bibfnamefont {L.}~\bibnamefont {H{\"o}rmander}},\ }\href {https://doi.org/10.1007/978-3-642-61497-2} {\emph {\bibinfo {title} {The {{Analysis}} of {{Linear Partial Differential Operators I}}}}},\ Classics in {{Mathematics}}\ (\bibinfo  {publisher} {Springer Berlin Heidelberg},\ \bibinfo {address} {Berlin, Heidelberg},\ \bibinfo {year} {2003})\BibitemShut {NoStop}%
\bibitem [{\citenamefont {Kanwal}(2004)}]{kanwalGeneralizedFunctions2004}%
  \BibitemOpen
  \bibfield  {author} {\bibinfo {author} {\bibfnamefont {R.~P.}\ \bibnamefont {Kanwal}},\ }\href {https://doi.org/10.1007/978-0-8176-8174-6} {\emph {\bibinfo {title} {Generalized {{Functions}}}}}\ (\bibinfo  {publisher} {Birkh{\"a}user Boston},\ \bibinfo {address} {Boston, MA},\ \bibinfo {year} {2004})\BibitemShut {NoStop}%
\bibitem [{\citenamefont {Baker}(1975)}]{bakerEssentialsPadeApproximants1975}%
  \BibitemOpen
  \bibfield  {author} {\bibinfo {author} {\bibfnamefont {G.~A.~J.}\ \bibnamefont {Baker}},\ }\href@noop {} {\emph {\bibinfo {title} {Essentials of {{Pad\'e Approximants}}}}}\ (\bibinfo  {publisher} {Academic Press},\ \bibinfo {address} {New York},\ \bibinfo {year} {1975})\BibitemShut {NoStop}%
\bibitem [{\citenamefont {Shi}\ \emph {et~al.}(2025)\citenamefont {Shi}, \citenamefont {Jiao},\ and\ \citenamefont {Torquato}}]{shiThreedimensionalConstructionHyperuniform2025}%
  \BibitemOpen
  \bibfield  {author} {\bibinfo {author} {\bibfnamefont {W.}~\bibnamefont {Shi}}, \bibinfo {author} {\bibfnamefont {Y.}~\bibnamefont {Jiao}},\ and\ \bibinfo {author} {\bibfnamefont {S.}~\bibnamefont {Torquato}},\ }\bibfield  {title} {\bibinfo {title} {Three-dimensional construction of hyperuniform, nonhyperuniform, and antihyperuniform disordered heterogeneous materials and their transport properties via spectral density functions},\ }\href {https://doi.org/10.1103/PhysRevE.111.035310} {\bibfield  {journal} {\bibinfo  {journal} {Physical Review E}\ }\textbf {\bibinfo {volume} {111}},\ \bibinfo {pages} {035310} (\bibinfo {year} {2025})}\BibitemShut {NoStop}%
\bibitem [{\citenamefont {Zhang}\ \emph {et~al.}(2016)\citenamefont {Zhang}, \citenamefont {Stillinger},\ and\ \citenamefont {Torquato}}]{zhangTransportGeometricalTopological2016a}%
  \BibitemOpen
  \bibfield  {author} {\bibinfo {author} {\bibfnamefont {G.}~\bibnamefont {Zhang}}, \bibinfo {author} {\bibfnamefont {F.~H.}\ \bibnamefont {Stillinger}},\ and\ \bibinfo {author} {\bibfnamefont {S.}~\bibnamefont {Torquato}},\ }\bibfield  {title} {\bibinfo {title} {Transport, geometrical, and topological properties of stealthy disordered hyperuniform two-phase systems},\ }\href {https://doi.org/10.1063/1.4972862} {\bibfield  {journal} {\bibinfo  {journal} {The Journal of Chemical Physics}\ }\textbf {\bibinfo {volume} {145}},\ \bibinfo {pages} {244109} (\bibinfo {year} {2016})}\BibitemShut {NoStop}%
\bibitem [{\citenamefont {Torquato}(2009)}]{torquatoInverseOptimizationTechniques2009}%
  \BibitemOpen
  \bibfield  {author} {\bibinfo {author} {\bibfnamefont {S.}~\bibnamefont {Torquato}},\ }\bibfield  {title} {\bibinfo {title} {Inverse optimization techniques for targeted self-assembly},\ }\href {https://doi.org/10.1039/b814211b} {\bibfield  {journal} {\bibinfo  {journal} {Soft Matter}\ }\textbf {\bibinfo {volume} {5}},\ \bibinfo {pages} {1157} (\bibinfo {year} {2009})}\BibitemShut {NoStop}%
\bibitem [{\citenamefont {Bhushan}\ and\ \citenamefont {Caspers}(2017)}]{bhushanOverviewAdditiveManufacturing2017}%
  \BibitemOpen
  \bibfield  {author} {\bibinfo {author} {\bibfnamefont {B.}~\bibnamefont {Bhushan}}\ and\ \bibinfo {author} {\bibfnamefont {M.}~\bibnamefont {Caspers}},\ }\bibfield  {title} {\bibinfo {title} {An overview of additive manufacturing ({{3D}} printing) for microfabrication},\ }\href {https://doi.org/10.1007/s00542-017-3342-8} {\bibfield  {journal} {\bibinfo  {journal} {Microsystem Technologies}\ }\textbf {\bibinfo {volume} {23}},\ \bibinfo {pages} {1117} (\bibinfo {year} {2017})}\BibitemShut {NoStop}%
\bibitem [{\citenamefont {Shirazi}\ \emph {et~al.}(2015)\citenamefont {Shirazi}, \citenamefont {Gharehkhani}, \citenamefont {Mehrali}, \citenamefont {Yarmand}, \citenamefont {Metselaar}, \citenamefont {Adib~Kadri},\ and\ \citenamefont {Osman}}]{shiraziReviewPowderbasedAdditive2015}%
  \BibitemOpen
  \bibfield  {author} {\bibinfo {author} {\bibfnamefont {S.~F.~S.}\ \bibnamefont {Shirazi}}, \bibinfo {author} {\bibfnamefont {S.}~\bibnamefont {Gharehkhani}}, \bibinfo {author} {\bibfnamefont {M.}~\bibnamefont {Mehrali}}, \bibinfo {author} {\bibfnamefont {H.}~\bibnamefont {Yarmand}}, \bibinfo {author} {\bibfnamefont {H.~S.~C.}\ \bibnamefont {Metselaar}}, \bibinfo {author} {\bibfnamefont {N.}~\bibnamefont {Adib~Kadri}},\ and\ \bibinfo {author} {\bibfnamefont {N.~A.~A.}\ \bibnamefont {Osman}},\ }\bibfield  {title} {\bibinfo {title} {A review on powder-based additive manufacturing for tissue engineering: Selective laser sintering and inkjet {{3D}} printing},\ }\href {https://doi.org/10.1088/1468-6996/16/3/033502} {\bibfield  {journal} {\bibinfo  {journal} {Science and Technology of Advanced Materials}\ }\textbf {\bibinfo {volume} {16}},\ \bibinfo {pages} {033502} (\bibinfo {year} {2015})}\BibitemShut {NoStop}%
\bibitem [{\citenamefont {Tumbleston}\ \emph {et~al.}(2015)\citenamefont {Tumbleston}, \citenamefont {Shirvanyants}, \citenamefont {Ermoshkin}, \citenamefont {Janusziewicz}, \citenamefont {Johnson}, \citenamefont {Kelly}, \citenamefont {Chen}, \citenamefont {Pinschmidt}, \citenamefont {Rolland}, \citenamefont {Ermoshkin}, \citenamefont {Samulski},\ and\ \citenamefont {DeSimone}}]{tumblestonContinuousLiquidInterface2015}%
  \BibitemOpen
  \bibfield  {author} {\bibinfo {author} {\bibfnamefont {J.~R.}\ \bibnamefont {Tumbleston}}, \bibinfo {author} {\bibfnamefont {D.}~\bibnamefont {Shirvanyants}}, \bibinfo {author} {\bibfnamefont {N.}~\bibnamefont {Ermoshkin}}, \bibinfo {author} {\bibfnamefont {R.}~\bibnamefont {Janusziewicz}}, \bibinfo {author} {\bibfnamefont {A.~R.}\ \bibnamefont {Johnson}}, \bibinfo {author} {\bibfnamefont {D.}~\bibnamefont {Kelly}}, \bibinfo {author} {\bibfnamefont {K.}~\bibnamefont {Chen}}, \bibinfo {author} {\bibfnamefont {R.}~\bibnamefont {Pinschmidt}}, \bibinfo {author} {\bibfnamefont {J.~P.}\ \bibnamefont {Rolland}}, \bibinfo {author} {\bibfnamefont {A.}~\bibnamefont {Ermoshkin}}, \bibinfo {author} {\bibfnamefont {E.~T.}\ \bibnamefont {Samulski}},\ and\ \bibinfo {author} {\bibfnamefont {J.~M.}\ \bibnamefont {DeSimone}},\ }\bibfield  {title} {\bibinfo {title} {Continuous liquid interface production of {{3D}} objects},\ }\href {https://doi.org/10.1126/science.aaa2397} {\bibfield  {journal} {\bibinfo  {journal} {Science}\ }\textbf {\bibinfo {volume} {347}},\ \bibinfo {pages} {1349} (\bibinfo {year} {2015})}\BibitemShut {NoStop}%
\bibitem [{\citenamefont {Wong}\ and\ \citenamefont {Hernandez}(2012)}]{wongReviewAdditiveManufacturing2012}%
  \BibitemOpen
  \bibfield  {author} {\bibinfo {author} {\bibfnamefont {K.~V.}\ \bibnamefont {Wong}}\ and\ \bibinfo {author} {\bibfnamefont {A.}~\bibnamefont {Hernandez}},\ }\bibfield  {title} {\bibinfo {title} {A {{Review}} of {{Additive Manufacturing}}},\ }\href {https://doi.org/10.5402/2012/208760} {\bibfield  {journal} {\bibinfo  {journal} {International Scholarly Research Notices}\ }\textbf {\bibinfo {volume} {2012}},\ \bibinfo {pages} {208760} (\bibinfo {year} {2012})}\BibitemShut {NoStop}%
\bibitem [{\citenamefont {Chen}\ \emph {et~al.}(2018)\citenamefont {Chen}, \citenamefont {Lomba},\ and\ \citenamefont {Torquato}}]{chenBinaryMixturesCharged2018}%
  \BibitemOpen
  \bibfield  {author} {\bibinfo {author} {\bibfnamefont {D.}~\bibnamefont {Chen}}, \bibinfo {author} {\bibfnamefont {E.}~\bibnamefont {Lomba}},\ and\ \bibinfo {author} {\bibfnamefont {S.}~\bibnamefont {Torquato}},\ }\bibfield  {title} {\bibinfo {title} {Binary mixtures of charged colloids: A potential route to synthesize disordered hyperuniform materials},\ }\href {https://doi.org/10.1039/C8CP02616E} {\bibfield  {journal} {\bibinfo  {journal} {Physical Chemistry Chemical Physics}\ }\textbf {\bibinfo {volume} {20}},\ \bibinfo {pages} {17557} (\bibinfo {year} {2018})}\BibitemShut {NoStop}%
\bibitem [{\citenamefont {Ma}\ \emph {et~al.}(2020)\citenamefont {Ma}, \citenamefont {Lomba},\ and\ \citenamefont {Torquato}}]{maOptimizedLargeHyperuniform2020}%
  \BibitemOpen
  \bibfield  {author} {\bibinfo {author} {\bibfnamefont {Z.}~\bibnamefont {Ma}}, \bibinfo {author} {\bibfnamefont {E.}~\bibnamefont {Lomba}},\ and\ \bibinfo {author} {\bibfnamefont {S.}~\bibnamefont {Torquato}},\ }\bibfield  {title} {\bibinfo {title} {Optimized {{Large Hyperuniform Binary Colloidal Suspensions}} in {{Two Dimensions}}},\ }\href {https://doi.org/10.1103/PhysRevLett.125.068002} {\bibfield  {journal} {\bibinfo  {journal} {Physical Review Letters}\ }\textbf {\bibinfo {volume} {125}},\ \bibinfo {pages} {068002} (\bibinfo {year} {2020})}\BibitemShut {NoStop}%
\bibitem [{\citenamefont {Yeong}\ and\ \citenamefont {Torquato}(1998)}]{yeongReconstructingRandomMedia1998a}%
  \BibitemOpen
  \bibfield  {author} {\bibinfo {author} {\bibfnamefont {C.~L.~Y.}\ \bibnamefont {Yeong}}\ and\ \bibinfo {author} {\bibfnamefont {S.}~\bibnamefont {Torquato}},\ }\bibfield  {title} {\bibinfo {title} {Reconstructing random media},\ }\href {https://doi.org/10.1103/PhysRevE.57.495} {\bibfield  {journal} {\bibinfo  {journal} {Physical Review E}\ }\textbf {\bibinfo {volume} {57}},\ \bibinfo {pages} {495} (\bibinfo {year} {1998})}\BibitemShut {NoStop}%
\end{thebibliography}
\end{document}